\documentclass[12pt,a4paper]{article}
\pdfoutput=1

\usepackage{enumerate}
\usepackage{graphicx,array}
\usepackage{hyperref}
\usepackage[normalem]{ulem} 
\usepackage{cite}
\usepackage{color}
\usepackage{appendix}
\usepackage{epsfig}
\usepackage{latexsym}
\usepackage{amsmath}
\usepackage{amssymb}
\usepackage{bm}
\usepackage{hyperref}
\usepackage{slashed}
\usepackage{bbold}
\usepackage{subfig}
\usepackage{multirow}
\usepackage{adjustbox}
\usepackage{colortbl}

\numberwithin{equation}{section}

\definecolor{darkblue}{cmyk}{1,0.3,0,0.2}
\definecolor{violet}{cmyk}{0,1,0,0.2}
\definecolor{orange}{rgb}{0.8,0.5,0.1}
\definecolor{gray}{rgb}{0.4,0.4,0.4}

\hypersetup{colorlinks, bookmarksnumbered, citecolor=darkblue, linkcolor=darkblue, pdfstartview=FitH, urlcolor=darkblue, linktocpage}

\newcommand{\be}{\begin{equation}}
\newcommand{\ee}{\end{equation}}
\newcommand{\bea}{\begin{eqnarray}}
\newcommand{\eea}{\end{eqnarray}}

\newcommand{\SU}{\textrm{SU}}

\newcommand{\gsim}{\lower.7ex\hbox{$\;\stackrel{\textstyle>}{\sim}\;$}}
\newcommand{\lsim}{\lower.7ex\hbox{$\;\stackrel{\textstyle<}{\sim}\;$}}

\newcommand{\LL}{\mathcal{L}}

\newcommand{\OO}{\mathcal{O}}

\def\lra#1{\overset{\text{\scriptsize$\leftrightarrow$}}{#1}}

\newcommand{\Dslash}{\!\not\!\!  D}

\newcommand{\ba} {\begin{eqnarray}}
\newcommand{\ea} {\end{eqnarray}}

\newcommand{\Dlr}{\stackrel{\leftrightarrow}{D}}

\begin{document}
 
 \hfill

\begin{flushright}
\hspace{3cm} 
SISSA 04/2020/FISI\\
TUM-HEP-1258-20
\end{flushright}

\vspace{1.0cm}

\begin{center}
{\LARGE\bf Matching scalar leptoquarks \\[0.4cm] to the SMEFT at one loop}
\\ 

\bigskip\vspace{1cm}{
{\large \mbox{Valerio Gherardi$^{a, b}$}, \mbox{David Marzocca$^{b}$}, \mbox{Elena Venturini$^{c}$} }
} \\[7mm]
{\em $(a)$ SISSA, Via Bonomea 265, 34136, Trieste, Italy}  \\ 
{\em $(b)$ INFN, Sezione di Trieste, SISSA, Via Bonomea 265, 34136, Trieste, Italy}  \\ 
{\em $(c)$ Technische Universit{\"a}t M{\"u}nchen, Physik-Department, 85748 Garching, Germany}  \\ 

\vspace*{0.5cm}
   
\end{center}
\vspace*{1.5cm}

\centerline{\large\bf Abstract}
\medskip\noindent 

In this paper we present the complete one-loop matching conditions, up to dimension-six operators of the Standard Model effective field theory, resulting by integrating out the two scalar leptoquarks $S_{1}\sim (\bar{\bf  3}, {\bf 1})_{\frac{1}{3}}$ and $S_{3}\sim ( \bar {\bf 3}, {\bf 3})_{\frac{1}{3}}$.
This allows a phenomenological study of low-energy constraints on this model at one-loop accuracy, which will be the focus of a subsequent work. Furthermore, it provides a rich comparison for functional and computational methods for one-loop matching, that are being developed.
As a corollary result, we derive a complete set of dimension-six operators independent under integration by parts, but not under equations of motions, called \emph{Green's basis}, as well as the complete reduction formulae from this set to the Warsaw basis. 

\vspace{0.3cm}

\newpage
\tableofcontents

\vspace{0.5cm}

\section{Introduction}

The Standard Model (SM) of Particle Physics still represents the best description of high-energy phenomena at our disposal. Nevertheless, several experimental and theoretical shortcomings require the presence of some New Physics (NP).
The lack of a NP discovery at the LHC and the SM's impressive phenomenological success could suggest the existence of a large separation between the Electroweak (EW) and NP scales. If this is the case, the SM should really be understood as the leading (renormalizable) approximation of a low energy Effective Field Theory (EFT), known as SMEFT \cite{Buchmuller:1985jz,Giudice:2007fh,Grzadkowski:2010es}.

Considering the SM as an EFT is not only conceptually, but also practically advantageous, for it allows to study a whole class of SM extensions in a model independent way. In fact, in the SMEFT framework, the effects of heavy NP are fully encoded in the Wilson Coefficients (WCs) of non-renormalizable operators, in terms of which low-energy observables can be computed without any reference to the specific ultraviolet (UV) model. The WCs for a given concrete UV SM extension can then be obtained by \textit{matching}.

In this work we present the complete one-loop matching conditions, up to dimension-six SMEFT operators, resulting by integrating out the two scalar leptoquarks (LQ) $S_1$ and $S_3$ \cite{Dorsner:2016wpm} with all admissible baryon and lepton number conserving couplings.
From the phenomenological point of view, such an effort is motivated by the recent interest received by the model in the context of the deviations from the SM observed in $B$-meson decays, for which it could provide a combined explanation \cite{Bauer:2015knc,Buttazzo:2017ixm,Crivellin:2017zlb,Marzocca:2018wcf,Arnan:2019olv,Yan:2019hpm,Bigaran:2019bqv,Crivellin:2019dwb}. The complete one-loop matching allows a thourough study of the model's phenomenology, which was indeed one of our initial goals, and will be reported in a separate contribution \cite{Gherardi:2020xyz}.

Aside from the phenomenological interest, this work represents one of the very few available examples of complete one-loop matching to the SMEFT. In \cite{Huo:2015nka,Wells:2017vla} the one-loop matching for bosonic SMEFT operators from integrating out sfermions in the MSSM is derived, Refs.~\cite{Jiang:2018pbd,Haisch:2020ahr} perform the complete one-loop matching for a singlet scalar (see also \cite{Boggia:2016asg}), and \cite{Chala:2020vqp} considers the SM with an additional light sterile neutrino and heavy fermions and a scalar singlet.
The model considered here, with two coloured and weakly-charged states coupled to all SM particles with non-trivial flavour structures, represents a very rich example of such a matching.
While functional \cite{Henning:2014wua,Drozd:2015rsp,Ellis:2016enq,Henning:2016lyp,Fuentes-Martin:2016uol,Zhang:2016pja,Ellis:2017jns,Kramer:2019fwz,Cohen:2019btp} and automated \cite{delAguila:2016zcb,Brivio:2019irc} methods for the task are currently under development, diagrammatic calculation is still the state of the art in this subject, and will represent an important cross-check when these more sophisticated methods will be available.

The matching conditions are obtained by equating EFT and UV theory one-light-particle irreducible (1LPI) off-shell Green's functions at the matching scale. This operation produces a set of operators which are independent under integration by parts (IBP), but possibly redundant under the SM renormalizable equations of motion (EOMs). A complete set of such operators has been called \emph{Green's} basis in \cite{Jiang:2018pbd}, and must then be suitably reduced to an operator basis for $S$-matrix elements by applying the SM EOMs (or field redefinitions).
As a byproduct of our work, we identify a complete Green's basis  of dimension-six SMEFT operators by extending the \emph{Warsaw} basis \cite{Grzadkowski:2010es}, and obtain the fully general reduction equations expressing Warsaw basis WCs in terms of Green's basis ones.

The paper is organized as follows: in Sec.~\ref{sec:model} we introduce the $S_1+S_3$ model; in Sec.~\ref{sec:WarsawMatching} we give the complete one-loop matching conditions in the Warsaw basis, which is the main result of this paper. We conclude in Sec.~\ref{sec:conclusions}.
Several results are contained in the appendices: in App.~\ref{app:GreenBasis} we discuss in full generality the Green's basis for the SMEFT; in App.~\ref{app:GreenToWarsaw} we provide the reduction equations from the Green's to the Warsaw basis, and in App.~\ref{app:MatchToGreen} we give the complete one-loop matching condition for the leptoquark theory in the Green's basis.

In the supplementary material we provide the complete one-loop matching in the Green's basis, the general reduction equations from the Green's to the Warsaw basis, as well as usage examples.

\section{The $S_1+S_3$ model}
\label{sec:model}

The UV model under consideration is defined by the SM gauge group and field content, with the addition of two colored scalar \emph{leptoquarks} 
\be
	S_{1}\sim (\bar{\bf  3}, {\bf 1})_{\frac{1}{3}} \quad \text{and} \quad
	S_{3}\sim ( \bar {\bf 3}, {\bf 3})_{\frac{1}{3}}~,
\ee
where in parenthesis we indicate the representation under $(\SU(3)_c, \SU(2)_L)_{U(1)_Y}$.
The combination of these two scalars has been considered in the literature as a possible simultaneous explanation of charged and neutral current $B$-anomalies \cite{Buttazzo:2017ixm,Crivellin:2017zlb,Marzocca:2018wcf,Arnan:2019olv,Yan:2019hpm,Bigaran:2019bqv,Crivellin:2019dwb}.
For such a purpose, both leptoquarks need to have $\text{TeV}$ scale masses and, consequently, negligibly small baryon number violating couplings. Enforcing baryon number conservation, the part of the Lagrangian involving $S_{1,3}$ is:
\be\begin{split}
	\LL_{\text{LQ}} &= |D_\mu S_1|^2 + |D_\mu S_3|^2 - M_{1}^2 |S_1|^2 - M_{3} ^2 |S_3|^2 + \\
		& + \left( (\lambda^{1L})_{i\alpha} \bar q^c _i \epsilon \ell _\alpha
			+ (\lambda^{1R})_{i\alpha} \bar u^c _i   e _\alpha  \right) S_1 
			+ (\lambda^{3L})_{i\alpha}\bar q^c _i \epsilon \sigma^I \ell _\alpha S_3^I + \text{h.c.} + \\
		& - \lambda_{H1} |H|^2 |S_1|^2 - \lambda_{H3} |H|^2 |S_3^I|^2 - \left(\lambda_{H13} (H^\dagger \sigma^I H) S_3^{I \dagger} S_1 + \text{h.c.}\right) + \\
		& - \lambda_{\epsilon H 3} i \epsilon^{IJK} (H^\dagger \sigma^I H) S_3^{J \dagger} S_3^{K}  -V_{S}(S_1, S_3),
\end{split}\ee
where $\epsilon=i\sigma _2$, $\lambda_{H1}, \lambda_{H3}, \lambda_{\epsilon H 3} \in \mathbb{R}$, $(\lambda^{1L})_{i\alpha}, (\lambda^{1R})_{i\alpha}, (\lambda^{3L})_{i\alpha}, \lambda_{H13} \in \mathbb{C}$, and the LQ self-interactions are described by
\be\begin{split}
V_{S}(S_1, S_3) & =\frac{c_{1}}{2}(S_{1}^{\dagger}S_{1})^{2}+\\
 & +c_{13}^{(1)}(S_{1}^{\dagger}S_{1})(S_{3}^{\dagger}S_{3})+c_{13}^{(8)}(S_{1}^{\dagger}T^{A}S_{1})(S_{3}^{\dagger}T^{A}S_{3})+\\
 & +\frac{c_{3}^{(1)}}{2}(S_{3}^{\dagger}S_{3})(S_{3}^{\dagger}S_{3})+\frac{c_{3}^{(3)}}{2}(S_{3}^{I\dagger}\epsilon^{IJK}S_{3}^{J})(S_{3}^{L\dagger}\epsilon^{LMK}S_{3}^{M})+\\
 & +\frac{c_{3}^{(5)}}{2}\left[\frac{(S_{3}^{I\dagger}S_{3}^{J})(S_{3}^{I\dagger}S_{3}^{J})+(S_{3}^{I\dagger}S_{3}^{J})(S_{3}^{J\dagger}S_{3}^{I})}{2}-\frac{1}{3}(S_{3}^{\dagger}S_{3})(S_{3}^{\dagger}S_{3})\right]~.
 	\label{eq:LQpotential}
\end{split}\ee

We denote SM quark and lepton fields by $q_i$, $u_i$, $d_i$, $\ell _\alpha$, and $e_\alpha$. We adopt latin letters ($i,\,j,\,k,\,\dots$) for quark flavor indices and greek letters ($\alpha,\,\beta,\,\gamma,\,\dots$) for lepton flavor indices.
We work in the down-quark and charged-lepton mass eigenstate basis, where
\be
	q_i = \left( \begin{array}{c} V^*_{ji} u^j_L \\ d^i_L  \end{array}  \right)\,, \qquad
	\ell_\alpha = \left( \begin{array}{c} \nu^\alpha_L \\ e^\alpha_L \end{array} \right)~,
\ee
and $V$ is the CKM matrix.
The Higgs field is denoted by $H$ and its hypercharge is normalized to $Y_H = \frac{1}{2}$. The covariant derivative of a generic field $\Phi\sim (\rho ^\Phi _3, \rho ^\Phi _2)_{Y_\Phi}$ is defined by
\be
D_\mu \Phi  = \left(\partial_\mu   - i g^\prime Y_\Phi B_\mu  -ig (t^\Phi _2) ^I W ^I _\mu -i g_s (t^\Phi _3 )^A G^A _\mu \right)\Phi,
\ee
and the corresponding field strengths read
\be\begin{split}
B_{\mu\nu} & =\partial_{\mu}B_{\nu}-\partial_{\nu}B_{\mu},\\
W_{\mu\nu}^{I} & =\partial_{\mu}W_{\nu}^{I}-\partial_{\nu}W_{\mu}^{I}+g\epsilon^{IJK}W_{\mu}^{J}W_{\nu}^{K},\\
G_{\mu\nu}^{A} & =\partial_{\mu}G_{\nu}^{A}-\partial_{\nu}G_{\mu}^{A}+g_{s}f^{ABC}G_{\mu}^{B}G_{\nu}^{C}.
\end{split}\ee
{
Dual field strengths are defined by:
\be 
\widetilde{X}_{\mu \nu} = \frac{1}{2} \varepsilon _{\mu \nu \rho \sigma} X ^{\rho \sigma}, 
\ee
where $\varepsilon$ is the Levi-Civita tensor, with $\varepsilon _{0123} = +1$.
}

The SM Yukawa lagrangian is defined by
\be
	\LL_{\rm Yuk} = -(y_E)_{\alpha\beta} \bar \ell_\alpha e_\beta H -(y_U)_{ij} \bar q_i u_j \widetilde H - (y_D)_{ij} \bar q_i d_j  H + \text{h.c.},
\ee
where $\widetilde H = i \sigma_2 H^*$, and the Higgs potential reads
\be V_{H}=-m^2 H^\dagger H+\frac{\lambda}{2}(H^\dagger H)^2. \ee
Finally, for future convenience, we define the bi-lateral derivatives:
\be\begin{split}
H^\dagger\overleftrightarrow{D}_\mu H & = H^\dagger (D_\mu H) - (D_\mu H)^\dagger H,\\ 
H^\dagger \overleftrightarrow{D}^I_\mu H & = H^\dagger \sigma^I (D_\mu H) - (D_\mu H)^\dagger \sigma^I H. 
\label{eq:Higgs currents}
\end{split}\ee
%

\subsection{Tree-level SMEFT matching conditions}

Since the two extra scalar fields are, by assumption, heavier than the electroweak scale, for the purpose of low energy phenomenology, we can integrate them out and work instead with a non-renormalizable SMEFT lagrangian.
This takes the form:
\be
	\LL_{\rm SMEFT} = \LL_{\rm SM} + \sum_i C_i \OO^{(6)}_i + \ldots ~,
\ee
where $\OO^{(6)}_i$ are dimension-six SMEFT operators, while the dots denote higher-dimension operators, which we neglect in the following.
We adopt the Warsaw basis \cite{Grzadkowski:2010es} for dimension-six operators.
Separating the contributions arising at tree level from the one-loop generated ones, we can write
\be
	C_i = C_i^{(0)} + \frac{1}{(4\pi)^2} C_i^{(1)} ~.
	\label{eq:Ci}
\ee
At tree level, only a set of semi-leptonic operators is generated, with WCs:
\be\begin{split}
	[C_{l q}^{(1)}]_{\alpha \beta i j}^{(0)} &= \frac{\lambda^{1L *}_{i \alpha} \lambda^{1L}_{j\beta}}{4 M_1^2} +  \frac{3 \lambda^{3L*}_{i \alpha} \lambda^{3L}_{j\beta}}{4 M_3^2} ~, \\
	[C_{l q}^{(3)}]_{\alpha \beta i j}^{(0)} &= - \frac{\lambda^{1L*}_{i \alpha} \lambda^{1L}_{j\beta}}{4 M_1^2} +  \frac{\lambda^{3L *}_{i \alpha} \lambda^{3L}_{j\beta}}{4 M_3^2}  ~, \\
	[C_{l equ}^{(1)}]_{\alpha \beta i j}^{(0)} &= \frac{\lambda^{1R}_{j\beta} \lambda_{i\alpha}^{1L*}}{2 M_1^2} ~, \\
	[C_{l equ}^{(3)}]_{\alpha \beta i j}^{(0)} &= - \frac{\lambda^{1R}_{j\beta} \lambda_{i\alpha}^{1L*}}{8 M_1^2} ~, \\
	[C_{eu}]_{\alpha \beta i j}^{(0)} &=  \frac{\lambda_{i\alpha}^{1R\,*} \lambda^{1R}_{j\beta}}{2 M_1^2}. ~
\label{eq:EFTS13treematch}
\end{split}\ee
For the operator definitions, see Tables~\ref{tab:GreenBosonic}, \ref{tab:GreenSingleFermion}, \ref{tab:4FermionBconserving} and \ref{tab:4FermionBviolating}.

\section{Complete one-loop SMEFT matching conditions}
\label{sec:WarsawMatching}

We report in this Section the complete one-loop SMEFT matching conditions for the $S_1 + S_3$ model introduced in the previous Section. 

The matching is performed diagrammatically, by equating 1LPI Green's functions in the UV and effective theory. As explained in the Introduction, this gives rise to a set of higher-dimensional effective operators in the Green's basis, which are then reduced to a minimal set of Warsaw basis operators by applying the SM equations of motion. This reduction is performed in full generality in App.~\ref{app:GreenToWarsaw}, following the complete classification of Green's basis operators in App.~\ref{app:GreenBasis}.
The one-loop matching to the Green's basis WCs for the $S_{1}+S_{3}$ model is given in App.~\ref{app:MatchToGreen}. Here we report the matching conditions in the Warsaw basis, which is the central result of the paper.

We performed our computations in a general $R_\xi$ gauge, adopting the $\overline{\text{MS}}$ subtraction scheme within Naive Dimensional Regularization (NDR). Whenever relevant, we explicitly checked the independence of matching conditions on the gauge fixing parameter $\xi$. The matching scale is $\mu_M$, which in practical applications should be taken of order of the leptoquark masses $\mu_M \sim M_{1}, M_{3}$, but is otherwise arbitrary. For physics to be independent of $\mu _M$, the resulting dependence of WCs from the matching scale should exactly correspond to their SMEFT renormalization group (RG) running \cite{Jenkins:2013zja,Jenkins:2013wua,Alonso:2013hga}, which we explicitly verified as a cross-check of our procedure.
In practice, for all the operators that are not already generated at the tree-level, the explicit (logarithmic) scale dependence we obtain from the matching computation corresponds to the one from the SMEFT RG equations.
For the operators listed in Eq.~\eqref{eq:EFTS13treematch}, instead, one should also consider the running of the leptoquark couplings and masses with $\mu _M$, schematically $[C_i]^{(0)}(\mu_M) \sim \lambda(\mu_M) \lambda(\mu_M) / M^2_{\rm LQ}(\mu_M)$. For instance, in the one-loop matching of these operators, there is a contribution from the quartic leptoquark couplings of Eq. \eqref{eq:LQpotential}: the logarithmic scale dependence of such contribution cancels the one arising from the RG evolution of the LQ masses in the tree-level matching.

In contrast to the aforementioned gauge fixing and matching scale independence, matching conditions may (and do) explicitly depend on the definition of evanescent operators \cite{Herrlich:1994kh}, \textit{i.e.} operators which vanish in $d=4$, but may be non-vanishing in $d\neq 4$. Two examples of evanescent Dirac structures relevant to us are: 
\begin{align}
\mathcal{E}_{1} & =P_{L}\gamma^{\mu}\gamma^{\nu}P_{L}\otimes P_{L}\gamma_{\mu}\gamma_{\nu}P_{L}-4P_{L}\otimes P_{L}+P_{L}\sigma^{\mu\nu}P_{L}\otimes P_{L}\sigma_{\mu\nu}P_{L},\\
\mathcal{E}_{2} & =P_{L}\gamma^{\mu}\gamma^{\nu}P_{L}\otimes P_{R}\gamma_{\mu}\gamma_{\nu}P_{R}-4P_{L}\otimes P_{R}.
\end{align}
The NDR defining equations in $d=4-2\epsilon$ dimensions, \textit{viz.}
\be
\left\lbrace \gamma ^\mu ,\gamma ^\nu \right\rbrace=2 \eta ^{\mu \nu}, \qquad \left\lbrace \gamma ^\mu ,\gamma ^5 \right\rbrace=0, \qquad \text {Tr}(\eta)=4-2\epsilon
\ee
imply $\mathcal E _1 =-2\epsilon P_L \otimes P_L$, but do not univocally fix $\mathcal E _2$. Following the notation of Ref. \cite{Dekens:2019ept}, we write: 
\be
\mathcal{E}_{2}=4a_{\text{ev}}\epsilon P_{L}\otimes P_{R}+E_{LR}^{(2)}(a_{\text{ev}}),
\ee
where the coefficient $a_{\text {ev}}$ can be regarded as the definition of the evanescent operator $E^{(2)}_{LR}$ (\textit{e.g.} for $a_{\text {ev}}=-1/2$ one gets $E_{LR}^{(2)}=P_{L}\sigma^{\mu\nu}P_{L}\otimes P_{R}\sigma_{\mu\nu}P_{R}$, which vanishes in four dimensions, cf. Ref. \cite{Dekens:2019ept}).

In order to facilitate result comparisons, we report the matching conditions for general $a_{\text {ev}}$ (the other scheme defining coefficients, $b_\text{ev}$, $c_\text{ev}$, etc., of Ref. \cite{Dekens:2019ept} do not enter in our one-loop computations). For practical calculations, Ref. \cite{buras_2020} recommends $a_{\text {ev}}=b_{\text {ev}}=\cdots = 1$, as in such scheme evanescent operators only affect two-loop anomalous dimensions.

We treat the Higgs mass term $m^2 H^\dagger H$ as an interaction (both in the SMEFT and UV theory) and work with a massless Higgs field propagator. By dimensional analysis, a diagram with internal Higgs lines and $n$ insertions of $m^2$ is suppressed by a factor $(m^2/M^2)^n$ (where $M^2=M^2_{1,3}$) relative to the same diagram with no insertions. Therefore, at dimension-six level, mass insertions can be relevant to the matching conditions for renormalizable operators (see below). However, in the present theory, one-loop diagrams with internal Higgs lines only give rise to dimension-six operators, so that $m^2$ does not contribute to the Green's basis matching conditions. It does, instead, contribute to the Warsaw basis matching conditions, where it makes its appearence through the Higgs EOM, see Eq. \eqref{eq:SMEOM}.

As a further check, we have also recomputed the one-loop Green's basis WCs of pure-Higgs operators belonging to classes $H^4 D^2$ and $H^6$ (see Table \ref{tab:GreenBosonic}) within the universal one-loop effective action (UOLEA) approach \cite{Drozd:2015rsp,Ellis:2016enq,Ellis:2017jns}, and we find agreement with our diagrammatic results.

Integrating out the leptoquarks at one loop also generates contributions to SM renormalizable operators and, in particular, fermion kinetic terms. Such modifications can be undone by suitable field and SM coupling redefinitions, which however also introduce additional contributions to tree-level generated WCs\footnote{Since field redefinitions arise at one loop in our model, only tree-level WCs are affected. In general, any tree-level shift in SM couplings and wave-function renormalizations that could influence loop-generated coefficients should be taken into account, see e.g. \cite{Jiang:2018pbd}.}. In our case only fermion kinetic terms (i.e. wave-functions renormalizations) are relevant, as the tree-level WCs in Eq.~\eqref{eq:EFTS13treematch} do not depend on any SM coupling. The one-loop formulas below include the contributions due to fermion field renormalization.

\subsection{Example}

In this Section we discuss in some details the matching of a specific Green's function, in order to illustrate some of the most relevant aspects of our computation.

\begin{figure}[t]
\centering
\includegraphics[width=0.85\textwidth]{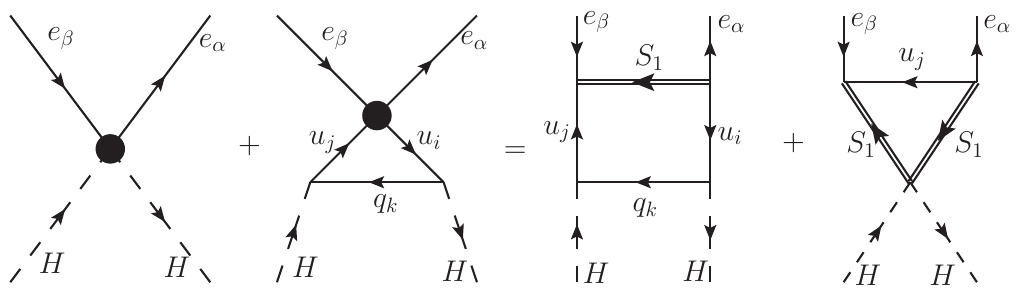}\hfill
\caption{\small Diagrams for the matching of the $\langle \bar e e H^\dagger H \rangle$ Green function.\label{fig:example}}
\end{figure}

Let us consider the off-shell Green's function $\mathcal G\equiv \langle e_\beta(p_1)  \bar e_\alpha(p_2)  H_b (q_1) H^\dagger_a (q_2) \rangle$, where all momenta are incoming and $a,b$ are $\SU(2)_L$ indices. The matching conditions for this correlator are depicted diagrammatically in Fig.\ref{fig:example}, where the left and right hand-side show the EFT and UV contributions, respectively. We briefly comment on the various steps of this computation.

We begin by listing the various contributions to $\mathcal G$, both in the SMEFT and the leptoquark model. The SMEFT operators which contribute at tree level to $\mathcal G$ are (c.f. Table~\ref{tab:GreenSingleFermion} for the notation):
\be\begin{split}
	[\OO_{He}]_{\alpha\beta} &= (\bar e_\alpha \gamma^\mu e_\beta) (H^\dagger i \overleftrightarrow{D}_\mu H)~, \\
	[\OO_{He}^{\prime}]_{\alpha\beta} &= (\bar e_\alpha i \overleftrightarrow{\slashed{D}} e_\beta) (H^\dagger H)~, \\
	[\OO_{He}^{\prime\prime}]_{\alpha\beta} &= (\bar e_\alpha \gamma^\mu e_\beta) \partial_\mu (H^\dagger H)~.
\end{split}\label{MatchingExampleOps}\ee

Moreover, we must take into account a one-loop contribution from $\OO_{eu}$, which is generated at the tree-level in our model according to Eq.~\eqref{eq:EFTS13treematch}. Since this tree-level WC is fixed, the matching of $\mathcal G$ allows us to fix the coefficients of the operators in \eqref{MatchingExampleOps}, see the left-hand side of Fig.\ref{fig:example}.
In the leptoquark model there are two diagrams contributing to $\mathcal G$, both mediated by $S_1$, shown in the right-hand side of Fig.\ref{fig:example}: a box diagram proportional to (schematically) $y_U y_U ^\dagger \lambda^{1R}\lambda^{1R}{}^\dagger $, and a triangle diagram proportional to $\lambda_{H1} \lambda^{1R}{}^\dagger \lambda^{1R}$.

By total momentum conservation, only three out of the four momenta $p_1,\,p_2,\,q_1,\,q_2$ are independent. Writing $(p_1, p_2, q_1, q_2)=(p-r,\,-p-r,\,q+r,\,-q+r),$ the tree-level contributions from the operators in Eq. \eqref{MatchingExampleOps} read: 
\be
[\mathcal{G} ^{\text {tree}} _{\text {EFT}}(\mu _M)]_{\alpha \beta}=2 \slashed{q}  [G_{He}(\mu_M)]_{\alpha\beta}+2 \slashed{p}  [G^\prime_{He}(\mu_M)]_{\alpha\beta} - 2 i \slashed{r}  [G^{\prime\prime }_{He}(\mu_M)]_{\alpha\beta},
\ee
where we drop here and below a global $\delta_{ab}$ factor, and we denote Green's basis WCs by $G_{i}$.
The UV and EFT one-loop contributions are more easily computed when only one of the independent momenta $p,\,q,\,r$ is non-vanishing, and yield respectively:
\begin{align}
\left[\mathcal{G}_{\text{UV}}^{\text{1-loop}}(\mu_{M})\right]_{\alpha\beta}^{q=r=0} & =	- \slashed{p} \frac{N_c (\lambda^{1R \dagger} y_U^T y_U^* \lambda^{1R})_{\alpha\beta} }{(4\pi)^2 2 M_1^2} 
		+ \slashed{p} \frac{N_c \lambda_{H1} (\lambda^{1R \dagger} \lambda^{1R})_{\alpha\beta} }{(4\pi)^2 2 M_1^2} ~,\\
\left[\mathcal{G}_{\text{UV}}^{\text{1-loop}}(\mu_{M})\right]_{\alpha\beta}^{p=r=0} & =- \slashed{q} \frac{N_c (\lambda^{1R \dagger} y_U^T y_U^* \lambda^{1R})_{\alpha\beta} }{(4\pi)^2 M_1^2} \log \frac{-q^2}{M_1^2} ~\label{eq:G_UV,p=r=0},\\
\left[\mathcal{G}_{\text{UV}}^{\text{1-loop}}(\mu_{M})\right]_{\alpha\beta}^{p=q=0} & =0 ~,
\end{align} 
and
\begin{align}
\left[\mathcal{G}_{\text{EFT}}^{\text{1-loop}}(\mu_{M})\right]_{\alpha\beta}^{q=r=0} & =0 ~,\\
\left[\mathcal{G}_{\text{EFT}}^{\text{1-loop}}(\mu_{M})\right]_{\alpha\beta}^{p=r=0} & =\slashed{q} \frac{N_c (\lambda^{1R \dagger} y_U^T y_U^* \lambda^{1R})_{\alpha\beta} }{(4\pi)^2 M_1^2} \left(1 + \log \frac{\mu_M^2}{-q^2} \right) ~\label{eq:G_EFT,p=r=0},\\
\left[\mathcal{G}_{\text{EFT}}^{\text{1-loop}}(\mu_{M})\right]_{\alpha\beta}^{p=q=0} & =0 ~,
\end{align}
where we employed the tree-level value of $[C_{eu}]^{(0)}_{\alpha \beta}$ given in Eq. \eqref{eq:EFTS13treematch}. Notice that the EFT computation presents an ultraviolet divergence, which we regulate in the $\overline{\text{MS}}$ scheme at renormalization scale $\mu_M$. On the other hand, on the basis of renormalizability, the UV contribution must be (and is) finite. Finally, both EFT and UV diagrams present an infrared divergence, corresponding to the $\log (- q^2)$ terms in Eqs. \eqref{eq:G_UV,p=r=0} and \eqref{eq:G_EFT,p=r=0}. The agreement of these two terms, which is guaranteed by the EFT construction, provides a further check of validity of the computation.

Requiring $\mathcal G_{\text{EFT}}(\mu _M) = \mathcal G_{\text{UV}}(\mu _M)$, we finally obtain the matching conditions:
\be\begin{split}
	[G_{He}(\mu_M)]_{\alpha\beta} &= - \frac{N_c (\lambda^{1R \dagger} y_U^T y_U^* \lambda^{1R})_{\alpha\beta} }{32 \pi^2 M_1^2} \left(1 + \log \frac{\mu_M^2}{M_1^2} \right) ~,\\
	[G_{He}^\prime (\mu_M)]_{\alpha\beta} &= - \frac{N_c (\lambda^{1R \dagger} y_U^T y_U^* \lambda^{1R})_{\alpha\beta} }{64 \pi^2 M_1^2} +
	\frac{N_c \lambda_{H1} (\lambda^{1R \dagger} \lambda^{1R})_{\alpha\beta} }{64 \pi^2 M_1^2}~, \\
	[G_{He}^{\prime \prime}(\mu_M)]_{\alpha\beta}&= 0~.
\end{split}\ee
As a cross-check, we observe that the $\mu_M$ dependence of $[G_{He}(\mu_M)]_{\alpha\beta}$ corresponds to the SMEFT RG running of $C_{He}$ due to  $C_{eu}$ \cite{Jenkins:2013wua},
\be
	(4\pi)^2 \mu \dfrac{d [C_{He}]_{\alpha\beta}}{d\mu} = - 2 N_c [C_{eu}]_{\alpha\beta i j} (y_U^T y_U^*)_{ij} ~,
\ee
once Eq.~\eqref{eq:EFTS13treematch} is taken into account.

\subsection{One-loop matching conditions in the Warsaw basis}\label{subsec:MatchToWarsaw}

In the following we report the complete one-loop matching conditions of the $S_1 + S_3$ model to dimension-six SMEFT operators in the Warsaw basis.
Definitions of the operators can be found in Tables.~\ref{tab:GreenBosonic}, \ref{tab:GreenSingleFermion}, \ref{tab:4FermionBconserving} and \ref{tab:4FermionBviolating}, while the $C_i^{(1)}$ coefficients are defined as in Eq.~(\ref{eq:Ci}). 
For convenience, we make the following definitions:
\begin{align}
	L_{1,3} & \equiv\ln(\frac{\mu_M^{2}}{M_{1,3}^{2}})~,\\
	h(M_{1},M_{3}) & \equiv\frac{M_{1}^{4}-M_{3}^{4}-2M_{1}^{2}M_{3}^{2}\log\frac{M_{1}^{2}}{M_{3}^{2}}}{(M_{1}^{2}-M_{3}^{2})^{3}}~,\nonumber\\
	 & \approx\frac{1}{3M_{1}^{2}}\left(1-\delta m+\frac{7}{10}\delta m^{2}+\mathcal{O}(\delta m^{3})\right)
	 	\qquad\left(\delta m\equiv M_{3}/M_{1}-1\right)~,\\
	n(M_1, M_3) &\equiv \frac{M_1^2-M_3^2 + M_3^2 \log \frac{M_3^2}{M_1^2}}{(M_3^2 - M_1^2)^2} \approx -\frac{1}{2 M_1^2} \left( 1 - \frac{2}{3} \delta m + \mathcal{O}(\delta m^2)\right)~,
\end{align}
\be\begin{array}{l l l}
	\Lambda_{q}^{(n)} \equiv \lambda^{nL*} \lambda^{nL \, T}, &\hspace{0.5cm}
	\Lambda_{q}^{(31)} \equiv \lambda^{3L*} \lambda^{1L \, T}, &\hspace{0.5cm}
	\Lambda_{u} \equiv \lambda^{1R*} \lambda^{1R \, T}, \\[1mm]
	\Lambda_{\ell}^{(n)}  \equiv\lambda^{nL\dagger}\lambda^{nL}, &\hspace{0.5cm}
	\Lambda_{\ell}^{(31)}  \equiv\lambda^{3L\dagger}\lambda^{1L}, &\hspace{0.5cm}
	\Lambda_{e} \equiv\lambda^{1R\dagger}\lambda^{1R},
\end{array}\ee
\be\begin{array}{l l}
	X_{1U}^{1L} \equiv\lambda^{1L\dagger}y_{U}^{*}\lambda^{1R}, &\hspace{0.5cm}
	X_{1E}^{1L}  \equiv(\lambda^{1R}y_{E}^{\dagger}\lambda^{1L\dagger})^{T}, \\[2mm]
	X_{1U}^{3L} \equiv\lambda^{3L\dagger}y_{U}^{*}\lambda^{1R}, &\hspace{0.5cm}
	X_{1E}^{3L}  \equiv(\lambda^{1R}y_{E}^{\dagger}\lambda^{3L\dagger})^{T}, \\[2mm]
	X_{2F}^{nL}  \equiv\lambda^{nL\dagger}y_{F}^{*}y_{F}^{T}\lambda^{nL}, &\hspace{0.5cm}
	X_{2U}^{1R}  \equiv\lambda^{1R\dagger}y_{U}^{T}y_{U}^{*}\lambda^{1R}, \\[2mm]
	X_{2E}^{nL}  \equiv(\lambda^{nL}y_{E}y_{E}^{\dagger}\lambda^{nL\dagger})^{T}, &\hspace{0.5cm}
	X_{2E}^{1R}  \equiv(\lambda^{1R}y_{E}^{\dagger}y_{E}\lambda^{1R\dagger})^{T}, \\[2mm]
	X_{3U}^{1L}  \equiv\lambda^{1L\dagger}y_{U}^{*}y_{U}^{T}y_{U}^{*}\lambda^{1R}, &\hspace{0.5cm}
	X_{3E}^{1L}  \equiv(\lambda^{1R}y_{E}^{\dagger}y_{E}y_{E}^{\dagger}\lambda^{1L\dagger})^{T},
\end{array}\ee
where $n=1,3$ and $F=U,D$, and the superscript \emph{T} stands for transpose.
The fermion wave function renormalizations, are given by $(Z_\psi)_{ij} = \delta_{ij} + \frac{1}{(4\pi)^2} (\delta Z_\psi)_{ij} $, where
\begin{align}
(\delta Z_{\ell})_{\alpha\beta} & =\frac{N_{c}}{2}\left[(\frac{1}{2}+L_{1})(\Lambda_{\ell}^{(1)})_{\alpha\beta}+3(\frac{1}{2}+L_{3})(\Lambda_{\ell}^{(3)})_{\alpha\beta}\right],\\
(\delta Z_{e})_{\alpha\beta} & =\frac{N_{c}}{2}(\frac{1}{2}+L_{1})(\Lambda_{e})_{\alpha\beta},\\
(\delta Z_{q}^{\text{ }})_{ij} & =\frac{1}{2}\left[(\frac{1}{2}+L_{1})(\Lambda_{q}^{(1)})_{ij}+3(\frac{1}{2}+L_{3})(\Lambda_{q}^{(3)})_{ij}\right],\\
(\delta Z_{u})_{ij} & =\frac{1}{2}(\frac{1}{2}+L_{1})(\Lambda_{u})_{ij}, \\
(\delta Z_{d})_{ij} & =0 .
\end{align}




\subsubsection{Renormalizable terms}

\begin{align}
(4\pi)^2 (\delta y_{E})_{\alpha\beta}= & -\frac{1}{2}\left[(\delta Z_{\ell})_{\alpha\gamma}\delta{}_{\delta\beta}+\delta_{\alpha\gamma}(\delta Z_{e})_{\delta\beta}\right](y_{E})_{\gamma\delta}+ \nonumber\\
 &-N_{c}\left[1+L_{1}+\frac{1}{2}\left(\frac{{3}}{2}+L_{1}\right)\frac{m^{2}}{M_{1}^{2}}\right](X_{1U}^{1L})_{\alpha\beta}~,\\
(4\pi)^2 (\delta y_{U})_{ij}= & -\frac{1}{2}\left[(\delta Z_{q})_{ik}\delta_{jl}+\delta_{ik}(\delta Z_{u})_{lj}\right](y_{U})_{kl}+\nonumber\\
 &-\left[1+L_{1}+\frac{1}{2}\left(\frac{{3}}{2}+L_{1}\right)\frac{m^{2}}{M_{1}^{2}}\right](X_{1E}^{1L})_{ij}~,\\
(4\pi)^2 (\delta y_{D})_{ij}= & 0~,
\end{align}
\begin{align}
(4\pi)^2 \delta\lambda= & -N_{c}\left[\lambda_{H1}^{2}L_{1}+(3\lambda_{H3}^{2}+2\lambda_{\epsilon H3}^{2})L_{3}+2\vert\lambda_{H13}\vert^{2}\left(1+\dfrac{M_{3}^{2}L_{3}-M_{1}^{2}L_{1}}{M_{3}^{2}-M_{1}^{2}}\right)\right] + \nonumber \\
	& - \frac{N_{c}}{15}g^{4}m^{2}\frac{1}{M_{3}^{2}}-4 N_c m^{4}\left(\frac{\lambda_{\epsilon H3}^{2}}{3M_{3}^{2}}+\left|\lambda_{H13}\right|{}^{2}h(M_{1},M_{3})\right) ~,\\
(4\pi)^2 \delta m^{2}= & N_{c}\left[\lambda_{H1}(1+L_{1})M_{1}^{2}+3\lambda_{H3}(1+L_{3})M_{3}^{2}\right]~ .
\end{align}

\subsubsection{Purely bosonic}

\paragraph{$X^{3}$}

\begin{align}
C^{(1)}_{3G} & =\frac{1}{360}g_{s}^{3}\left(\frac{3}{M_{3}^{2}}+\frac{1}{M_{1}^{2}}\right),\\
C^{(1)}_{3W} & =\frac{N_{c}}{90}g^{3}\frac{1}{M_{3}^{2}}, \\
C^{(1)}_{3\widetilde G} & = C^{(1)}_{3 \widetilde W}  = 0.
\end{align}

\paragraph{$X^{2}H^{2}$}

\begin{align}
C^{(1)}_{HG} & =\frac{g_{s}^{2}}{{24}}\left(\frac{3\lambda_{H3}}{M_{3}^{2}}+\frac{\lambda_{H1}}{M_{1}^{2}}\right)~,\\
C^{(1)}_{HW} & =\frac{N_{c}}{{6}}g^{2}\frac{\lambda_{H3}}{M_{3}^{2}}~,\\
C^{(1)}_{HB} & =\frac{N_{c}}{{12}}g^{\prime2}\left(3\frac{\lambda_{H3}Y_{S_{3}}^{2}}{M_{3}^{2}}+\frac{\lambda_{H1}Y_{S_{1}}^{2}}{M_{1}^{2}}\right)~,\\
C^{(1)}_{HWB} & =-N_{c}\frac{gg^{\prime}Y_{S_{3}}\lambda_{\epsilon H3}}{3M_{3}^{2}}~, \\
C^{(1)}_{H \widetilde G} & = C^{(1)}_{H \widetilde W} = C^{(1)}_{H \widetilde B} = C^{(1)}_{H \widetilde W B} = 0~.
\end{align}

\paragraph{$H^{4}D^{2}$}

\begin{align}
C^{(1)}_{H\Box} =& -N_{c}\left(\frac{1}{40}g^{4}+\frac{1}{20}g^{\prime4}Y_{H}^{2}Y_{S_{3}}^{2}+\frac{3\lambda_{H3}^{2}-2\lambda_{\epsilon H3}^{2}}{12}\right)\frac{1}{M_{3}^{2}}-N_{c}\left(\frac{1}{60}g^{\prime4}Y_{H}^{2}Y_{S_{1}}^{2}+\frac{\lambda_{H1}^{2}}{12}\right)\frac{1}{M_{1}^{2}}+ \nonumber\\
 & +\frac{N_{c}}{2}\left|\lambda_{H13}\right|{}^{2}h(M_{1},M_{3})~,\\
C^{(1)}_{HD} & =-\frac{N_{c}}{15}g^{\prime4}Y_{H}^{2}\left(\frac{3Y_{S_{3}}^{2}}{M_{3}^{2}}+\frac{Y_{S_{1}}^{2}}{M_{1}^{2}}\right)-2N_{c}\left(\frac{\lambda_{\epsilon H3}^{2}}{3M_{3}^{2}}+\left|\lambda_{H13}\right|{}^{2}h(M_{1},M_{3})\right)~.
\end{align}

\paragraph{$H^{6}$}

\begin{align}
C^{(1)}_{H}  =& -N_{c}\left(\frac{1}{30}g^{4}\lambda+\frac{1}{2}\lambda_{H3}^{3}+\lambda_{H3}\lambda_{\epsilon H3}^{2}\right)\frac{1}{M_{3}^{2}}-N_{c}\frac{\lambda_{H1}^{3}}{6 M_{1}^{2}}+2N_{c}\lambda\left(\frac{\lambda_{\epsilon H3}^{2}}{3M_{3}^{2}}+\left|\lambda_{H13}\right|{}^{2}h(M_{1},M_{3})\right)\nonumber\\
 & +\frac{N_{c}\left|\lambda_{H13}\right|{}^{2}}{M_{1}^{2}-M_{3}^{2}}\left(\lambda_{H3}-\lambda_{H1}+\frac{\ln(\frac{M_{1}^{2}}{M_{3}^{2}})}{M_{1}^{2}-M_{3}^{2}}(\lambda_{H1}M_{3}^{2}-\lambda_{H3}M_{1}^{2})\right)~.
\end{align}

\subsubsection{Two-fermion operators}

\paragraph{$\psi^{2}XH$}

\begin{align}
[C_{uG}]^{(1)}_{ij}= & -\frac{1}{4} g_{s} \frac{(X_{1E}^{1L})_{ij}}{M_{1}^{2}}
-\frac{1}{24}g_{s}\left(3\frac{(\Lambda_{q}^{(3)}y_{U})_{ij}}{M_{3}^{2}}+\frac{(\Lambda_{q}^{(1)}y_{U})_{ij}}{M_{1}^{2}}\right)
-\frac{1}{24}g_{s}\frac{(y_{U}\Lambda_{u})_{ij}}{M_{1}^{2}},\\
[C_{uW}]^{(1)}_{ij}= & -\frac{1}{8}g\left(\frac{{3}}{2}+L_{1}\right)\frac{(X_{1E}^{1L})_{ij}}{M_{1}^{2}}
-\frac{1}{24}g\left(3\frac{(\Lambda_{q}^{(3)}y_{U})_{ij}}{M_{3}^{2}}-\frac{(\Lambda_{q}^{(1)}y_{U})_{ij}}{M_{1}^{2}}\right),\\
[C_{uB}]^{(1)}_{ij}= &  +\frac{1}{4}g^{\prime}\left[(Y_{l}+Y_{e})L_{1}+\frac{1}{2}Y_{l}+\frac{3}{2}Y_{e}-Y_{u} {+ Y_H} \right]\frac{(X_{1E}^{1L})_{ij}}{M_{1}^{2}} +\nonumber\\
& {-\frac{1}{24}g^{\prime}(Y_{q}+3Y_{\ell})\left(3\frac{(\Lambda_{q}^{(3)}y_{U})_{ij}}{M_{3}^{2}}+\frac{(\Lambda_{q}^{(1)}y_{U})_{ij}}{M_{1}^{2}}\right)} 
- \frac{1}{24}g^{\prime}(Y_{u} {+} 3Y_{e})\frac{(y_{U}\Lambda_{u})_{ij}}{M_{1}^{2}},
\end{align}
\begin{align}
[C_{dG}]^{(1)}_{ij}= & -\frac{1}{24}g_{s}\left(3\frac{(\Lambda_{q}^{(3)}y_{D})_{ij}}{M_{3}^{2}}+\frac{(\Lambda_{q}^{(1)}y_{D})_{ij}}{M_{1}^{2}}\right),\\{}
[C_{dW}]^{(1)}_{ij}= & -\frac{1}{24}g\left(3\frac{(\Lambda_{q}^{(3)}y_{D})_{ij}}{M_{3}^{2}}-\frac{(\Lambda_{q}^{(1)}y_{D})_{ij}}{M_{1}^{2}}\right),\\{}
[C_{dB}]^{(1)}_{ij}= & {-\frac{1}{24}g'(Y_{q}+3Y_{\ell})\left(3\frac{(\Lambda_{q}^{(3)}y_{D})_{ij}}{M_{3}^{2}}+\frac{(\Lambda_{q}^{(1)}y_{D})_{ij}}{M_{1}^{2}}\right) } ,
\end{align}
\begin{align}
[C_{eW}]^{(1)}_{\alpha\beta}= & -\frac{N_{c}}{8}g\left(\frac{{3}}{2}+L_{1}\right)\frac{(X_{1U}^{1L})_{\alpha\beta}}{M_{1}^{2}}
-\frac{N_{c}}{24}g\left(3\frac{(\Lambda_{\ell}^{(3)}y_{E})_{\alpha\beta}}{M_{3}^{2}} -\frac{(\Lambda_{\ell}^{(1)}y_{E})_{\alpha\beta}}{M_{1}^{2}}\right),\\{}
[C_{eB}]^{(1)}_{\alpha\beta}= & \frac{N_{c}}{4}g^{\prime}\left[(Y_{q}+Y_{u})L_{1} + \frac{1}{2}Y_{q}+\frac{3}{2}Y_{u}-Y_{e} { - Y_H} \right]\frac{(X_{1U}^{1L})_{\alpha\beta}}{M_{1}^{2}} + \nonumber\\
& - \frac{N_{c}}{24}g^{\prime}(Y_{e}{+}3Y_{u})\frac{(y_{E}\Lambda_{e})_{\alpha\beta}}{M_{1}^{2}}.
\end{align}

\paragraph{$\psi^{2}H^{2}D$}

\begin{align}
[C_{Hq}^{(1)}]^{(1)}_{ij}= & -\frac{N_{c}}{30}g^{\prime4}Y_{H}Y_{q}\delta_{ij}\left(\frac{3 Y_{S_3}^{2}}{M_{3}^{2}}+\frac{Y_{S_1}^{2}}{M_{1}^{2}}\right)
 - \frac{1}{24}\frac{(y_{U}\Lambda_{u}y_{U}^{\dagger})_{ij}}{M_{1}^{2}}+\nonumber\\
 & +\frac{g^{\prime2}Y_{H}}{3} \left\{ 3 \left(\frac{{8 Y_\ell -Y_{S_3}}}{6}+Y_{\ell} L_3 \right)\frac{(\Lambda_{q}^{(3)})_{ij}}{M_{3}^{2}} 
 + \left(\frac{{8 Y_\ell -Y_{S_1}}}{6}+Y_{\ell} L_1\right)\frac{(\Lambda_{q}^{(1)})_{ij}}{M_{1}^{2}} \right\} +\nonumber\\
 & -\frac{1}{4}\left(3(1+L_{3})\frac{(X_{2E}^{3L})_{ij}}{M_{3}^{2}}+(1+L_{1})\frac{(X_{2E}^{1L})_{ij}}{M_{1}^{2}}\right) 
  {+ \frac{( X_{1E}^{1L} y_U^\dagger)_{ij} + (y_U X_{1E}^{1L \dagger} )_{ij}}{8 M_1^2} } ,\\
[C_{Hq}^{(3)}]^{(1)}_{ij}= & -\frac{N_{c}}{60}g^{4}\delta_{ij}\frac{1}{M_{3}^{2}}+\frac{1}{24}\frac{(y_{U}\Lambda_{u}y_{U}^{\dagger})_{ij}}{M_{1}^{2}}+\nonumber\\
 & +\frac{1}{12}g^{2}\left(\left({2}+L_{3}\right)\frac{(\Lambda_{q}^{(3)})_{ij}}{M_{3}^{2}}-\left({\frac{4}{3}}+L_{1}\right)\frac{(\Lambda_{q}^{(1)})_{ij}}{M_{1}^{2}}\right)+\nonumber\\
 & -\frac{1}{4}\left((1+L_{3})\frac{(X_{2E}^{3L})_{ij}}{M_{3}^{2}}-(1+L_{1})\frac{(X_{2E}^{1L})_{ij}}{M_{1}^{2}}\right)  
{- \frac{( X_{1E}^{1L} y_U^\dagger)_{ij} + (y_U X_{1E}^{1L \dagger} )_{ij}}{8 M_1^2}},
\end{align}
\begin{align}
[C_{Hu}]^{(1)}_{ij}= & -\frac{N_{c}}{30}g^{\prime4}Y_{H}Y_{u}\delta_{ij}\left(\frac{3 Y_{S_3}^{2}}{M_{3}^{2}}+\frac{Y_{S_1}^{2}}{M_{1}^{2}}\right)+\frac{1}{12}\left(3\frac{(y_{U}^{\dagger}\Lambda_{q}^{(3)}y_{U})_{ij}}{M_{3}^{2}}+\frac{(y_{U}^{\dagger}\Lambda_{q}^{(1)}y_{U})_{ij}}{M_{1}^{2}}\right)+\nonumber\\
 & +\frac{1}{3}g^{\prime2}Y_{H}\left(- \frac{Y_{S_{1}}}{6}+Y_{e} \left({-\frac{1}{6} } + L_{1}\right)\right)\frac{(\Lambda_{u})_{ij}}{M_{1}^{2}}+\frac{1}{2}(1+L_{1})\frac{(X_{2E}^{1R})_{ij}}{M_{1}^{2}} + \nonumber\\
 &  {- \frac{( y_U^\dagger X_{1E}^{1L} )_{ij} + (X_{1E}^{1L \dagger} y_U)_{ij}}{4 M_1^2} },\\
[C_{Hd}]^{(1)}_{ij}= & -\frac{N_{c}}{30}g^{\prime4}Y_{H}Y_{d}\delta_{ij}\left(\frac{3 Y_{S_3}^{2}}{M_{3}^{2}}+\frac{Y_{S_1}^{2}}{M_{1}^{2}}\right)+\frac{1}{12}\left(3\frac{(y_{D}^{\dagger}\Lambda_{q}^{(3)}y_{D})_{ij}}{M_{3}^{2}}+\frac{(y_{D}^{\dagger}\Lambda_{q}^{(1)}y_{D})_{ij}}{M_{1}^{2}}\right),\\
[C_{Hud}]^{(1)}_{ij}= & \frac{1}{12}\left(3\frac{(y_{U}^{\dagger}\Lambda_{q}^{(3)}y_{D})_{ij}}{M_{3}^{2}}+\frac{(y_{U}^{\dagger}\Lambda_{q}^{(1)}y_{D})_{ij}}{M_{1}^{2}}\right)
 {- \frac{( X_{1E}^{1L \dagger} y_D)_{ij} }{2 M_1^2}},
\end{align}
\begin{align}
[C_{H\ell}^{(1)}]^{(1)}_{\alpha\beta}= & -\frac{N_{c}}{30}g^{\prime4}Y_{H}Y_{\ell}\delta_{\alpha\beta}\left(\frac{3 Y_{S_3}^{2}}{M_{3}^{2}}+\frac{Y_{S_1}^{2}}{M_{1}^{2}}\right)+\frac{N_{c}}{24}\frac{(y_{E}\Lambda_{e}y_{E}^{\dagger})_{\alpha\beta}}{M_{1}^{2}} 
{- \frac{( X_{1U}^{1L} y_E^\dagger)_{\alpha\beta} + ( y_E X_{1U}^{1L \dagger})_{\alpha\beta} }{8 M_1^2}} +\nonumber\\
 & + N_{c} g^{\prime2}Y_{H}\left(\frac{{8 Y_q -Y_{S_3} } }{6}+Y_{q} L_{3}\right)\frac{(\Lambda_{\ell}^{(3)})_{\alpha\beta}}{M_{3}^{2}}+\nonumber\\
 & +\frac{N_{c} g^{\prime2}Y_{H}}{3}\left(\frac{{8 Y_q -Y_{S_1}}}{6}+Y_{q} L_{1}\right)\frac{(\Lambda_{\ell}^{(1)})_{\alpha\beta}}{M_{1}^{2}} +\nonumber\\
 & -\frac{N_{c}}{4}\left\{ 3(1+L_{3})\frac{-(X_{2U}^{3L})_{\alpha\beta}+(X_{2D}^{3L})_{\alpha\beta}}{M_{3}^{2}}+(1+L_{1})\frac{-(X_{2U}^{1L})_{\alpha\beta}+(X_{2D}^{1L})_{\alpha\beta}}{M_{1}^{2}}\right\} ,\\
[C_{H\ell}^{(3)}]^{(1)}_{\alpha\beta}= & -\frac{N_{c}}{60}g^{4}\delta_{\alpha\beta}\frac{1}{M_{3}^{2}}+\frac{N_{c}}{24}\frac{(y_{E}\Lambda_{e}y_{E}^{\dagger})_{\alpha\beta}}{M_{1}^{2}}  
{- \frac{( X_{1U}^{1L} y_E^\dagger)_{\alpha\beta} + ( y_E X_{1U}^{1L \dagger})_{\alpha\beta} }{8 M_1^2}} +\nonumber\\
 & +\frac{N_{c}}{12}g^{2}\left(\left({2}+L_{3}\right)\frac{(\Lambda_{\ell}^{(3)})_{\alpha\beta}}{M_{3}^{2}}-\left({\frac{4}{3}}+L_{1}\right)\frac{(\Lambda_{\ell}^{(1)})_{\alpha\beta}}{M_{1}^{2}}\right) + \nonumber\\
 & -\frac{N_{c}}{4}\left\{ (1+L_{3})\frac{(X_{2U}^{3L})_{\alpha\beta}+(X_{2D}^{3L})_{\alpha\beta}}{M_{3}^{2}}-(1+L_{1})\frac{(X_{2U}^{1L})_{\alpha\beta}+(X_{2D}^{1L})_{\alpha\beta}}{M_{1}^{2}}\right\} ,\\{}
[C_{He}]^{(1)}_{\alpha\beta}= & -\frac{N_{c}}{30}g^{\prime4}Y_{H}Y_{e}\delta_{\alpha\beta}\left(\frac{3 Y_{S_3}^{2}}{M_{3}^{2}}+\frac{Y_{S_1}^{2}}{M_{1}^{2}}\right)+\frac{N_{c}}{12}\left(3\frac{(y_{E}^{\dagger}\Lambda_{\ell}^{(3)}y_{E})_{\alpha\beta}}{M_{3}^{2}}+\frac{(y_{E}^{\dagger}\Lambda_{\ell}^{(1)}y_{E})_{\alpha\beta}}{M_{1}^{2}}\right)+\nonumber\\
 & +\frac{N_{c}}{3}g^{\prime2}Y_{H}\left(- \frac{Y_{S_{1}}}{6}+Y_{u} \left({-\frac{1}{6}} + L_{1} \right)\right)\frac{(\Lambda_{e})_{\alpha\beta}}{M_{1}^{2}}-\frac{N_{c}}{2}(1+L_{1})\frac{(X_{2U}^{1R})_{\alpha\beta}}{M_{1}^{2}} + \nonumber\\
 & {\frac{( y_E^\dagger X_{1U}^{1L} )_{\alpha\beta} + ( X_{1U}^{1L \dagger} y_E)_{\alpha\beta} }{4 M_1^2}}  .
\end{align}

\paragraph{$\psi^{2}H^{3}$}

\begin{align}
[C_{uH}]^{(1)}_{ij}= & -\frac{N_{c}}{60}g^{4}(y_{U})_{ij}\frac{1}{M_{3}^{2}}+N_{c}(y_{U})_{ij}\left(\frac{\lambda_{\epsilon H3}^{2}}{3M_{{3}}^{2}}+\left|\lambda_{H13}\right|{}^{2}h(M_{1},M_{3})\right)+\nonumber\\
 & +\frac{1}{12}\left(3\frac{(y_{U}y_{U}^{\dagger}\Lambda_{q}^{(3)}y_{U})_{ij}}{M_{3}^{2}}+\frac{(y_{U}y_{U}^{\dagger}\Lambda_{q}^{(1)}y_{U})_{ij}}{M_{1}^{2}}\right)+\frac{1}{12}\frac{(y_{U}\Lambda_{u}y_{U}^{\dagger}y_{U})_{ij}}{M_{1}^{2}}+\nonumber\\
 & -\frac{\lambda}{2} \left(L_{1}+\frac{{3}}{2}\right) \frac{(X_{1E}^{1L})_{ij}}{M_{1}^{2}} + {\frac{1}{4}\frac{(X_{1E}^{1L} y_U^\dagger y_{U})_{ij} + (y_{U} y_U^\dagger X_{1E}^{1L} )_{ij}}{M_{1}^{2}}}+\nonumber\\
 & -\frac{1}{4}\left(\frac{(X_{2E}^{3L}y_{U})_{ij}+(3\lambda_{H3}-2\lambda_{\epsilon H3})(\Lambda_{q}^{(3)}y_{U})_{ij}}{M_{3}^{2}}+\frac{(X_{2E}^{1L}y_{U})_{ij}+\lambda_{H1}(\Lambda_{q}^{(1)}y_{U})_{ij}}{M_{1}^{2}}\right)+\nonumber\\
 & - \frac{1}{2} \left[ \left(\lambda_{H13}^{*}\Lambda_{q}^{(31)} y_U\right)_{ij} n(M_1, M_3) + \left(\lambda_{H13}\Lambda_{q}^{(31)\dagger} y_U \right)_{ij} n(M_3, M_1) \right] + \nonumber\\
 & {-}\frac{1}{4}\frac{(y_{U}X_{2E}^{1R})_{ij}+\lambda_{H1}(y_{U}\Lambda_{u})_{ij}}{M_{1}^{2}}+\frac{(1+L_{1})(X_{3E}^{1L})_{ij}-\lambda_{H1}(X_{1E}^{1L})_{ij}}{M_{1}^{2}}-\frac{\lambda_{H13}^{*}(X_{1E}^{3L})_{ij}\log\frac{M_{1}^{2}}{M_{3}^{2}}}{M_{1}^{2}-M_{3}^{2}},
\end{align}
\begin{align}
[C_{dH}]^{(1)}_{ij}= & -\frac{N_{c}}{60}g^{4}(y_{D})_{ij}\frac{1}{M_{3}^{2}}+N_{c}(y_{D})_{ij}\left(\frac{\lambda_{\epsilon H3}^{2}}{3M_{{3}}^{2}}+\left|\lambda_{H13}\right|{}^{2}h(M_{1},M_{3})\right)+\nonumber\\
 & +\frac{1}{12}\left(3\frac{(y_{D}y_{D}^{\dagger}\Lambda_{q}^{(3)}y_{D})_{ij}}{M_{3}^{2}}+\frac{(y_{D}y_{D}^{\dagger}\Lambda_{q}^{(1)}y_{D})_{ij}}{M_{1}^{2}}\right)+\nonumber\\
 & -\frac{1}{4}\left(\frac{2(X_{2E}^{3L}y_{D})_{ij}+(3\lambda_{H3}+2\lambda_{\epsilon H3})(\Lambda_{q}^{(3)}y_{D})_{ij}}{M_{3}^{2}}+\frac{\lambda_{H1}(\Lambda_{q}^{(1)}y_{D})_{ij}}{M_{1}^{2}}\right)+\nonumber\\
 &+\frac{1}{2} \left[ \left(\lambda_{H13}^{*}\Lambda_{q}^{(31)} y_D\right)_{ij} n(M_1, M_3) + \left(\lambda_{H13}\Lambda_{q}^{(31)\dagger} y_D \right)_{ij} n(M_3, M_1) \right] ,
\end{align}
\begin{align}
[C_{eH}]^{(1)}_{\alpha\beta}= & -\frac{N_{c}}{60}g^{4}(y_{E})_{\alpha\beta}\frac{1}{M_{3}^{2}}+N_{c}(y_{E}){}_{\alpha\beta}\left(\frac{\lambda_{\epsilon H3}^{2}}{3M_{{3}}^{2}}+\left|\lambda_{H13}\right|{}^{2}h(M_{1},M_{3})\right)+\nonumber\\
 & +\frac{N_{c}}{12}\left(3\frac{(y_{E}y_{E}^{\dagger}\Lambda_{\ell}^{(3)}y_{E})_{\alpha\beta}}{M_{3}^{2}}+\frac{(y_{E}y_{E}^{\dagger}\Lambda_{\ell}^{(1)}y_{E})_{\alpha\beta}}{M_{1}^{2}}\right)+\frac{N_{c}}{12}\frac{(y_{E}\Lambda_{e}y_{E}^{\dagger}y_{E})_{\alpha\beta}}{M_{1}^{2}}+\nonumber\\
 & -\frac{N_{c}}{2}\left(\frac{{3}}{2}+L_{1}\right)\lambda\frac{(X_{1U}^{1L})_{\alpha\beta}}{M_{1}^{2}}+
 {\frac{1}{4}\frac{(X_{1U}^{1L} y_E^\dagger y_{E})_{ij} + (y_{E} y_E^\dagger X_{1U}^{1L} )_{\alpha\beta}}{M_{1}^{2}}} +\nonumber\\
 & -\frac{N_{c}}{4} \frac{(X_{2U}^{3L}y_{E})_{\alpha\beta}+2(X_{2D}^{3L}y_{E})_{\alpha\beta}+(3\lambda_{H3}+2\lambda_{\epsilon H3})(\Lambda_{\ell}^{(3)}y_{E})_{\alpha\beta}}{M_{3}^{2}} + \nonumber \\
 & -\frac{N_{c}}{4} \frac{(X_{2U}^{1L}y_{E})_{\alpha\beta}+\lambda_{H1}(\Lambda_{\ell}^{(1)}y_{E})_{\alpha\beta}}{M_{1}^{2}} -\frac{N_{c}}{4}\frac{(y_{E}X_{2U}^{1R})_{\alpha\beta}+\lambda_{H1}(y_{E}\Lambda_{e})_{\alpha\beta}}{M_{1}^{2}} +\nonumber\\
 & -\frac{N_c}{2} \left[ \left( \lambda_{H13}^{*}\Lambda_{\ell}^{(31)} y_E\right)_{\alpha\beta} n(M_1, M_3) + \left(\lambda_{H13}\Lambda_{\ell}^{(31)\dagger} y_E \right)_{\alpha\beta} n(M_3, M_1) \right]+ \nonumber \\
 & +N_{c}\frac{(1+L_{1})(X_{3U}^{1L})_{\alpha\beta}-\lambda_{H1}(X_{1U}^{1L})_{\alpha\beta}}{M_{1}^{2}}-N_{c}\frac{\lambda_{H13}^{*}(X_{1U}^{3L})_{\alpha\beta}\log\frac{M_{1}^{2}}{M_{3}^{2}}}{M_{1}^{2}-M_{3}^{2}}.
\end{align}
%

\subsubsection{Four-fermion operators}

\paragraph{Four-quark}\footnote{Box diagram contributions to four-quark operators from $S_1$ and $S_3$, taken separately, have also been computed in \cite{Bobeth:2017ecx}. We find agreement except for $C_{qq}^{(1)}$ and $C_{qq}^{(3)}$, where we found an inconsistency in  \cite{Bobeth:2017ecx}. We thank the authors for clarifications about this point.}
\begin{align}
[C_{qq}^{(1)}]_{ijkl}^{(1)}= & -\frac{1}{240}g_{s}^{4}\left(\frac{1}{2}\delta_{il}\delta_{kj}-\frac{1}{3}\delta_{ij}\delta_{kl}\right)\left(\frac{3}{M_{3}^{2}}+\frac{1}{M_{1}^{2}}\right)-\frac{N_{c}}{60}g^{\prime4}Y_{q}^{2}\delta_{ij}\delta_{kl}\left(\frac{3Y_{S_{3}}^{2}}{M_{3}^{2}}+\frac{Y_{S_{1}}^{2}}{M_{1}^{2}}\right)+\nonumber \\
 & +\frac{1}{72}g_{s}^{2}\left(\frac{1}{2}\delta_{kj}\delta_{im}\delta_{ln}+\frac{1}{2}\delta_{il}\delta_{km}\delta_{jn}-\frac{1}{3}\delta_{kl}\delta_{im}\delta_{jn}-\frac{1}{3}\delta_{ij}\delta_{km}\delta_{ln}\right)\left(3\frac{(\Lambda_{q}^{(3)})_{mn}}{M_{3}^{2}}+\frac{(\Lambda_{q}^{(1)})_{mn}}{M_{1}^{2}}\right)+\nonumber \\
 & +\frac{1}{6}g^{\prime2}Y_{q}\left(\delta_{kl}\delta_{im}\delta_{jn}+\delta_{ij}\delta_{km}\delta_{ln}\right)\left\{ 3\left( \frac{{8 Y_\ell - Y_{S_3}}}{6}+ Y_\ell L_3 \right)\frac{(\Lambda_{q}^{(3)})_{mn}}{M_{3}^{2}} + \right. \nonumber \\
 &\left. +\left( \frac{{8 Y_\ell - Y_{S_1}}}{6}+ Y_\ell L_1 \right)\frac{(\Lambda_{q}^{(1)})_{mn}}{M_{1}^{2}}\right\} +\nonumber \\
 & -\frac{1}{16}\left(9\frac{(\Lambda_{q}^{(3)})_{il}(\Lambda_{q}^{(3)})_{kj}}{M_{3}^{2}}+\frac{(\Lambda_{q}^{(1)})_{il}(\Lambda_{q}^{(1)})_{kj}}{M_{1}^{2}}+3\frac{\log\frac{M_{3}^{2}}{M_{1}^{2}}\left[(\Lambda_{q}^{31})_{il}(\Lambda_{q}^{31\dagger})_{kj}+(\Lambda_{q}^{31\dagger})_{il}(\Lambda_{q}^{31})_{kj}\right]}{M_{3}^{2}-M_{1}^{2}}\right),\\{}
[C_{qq}^{(3)}]_{ijkl}^{(1)}= & -\frac{1}{480}g_{s}^{4}\delta_{il}\delta_{kj}\left(\frac{3}{M_{3}^{2}}+\frac{1}{M_{1}^{2}}\right)-\frac{N_{c}}{120}g^{4}\delta_{ij}\delta_{kl}\frac{1}{M_{3}^{2}}+\nonumber \\
 & +\frac{1}{144}g_{s}^{2}\left(\delta_{kj}\delta_{im}\delta_{ln}+\delta_{il}\delta_{km}\delta_{jn}\right)\left(3\frac{(\Lambda_{q}^{(3)})_{mn}}{M_{3}^{2}}+\frac{(\Lambda_{q}^{(1)})_{mn}}{M_{1}^{2}}\right)+\nonumber \\
 & +\frac{1}{24}g^{2}\left(\delta_{kl}\delta_{im}\delta_{jn}+\delta_{ij}\delta_{km}\delta_{ln}\right)\left( \left({2}+L_{3}\right) \frac{(\Lambda_{q}^{(3)})_{mn}}{M_{3}^{2}}-\left({\frac{4}{3}}+L_{1}\right)\frac{(\Lambda_{q}^{(1)})_{mn}}{M_{1}^{2}}\right)+\nonumber \\
 & -\frac{1}{16}\left(\frac{(\Lambda_{q}^{(3)})_{il}(\Lambda_{q}^{(3)})_{kj}}{M_{3}^{2}}+\frac{(\Lambda_{q}^{(1)})_{il}(\Lambda_{q}^{(1)})_{kj}}{M_{1}^{2}}-\frac{\log\frac{M_{3}^{2}}{M_{1}^{2}}\left[(\Lambda_{q}^{31})_{il}(\Lambda_{q}^{31\dagger})_{kj}+(\Lambda_{q}^{31\dagger})_{il}(\Lambda_{q}^{31})_{kj}\right]}{M_{3}^{2}-M_{1}^{2}}\right),
\end{align}
\begin{align}
[C_{uu}]_{ijkl}^{(1)}= & -\frac{g_{s}^{4}}{240}\left(\delta_{il}\delta_{kj}-\frac{1}{3}\delta_{ij}\delta_{kl}\right)\left(\frac{3}{M_{3}^{2}}+\frac{1}{M_{1}^{2}}\right)-\frac{N_{c}}{60}g^{\prime4}Y_{u}^{2}\delta_{ij}\delta_{kl}\left(\frac{3Y_{S_{3}}^{2}}{M_{3}^{2}}+\frac{Y_{S_{1}}^{2}}{M_{1}^{2}}\right)+\nonumber \\
 & +\frac{1}{6}g^{\prime2}Y_{u}\left(\delta_{kl}\delta_{im}\delta_{jn}+\delta_{ij}\delta_{km}\delta_{ln}\right)\left(\frac{{-} Y_{e}-Y_{S_{1}}}{6}+Y_{e}L_{1}\right)\frac{(\Lambda_{u})_{mn}}{M_{1}^{2}}+\nonumber \\
 & +\frac{1}{72}g_{s}^{2}\left(-\frac{1}{3}\delta_{kl}\delta_{im}\delta_{jn}-\frac{1}{3}\delta_{ij}\delta_{km}\delta_{ln}+\delta_{kj}\delta_{im}\delta_{ln}+\delta_{il}\delta_{km}\delta_{jn}\right)\frac{(\Lambda_{u})_{mn}}{M_{1}^{2}}+\nonumber \\
 & -\frac{1}{8}\frac{(\Lambda_{u})_{il}(\Lambda_{u})_{kj}}{M_{1}^{2}},\\{}
[C_{dd}]_{ijkl}^{(1)}= & -\frac{g_{s}^{4}}{240}(\delta_{il}\delta_{kj}-\frac{1}{3}\delta_{ij}\delta_{kl})\left(\frac{3}{M_{3}^{2}}+\frac{1}{M_{1}^{2}}\right)-\frac{N_{c}}{60}g^{\prime4}Y_{d}^{2}\delta_{ij}\delta_{kl}\left(\frac{3Y_{S_{3}}^{2}}{M_{3}^{2}}+\frac{Y_{S_{1}}^{2}}{M_{1}^{2}}\right),
\end{align}
\begin{align}
[C_{ud}^{(1)}]_{ijkl}^{(1)}= & -\frac{N_{c}}{30}g^{\prime4}Y_{d}Y_{d}\delta_{ij}\delta_{kl}\left(\frac{3Y_{S_{3}}^{2}}{M_{3}^{2}}+\frac{Y_{S_{1}}^{2}}{M_{1}^{2}}\right)+\nonumber \\
 & +\frac{1}{3}g^{\prime2}Y_{d}\left(\frac{{-} Y_{e}-Y_{S_{1}}}{6}+Y_{e}L_{1}\right)\frac{\delta_{kl}(\Lambda_{u})_{ij}}{M_{1}^{2}},\\{}
[C_{ud}^{(8)}]_{ijkl}^{(1)}= & -\frac{g_{s}^{4}}{60}\delta_{ij}\delta_{kl}\left(\frac{3}{M_{3}^{2}}+\frac{1}{M_{1}^{2}}\right)+\frac{1}{18}g_{s}^{2}\frac{\delta_{kl}(\Lambda_{u})_{ij}}{M_{1}^{2}},\\
[C_{qu}^{(1)}]_{ijkl}^{(1)}= & -\frac{N_{c}}{30}g^{\prime4}Y_{q}Y_{u}\delta_{ij}\delta_{kl}\left(\frac{3Y_{S_{3}}^{2}}{M_{3}^{2}}+\frac{Y_{S_{1}}^{2}}{M_{1}^{2}}\right)+\nonumber \\
 & +\frac{1}{3}g^{\prime2}Y_{u}\left\{ 3\left({\frac{8 Y_\ell -Y_{S_3}}{6}}+Y_{\ell}L_{3}\right)\frac{\delta_{kl}(\Lambda_{q}^{(3)})_{ij}}{M_{3}^{2}}+\left({\frac{8 Y_\ell -Y_{S_1}}{6}}+Y_{\ell}L_{1}\right)\frac{\delta_{kl}(\Lambda_{q}^{(1)})_{ij}}{M_{1}^{2}}\right\} +\nonumber \\
 & +\frac{1}{3}g^{\prime2}Y_{q}\left(\frac{{-}Y_{e}-Y_{S_{1}}}{6}+Y_{e}L_{1}\right)\frac{\delta_{ij}(\Lambda_{u})_{kl}}{M_{1}^{2}}+\nonumber \\
 & +\frac{1}{12}\left(\frac{{3}}{2}+L_{1}\right)\frac{(y_{U}^{\dagger})_{kj}(X_{1E}^{1L})_{il}+(y_{U})_{il}(X_{1E}^{1L\dagger})_{kj}}{M_{1}^{2}}-\frac{1}{12}\frac{(\Lambda_{q}^{(1)})_{ij}(\Lambda_{u})_{kl}}{M_{1}^{2}},\\{}
[C_{qu}^{(8)}]_{ijkl}^{(1)}= & -\frac{1}{60}g_{s}^{4}\delta_{ij}\delta_{kl}\left(\frac{3}{M_{3}^{2}}+\frac{1}{M_{1}^{2}}\right)+\nonumber \\
 & +\frac{1}{18}g_{s}^{2}\left(3\frac{(\Lambda_{q}^{(3)})_{ij}\delta_{kl}}{M_{3}^{2}}+\frac{(\Lambda_{q}^{(1)})_{ij}\delta_{kl}}{M_{1}^{2}}\right)+\frac{1}{18}g_{s}^{2}\frac{\delta_{ij}(\Lambda_{u})_{kl}}{M_{1}^{2}}+\nonumber \\
 & +\frac{1}{2}\left(\frac{{3}}{2}+L_{1}\right)\frac{(X_{1E}^{1L})_{il}(y_{U}^{\dagger})_{kj}+(y_{U})_{il}(X_{1E}^{1L\dagger})_{kj}}{M_{1}^{2}}-\frac{1}{2}\frac{(\Lambda_{q}^{(1)})_{ij}(\Lambda_{u})_{kl}}{M_{1}^{2}},\\{}
[C_{qd}^{(1)}]_{ijkl}^{(1)}= & -\frac{N_{c}}{30}g^{\prime4}Y_{q}Y_{d}\delta_{ij}\delta_{kl}\left(\frac{3Y_{S_{3}}^{2}}{M_{3}^{2}}+\frac{Y_{S_{1}}^{2}}{M_{1}^{2}}\right)+\nonumber \\
 & +\frac{1}{3}g^{\prime2}Y_{d}\left\{ 3\left({\frac{8Y_\ell-Y_{S_3}}{6}}+Y_{\ell}L_{3}\right)\frac{\delta_{kl}(\Lambda_{q}^{(3)})_{ij}}{M_{3}^{2}}+\left({\frac{8Y_\ell-Y_{S_1}}{6}}+Y_{\ell}L_{1}\right)\frac{\delta_{kl}(\Lambda_{q}^{(1)})_{ij}}{M_{1}^{2}}\right\} ,\\{}
[C_{qd}^{(8)}]_{ijkl}^{(1)}= & -\frac{1}{60}g_{s}^{4}\delta_{ij}\delta_{kl}\left(\frac{3}{M_{3}^{2}}+\frac{1}{M_{1}^{2}}\right)+\frac{1}{18}g_{s}^{2}\left(3\frac{(\Lambda_{q}^{(3)})_{ij}\delta_{kl}}{M_{3}^{2}}+\frac{(\Lambda_{q}^{(1)})_{ij}\delta_{kl}}{M_{1}^{2}}\right),\\
	[C_{quqd}^{(1)}]_{ijkl}^{(1)}=& -\frac{1}{2}\left(\frac{{3}}{2} + L_{1}\right)\frac{(X_{1E}^{1L})_{ij}(y_{D})_{kl}}{M_{1}^{2}}, \\
	[C_{quqd}^{(8)}]_{ijkl}^{(1)}=& 0.
\end{align}

\paragraph{Four-lepton}

\begin{align}
[C_{\ell\ell}]_{\alpha\beta\gamma\delta}^{(1)}= & -\frac{N_{c}}{120}g^{4}(2\delta_{\alpha\delta}\delta_{\gamma\beta}-\delta_{\alpha\beta}\delta_{\gamma\delta})\frac{1}{M_{3}^{2}}-\frac{N_{c}}{60}g^{\prime4}Y_{\ell}^{2}\delta_{\alpha\beta}\delta_{\gamma\delta}\left(\frac{3Y_{S_{3}}^{2}}{M_{3}^{2}}+\frac{Y_{S_{1}}^{2}}{M_{1}^{2}}\right)+\nonumber \\
 & +\frac{N_{c}}{24}g^{2}\left(-\delta_{\gamma\delta}\delta_{\alpha\rho}\delta_{\beta\sigma}+2\delta_{\gamma\beta}\delta_{\alpha\rho}\delta_{\delta\sigma}-\delta_{\alpha\beta}\delta_{\gamma\rho}\delta_{\delta\sigma}+2\delta_{\alpha\delta}\delta_{\gamma\rho}\delta_{\beta\sigma}\right)\times\nonumber \\
 & \times\left(\left(L_{3}+{2}\right)\frac{(\Lambda_{\ell}^{(3)})_{\rho\sigma}}{M_{3}^{2}}-\left(L_{1}{+\frac{4}{3}}\right)\frac{(\Lambda_{\ell}^{(1)})_{\rho\sigma}}{M_{1}^{2}}\right)+\nonumber \\
 & +\frac{N_{c}}{6}g^{\prime2}Y_{\ell}\left(\delta_{\gamma\delta}\delta_{\alpha\rho}\delta_{\beta\sigma}+\delta_{\alpha\beta}\delta_{\gamma\rho}\delta_{\delta\sigma}\right)\times\nonumber \\
 & \times\left[3\left({\frac{8Y_q-Y_{S_3}}{6}}+Y_{q}L_{3}\right)\frac{(\Lambda_{\ell}^{(3)})_{\rho\sigma}}{M_{3}^{2}}+\left({\frac{8Y_q-Y_{S_1}}{6}}+Y_{q}L_{1}\right)\frac{(\Lambda_{\ell}^{(1)})_{\rho\sigma}}{M_{1}^{2}}\right]+\nonumber \\
 & -\frac{N_{c}}{8}\left\{ \frac{(\Lambda_{\ell}^{(3)})_{\alpha\beta}(\Lambda_{\ell}^{(3)})_{\gamma\delta}}{M_{3}^{2}}+4\frac{(\Lambda_{\ell}^{(3)})_{\alpha\delta}(\Lambda_{\ell}^{(3)})_{\gamma\beta}}{M_{3}^{2}}+\frac{(\Lambda_{\ell}^{(1)})_{\alpha\beta}(\Lambda_{\ell}^{(1)})_{\gamma\delta}}{M_{1}^{2}}+\right.\nonumber \\
 & -\frac{\log\frac{M_{3}^{2}}{M_{1}^{2}}}{M_{3}^{2}-M_{1}^{2}}\left[(\Lambda_{\ell}^{(31)})_{\alpha\beta}(\Lambda_{\ell}^{(31)\dagger})_{\gamma\delta}+(\Lambda_{\ell}^{(31)\dagger})_{\alpha\beta}(\Lambda_{\ell}^{(31)})_{\gamma\delta}\right.+\nonumber \\
 & \left.\left.-2(\Lambda_{\ell}^{(31)})_{\alpha\delta}(\Lambda_{\ell}^{(31)\dagger})_{\gamma\beta}-2(\Lambda_{\ell}^{(31)\dagger})_{\alpha\delta}(\Lambda_{\ell}^{(31)})_{\gamma\beta}\right]\right\} ,
\end{align}
\begin{align}
[C_{ee}]_{\alpha\beta\gamma\delta}^{(1)}= & -\frac{N_{c}}{60}g^{\prime4}Y_{e}^{2}\delta_{\alpha\beta}\delta_{\gamma\delta}\left(\frac{3Y_{S_{3}}^{2}}{M_{3}^{2}}+\frac{Y_{S_{1}}^{2}}{M_{1}^{2}}\right)+\nonumber \\
 & +\frac{N_{c}}{6}g^{\prime2}Y_{\ell}\left(\frac{{-}Y_{u}-Y_{S_{1}}}{6}+Y_{u}L_{1}\right)\left(\delta_{\gamma\delta}\delta_{\alpha\rho}\delta_{\beta\sigma}+\delta_{\alpha\beta}\delta_{\gamma\rho}\delta_{\delta\sigma}\right)\frac{(\Lambda_{e})_{\rho\sigma}}{M_{1}^{2}}+\nonumber \\
 & -\frac{N_{c}}{8}\frac{(\Lambda_{e})_{\alpha\beta}(\Lambda_{e})_{\gamma\delta}}{M_{1}^{2}},
\end{align}
\begin{align}
[C_{\ell e}]_{\alpha\beta\gamma\delta}^{(1)}= & -\frac{N_{c}}{30}g^{\prime4}Y_{\ell}Y_{e}\delta_{\alpha\beta}\delta_{\gamma\delta}\left(\frac{3Y_{S_{3}}^{2}}{M_{3}^{2}}+\frac{Y_{S_{1}}^{2}}{M_{1}^{2}}\right)+\nonumber \\
 & +\frac{N_{c}}{3}g^{\prime2}Y_{e}\left(3\left({\frac{8Y_q-Y_{S_3}}{6}}+Y_{q}L_{3}\right)\frac{(\Lambda_{\ell}^{(3)})_{\alpha\beta}\delta_{\gamma\delta}}{M_{3}^{2}}+\left({\frac{8Y_q-Y_{S_3}}{6}}+Y_{q}L_{1}\right)\frac{(\Lambda_{\ell}^{(1)})_{\alpha\beta}\delta_{\gamma\delta}}{M_{1}^{2}}\right)+\nonumber \\
 & +\frac{N_{c}}{3}g^{\prime2}Y_{\ell}\left(\frac{{-}Y_{u}-Y_{S_{1}}}{6}+Y_{u}L_{1}\right)\frac{\delta_{\alpha\beta}(\Lambda_{e})_{\gamma\delta}}{M_{1}^{2}}+\nonumber \\
 & +\frac{N_{c}}{4}\left(\frac{{3}}{2}+L_{1}\right)\frac{(X_{1U}^{1L})_{\alpha\delta}(y_{E}^{\dagger})_{\gamma\beta}+(X_{1U}^{1L\dagger})_{\gamma\beta}(y_{E})_{\alpha\delta}}{M_{1}^{2}}-\frac{N_{c}}{4}\dfrac{(\Lambda_{\ell}^{(1)})_{\alpha\beta}(\Lambda_{e})_{\gamma\delta}}{M_{1}^{2}}.
\end{align}

\paragraph{Semileptonic}

\begin{align}
[C_{\ell q}^{(1)}]_{\alpha\beta ij}^{(1)}= & \frac{1}{4}\left(\frac{1}{2}+a_{\text{ev}}\right)\left[g_{s}^{2}\dfrac{N_c^2-1}{2N_c}+g^{\prime2}(Y_{q}-Y_{\ell})^{2}\right]\left(\frac{3\lambda_{i\alpha}^{3L*}\lambda_{j\beta}^{3L}}{M_{3}^{2}}+\frac{\lambda_{i\alpha}^{1L*}\lambda_{j\beta}^{1L}}{M_{1}^{2}}\right)+\nonumber\\
&+\frac{1}{4}\left(\frac{1}{2}+a_{\text{ev}}\right) g^2 \, 3\, \left(\frac{\lambda_{i\alpha}^{3L*}\lambda_{j\beta}^{3L}}{M_{3}^{2}}+\frac{\lambda_{i\alpha}^{1L*}\lambda_{j\beta}^{1L}}{M_{1}^{2}}\right)+\nonumber\\
&-\left[\frac{1}{2}(\delta Z_{\ell})_{\alpha\gamma}\delta_{\beta\delta}\delta_{ik}\delta_{lj}+\frac{1}{2}\delta{}_{\alpha\gamma}(\delta Z_{\ell})_{\delta\beta}\delta_{ik}\delta_{lj}+\right.\nonumber \\
 & \left.+\frac{1}{2}\delta_{\alpha\gamma}\delta_{\beta\delta}(\delta Z_{q})_{ik}\delta_{jl}+\frac{1}{2}\delta_{\alpha\gamma}\delta_{\beta\delta}\delta_{ik}(\delta Z_{q})_{lj}\right][C_{\ell q}^{(1)}]_{\gamma\delta kl}^{(0)} -\frac{N_{c}}{30}g^{\prime4}Y_{\ell}Y_{q}\delta_{\alpha\beta}\delta_{ij}\left(\frac{3Y_{S_{3}}^{2}}{M_{3}^{2}}+\frac{Y_{S_{1}}^{2}}{M_{1}^{2}}\right)+\nonumber \\
 & +\frac{1}{3}g^{\prime2}Y_{\ell}\delta_{\alpha\beta}\left\{ 3\left({\frac{8Y_\ell-Y_{S_3}}{6}}+Y_{\ell}L_{3}\right)\frac{(\Lambda_{q}^{(3)})_{ij}}{M_{3}^{2}}+\left({\frac{8Y_\ell-Y_{S_1}}{6}}+Y_{\ell}L_{1}\right)\frac{(\Lambda_{q}^{(1)})_{ij}}{M_{1}^{2}}\right\} +\nonumber \\
 & +\frac{N_{c}}{3}g^{\prime2}Y_{q}\delta_{ij}\left(3\left({\frac{8Y_q-Y_{S_3}}{6}}+Y_{q}L_{3}\right)\frac{(\Lambda_{\ell}^{(3)})_{\alpha\beta}}{M_{3}^{2}}+\left({\frac{8Y_q-Y_{S_1}}{6}}+Y_{q}L_{1}\right)\frac{(\Lambda_{\ell}^{(1)})_{\alpha\beta}}{M_{1}^{2}}\right)+\nonumber \\
 & -\frac{1}{4}\left(3\frac{(\Lambda_{\ell}^{(3)})_{\alpha\beta}(\Lambda_{q}^{(3)})_{ij}}{M_{3}^{2}}+\frac{(\Lambda_{\ell}^{(1)})_{\alpha\beta}(\Lambda_{q}^{(1)})_{ij}}{M_{1}^{2}}\right)+\nonumber\\
& +\left(c_{1}(1+L_{1})+\frac{9}{4}(1+L_{3})c_{13}^{(1)}\frac{M_{3}^{2}}{M_{1}^{2}}\right)\frac{\lambda_{\alpha i}^{1L\dagger}\lambda_{j\beta}^{1L}}{M_{1}^{2}}+\nonumber\\
 & +\left(\frac{9}{4}(1+L_{1})c_{13}^{(1)}\frac{M_{1}^{2}}{M_{3}^{2}}+\frac{3}{2}(1+L_{3})\left[5c_{3}^{(1)}-c_{3}^{(3)}+\frac{5}{6}c_{3}^{(5)}\right]\right)\frac{\lambda_{\alpha i}^{3L\dagger}\lambda_{j\beta}^{3L}}{M_{3}^{2}},\\{}
[C_{\ell q}^{(3)}]_{\alpha\beta ij}^{(1)}= & \frac{1}{4}\left(\frac{1}{2}+a_{\text{ev}}\right)\left[g_{s}^{2}\dfrac{N_c^2-1}{2N_c}+g^{\prime2}(Y_{q}-Y_{\ell})^{2}\right]\left(\frac{\lambda_{i\alpha}^{3L*}\lambda_{j\beta}^{3L}}{M_{3}^{2}}-\frac{\lambda_{i\alpha}^{1L*}\lambda_{j\beta}^{1L}}{M_{1}^{2}}\right)+\nonumber\\
&+\frac{1}{4}\left(\frac{1}{2}+a_{\text{ev}}\right) g^2 \left(-\frac{3\lambda_{i\alpha}^{3L*}\lambda_{j\beta}^{3L}}{M_{3}^{2}}+\frac{\lambda_{i\alpha}^{1L*}\lambda_{j\beta}^{1L}}{M_{1}^{2}}\right)+\nonumber\\
&-\left[\frac{1}{2}(\delta Z_{\ell})_{\alpha\gamma}\delta_{\beta\delta}\delta_{ik}\delta_{lj}+\frac{1}{2}\delta{}_{\alpha\gamma}(\delta Z_{\ell})_{\delta\beta}\delta_{ik}\delta_{lj}+\right.\nonumber \\
 & \left.+\frac{1}{2}\delta_{\alpha\gamma}\delta_{\beta\delta}(\delta Z_{q})_{ik}\delta_{jl}+\frac{1}{2}\delta_{\alpha\gamma}\delta_{\beta\delta}\delta_{ik}(\delta Z_{q})_{lj}\right][C_{\ell q}^{(3)}]_{\gamma\delta kl}^{(0)} -\frac{N_{c}}{60}g^{4}\delta_{\alpha\beta}\delta_{ij}\frac{1}{M_{3}^{2}}+\nonumber \\
 & +\frac{1}{12}g^{2}\delta_{\alpha\beta}\left(\left({2}+L_{3}\right)\frac{(\Lambda_{q}^{(3)})_{ij}}{M_{3}^{2}}-\left({\frac{4}{3}}+L_{1}\right)\frac{(\Lambda_{q}^{(1)})_{ij}}{M_{1}^{2}}\right)+\nonumber \\
 & +\frac{N_{c}}{12}g^{2}\delta_{ij}\left(\left({2}+L_{3}\right)\frac{(\Lambda_{\ell}^{(3)})_{\alpha\beta}}{M_{3}^{2}}-\left({\frac{4}{3}}+L_{1}\right)\frac{(\Lambda_{\ell}^{(1)})_{\alpha\beta}}{M_{1}^{2}}\right)+\nonumber \\
 & -\frac{1}{4}\left(2\frac{(\Lambda_{\ell}^{(3)})_{\alpha\beta}(\Lambda_{q}^{(3)})_{ij}}{M_{3}^{2}}+\frac{\log\frac{M_{3}^{2}}{M_{1}^{2}}\left[(\Lambda_{\ell}^{(31)})_{\alpha\beta}(\Lambda_{q}^{(31)\dagger})_{ij}+(\Lambda_{\ell}^{(31)\dagger})_{\alpha\beta}(\Lambda_{q}^{(31)})_{ij}\right]}{M_{3}^{2}-M_{1}^{2}}\right)+\nonumber\\
& -\left(c_{1}(1+L_{1})+\frac{9}{4}(1+L_{3})c_{13}^{(1)}\frac{M_{3}^{2}}{M_{1}^{2}}\right)\frac{\lambda_{\alpha i}^{1L\dagger}\lambda_{j\beta}^{1L}}{M_{1}^{2}}+\nonumber\\
 &+\left(\frac{3}{4}(1+L_{1})c_{13}^{(1)}\frac{M_{1}^{2}}{M_{3}^{2}}+\frac{1}{2}(1+L_{3})\left[5c_{3}^{(1)}-c_{3}^{(3)}+\frac{5}{6}c_{3}^{(5)}\right]\right)\frac{\lambda_{\alpha i}^{3L\dagger}\lambda_{j\beta}^{3L}}{M_{3}^{2}},
\end{align}
\begin{align}
[C_{eu}]_{\alpha\beta ij}^{(1)}= & \frac{1}{2}\left(\frac{1}{2}+a_{\text{ev}}\right)\left[g_{s}^{2}\dfrac{N_c^2-1}{2N_c}+g^{\prime2}(Y_{u}-Y_{e})^{2}\right]\frac{\lambda_{i\alpha}^{1R*}\lambda_{j\beta}^{1R}}{M_{1}^{2}}+\nonumber\\
&-\left[\frac{1}{2}(\delta Z_{e})_{\alpha\gamma}\delta_{\beta\delta}\delta_{ik}\delta_{lj}+\frac{1}{2}\delta{}_{\alpha\gamma}(\delta Z_{e})_{\delta\beta}\delta_{ik}\delta_{lj}+\right.\nonumber \\
 & \left.+\frac{1}{2}\delta_{\alpha\gamma}\delta_{\beta\delta}(\delta Z_{u})_{ik}\delta_{jl}+\frac{1}{2}\delta_{\alpha\gamma}\delta_{\beta\delta}\delta_{ik}(\delta Z_{u})_{lj}\right][C_{eu}]_{\gamma\delta kl}^{(0)}+\nonumber\\
 & -\frac{N_{c}}{30}g^{\prime4}Y_{e}Y_{u}\delta_{\alpha\beta}\delta_{ij}\left(\frac{3Y_{S_{3}}^{2}}{M_{3}^{2}}+\frac{Y_{S_{1}}^{2}}{M_{1}^{2}}\right)+\nonumber \\
 & +\frac{1}{3}g^{\prime2}Y_{e}\left(\frac{{-}Y_{e}-Y_{S_{1}}}{6}+Y_{e}L_{1}\right)\frac{\delta_{\alpha\beta}(\Lambda_{u})_{ij}}{M_{1}^{2}}+\nonumber \\
 & +\frac{N_{c}}{3}g^{\prime2}Y_{u}\left(\frac{{-}Y_{u}-Y_{S_{1}}}{6}+Y_{u}L_{1}\right)\frac{\delta_{ij}(\Lambda_{e})_{\alpha\beta}}{M_{1}^{2}}+\nonumber \\
 & -\frac{1}{4}\frac{(\Lambda_{e})_{\alpha\beta}(\Lambda_{u})_{ij}}{M_{1}^{2}}+\left(2c_{1}(1+L_{1})+\frac{9}{2}(1+L_{3})c_{13}^{(1)}\frac{M_{3}^{2}}{M_{1}^{2}}\right)\frac{\lambda_{\alpha i}^{1R\dagger}\lambda_{j\beta}^{1R}}{M_{1}^{2}}, \\
[C_{ed}]_{\alpha\beta ij}^{(1)}= & -\frac{N_{c}}{30}g^{\prime4}Y_{e}Y_{d}\delta_{\alpha\beta}\delta_{ij}\left(\frac{3Y_{S_{3}}^{2}}{M_{3}^{2}}+\frac{Y_{S_{1}}^{2}}{M_{1}^{2}}\right)+\nonumber \\
 & +\frac{N_{c}}{3}g^{\prime2}Y_{d}\left(\frac{{-}Y_{u}-Y_{S_{1}}}{6}+Y_{u}L_{1}\right)\frac{\delta_{ij}(\Lambda_{e})_{\alpha\beta}}{M_{1}^{2}}, \\
[C_{qe}]_{ij\alpha\beta}^{(1)}= & -\frac{N_{c}}{30}g^{\prime4}Y_{q}Y_{e}\delta_{ij}\delta_{\alpha\beta}\left(\frac{3Y_{S_{3}}^{2}}{M_{3}^{2}}+\frac{Y_{S_{1}}^{2}}{M_{1}^{2}}\right)+\nonumber \\
 & +\frac{1}{3}g^{\prime2}Y_{e}\left\{ 3\left(\frac{{8 Y_\ell - Y_{S_3}}}{6}+Y_{\ell}L_{3}\right)\frac{(\Lambda_{q}^{(3)})_{ij}\delta_{\alpha\beta}}{M_{3}^{2}}+\left(\frac{{8 Y_\ell - Y_{S_1}}}{6}+Y_{\ell}L_{1}\right)\frac{(\Lambda_{q}^{(1)})_{ij}\delta_{\alpha\beta}}{M_{1}^{2}}\right\} +\nonumber \\
 & +\frac{N_{c}}{3}g^{\prime2}Y_{q}\left(\frac{{-}Y_{u}-Y_{S_{1}}}{6}+Y_{u}L_{1}\right)\frac{\delta_{ij}(\Lambda_{e})_{\alpha\beta}}{M_{1}^{2}}+\nonumber \\
 & -\frac{1}{4}\frac{(\Lambda_{q}^{(1)})_{ij}(\Lambda_{e})_{\alpha\beta}}{M_{1}^{2}}-\frac{3}{4}\left(\frac{3}{2}+L_{3}\right)\frac{(\lambda^{3L*}y_{E}^{*})_{i\alpha}(\lambda^{3L}y_{E})_{j\beta}}{M_{3}^{2}}+\nonumber \\
 & -\frac{1}{4}\left(\frac{3}{2}+L_{1}\right)\frac{(\lambda^{1L*}y_{E}^{*}-y_{U}\lambda^{1R*})_{i\alpha}(\lambda^{1L}y_{E}-y_{U}^{*}\lambda^{1R})_{j\beta}}{M_{1}^{2}},
 \end{align}
\begin{align}
[C_{\ell u}]_{\alpha\beta ij}^{(1)}= & -\frac{N_{c}}{30}g^{\prime4}Y_{\ell}Y_{u}\delta_{ij}\delta_{\alpha\beta}\left(\frac{3Y_{S_{3}}^{2}}{M_{3}^{2}}+\frac{Y_{S_{1}}^{2}}{M_{1}^{2}}\right)+\nonumber \\
 & +\frac{1}{3}g^{\prime2}Y_{\ell}\left(\frac{{-}Y_{e}-Y_{S_{1}}}{6}+Y_{e}L_{1}\right)\frac{\delta_{\alpha\beta}(\Lambda_{u})_{ij}}{M_{1}^{2}}+\nonumber \\
 & +\frac{N_{c}}{3}g^{\prime2}Y_{u}\left(3\left(\frac{{8 Y_q - Y_{S_3}}}{6}+Y_{q}L_{3}\right)\frac{(\Lambda_{\ell}^{(3)})_{\alpha\beta}\delta_{ij}}{M_{3}^{2}}+\left(\frac{{8 Y_q - Y_{S_1}}}{6}+Y_{q}L_{1}\right)\frac{(\Lambda_{\ell}^{(1)})_{\alpha\beta}\delta_{ij}}{M_{1}^{2}}\right)+\nonumber \\
 & -\frac{1}{4}\frac{(\Lambda_{\ell}^{(1)})_{\alpha\beta}(\Lambda_{u})_{ij}}{M_{1}^{2}}-\frac{3}{4}\left(\frac{3}{2}+L_{3}\right)\frac{(\lambda^{3L\dagger}y_{U}^{*})_{\alpha i}(\lambda^{3L\,T}y_{U})_{\beta j}}{M_{3}^{2}}+\nonumber \\
 & -\frac{1}{4}\left(\frac{3}{2}+L_{1}\right)\dfrac{(\lambda^{1L\,\dagger}y_{U}^{*}-y_{E}\lambda^{1R\dagger})_{\alpha i}(\lambda^{1L\,T}y_{U}-y_{E}^{*}\lambda^{1RT})_{\beta j}}{M_{1}^{2}},\\{}
[C_{\ell d}]_{\alpha\beta ij}^{(1)}= & -\frac{N_{c}}{30}g^{\prime4}Y_{\ell}Y_{d}\delta_{ij}\delta_{\alpha\beta}\left(\frac{3Y_{S_{3}}^{2}}{M_{3}^{2}}+\frac{Y_{S_{1}}^{2}}{M_{1}^{2}}\right)+\nonumber \\
 & +\frac{N_{c}}{3}g^{\prime2}Y_{d}\left(3\left(\frac{{8 Y_q - Y_{S_3}}}{6} +Y_{q}L_{3}\right)\frac{(\Lambda_{\ell}^{(3)})_{\alpha\beta}\delta_{ij}}{M_{3}^{2}}+\left(\frac{{8 Y_q - Y_{S_1}}}{6}+Y_{q}L_{1}\right)\frac{(\Lambda_{\ell}^{(1)})_{\alpha\beta}\delta_{ij}}{M_{1}^{2}}\right)+\nonumber \\
 & -\frac{1}{4}\left(3\left(\frac{3}{2}+L_{3}\right)\frac{(\lambda^{3L\dagger}y_{D}^{*})_{\alpha i}(\lambda^{3L\,T}y_{D})_{\beta j}}{M_{3}^{2}}+\left(\frac{3}{2}+L_{1}\right)\frac{(\lambda^{1L\,\dagger}y_{D}^{*})_{\alpha i}(\lambda^{1L\,T}y_{D})_{\beta j}}{M_{1}^{2}}\right),
\end{align}

\begin{align}
[C_{\ell edq}]_{\alpha\beta ij}^{(1)}= & -\frac{N_{c}}{2}\left(\frac{{3}}{2}+L_{1}\right)\frac{(y_{D}^{\dagger})_{ij}(X_{1U}^{1L})_{\alpha\beta}}{M_{1}^{2}}+\nonumber \\
 & -\frac{1}{2}\left(-3\left(\frac{3}{2}+L_{3}\right)\frac{(\lambda^{3L\,\dagger}y_{D}^{*})_{\alpha i}(\lambda^{3L}y_{E})_{j\beta}}{M_{3}^{2}}+\left(\frac{3}{2}+L_{1}\right)\frac{(\lambda^{1L\,\dagger}y_{D}^{*})_{\alpha i}(\lambda^{1L}y_{E})_{j\beta}}{M_{1}^{2}}\right)+\nonumber \\
 & +\frac{1}{2}\left(\frac{3}{2}+L_{1}\right)\frac{(\lambda^{1L\,\dagger}y_{D}^{*})_{\alpha i}(y_{U}^{*}\lambda^{1R})_{j\beta}}{M_{1}^{2}},
\end{align}

\begin{align}
[C_{\ell equ}^{(1)}]_{\alpha\beta ij}^{(1)}= & -\left[\frac{1}{2}(\delta Z_{\ell})_{\alpha\gamma}\delta_{\beta\delta}\delta_{ik}\delta_{lj}+\frac{1}{2}\delta{}_{\alpha\gamma}(\delta Z_{e})_{\delta\beta}\delta_{ik}\delta_{lj}+\right.\nonumber \\
 & \left.+\frac{1}{2}\delta_{\alpha\gamma}\delta_{\beta\delta}(\delta Z_{q})_{ik}\delta_{jl}+\frac{1}{2}\delta_{\alpha\gamma}\delta_{\beta\delta}\delta_{ik}(\delta Z_{u})_{lj}\right][C_{\ell equ}^{(1)}]_{\gamma\delta kl}^{(0)}+\nonumber\\
 & +\frac{1}{2}\left({\frac{3}{2}} + L_{1}\right)\frac{(y_{E})_{\alpha\beta}(X_{1E}^{1L})_{ij}}{M_{1}^{2}}+\frac{N_{c}}{2}\left({\frac{3}{2}}+L_{1}\right)\frac{(y_{U})_{ij}(X_{1U}^{1L})_{\alpha\beta}}{M_{1}^{2}}+\nonumber \\
+ & \frac{\lambda_{\alpha i}^{1L\dagger}\lambda_{j\beta}^{1R}}{M_{1}^{2}}\left\{ 2(1+L_{1})c_{1}+\frac{9}{2}(1+L_{3})c_{13}^{(1)}\frac{M_{3}^{2}}{M_{1}^{2}}+\right.\nonumber \\
 & \left.-\frac{3}{2}\left(\frac{3}{2}+L_{1}\right)\left[(Y_{q}-Y_{\ell})(Y_{u}-Y_{e})g^{\prime2}+\frac{N_c^2-1}{2 N_c}g_{s}^{2}\right]\right\},
 \end{align}
\begin{align}
[C_{\ell equ}^{(3)}]_{\alpha\beta ij}^{(1)}= & -\left[\frac{1}{2}(\delta Z_{\ell})_{\alpha\gamma}\delta_{\beta\delta}\delta_{ik}\delta_{lj}+\frac{1}{2}\delta{}_{\alpha\gamma}(\delta Z_{e})_{\delta\beta}\delta_{ik}\delta_{lj}\right.\nonumber \\
 & \left.+\frac{1}{2}\delta_{\alpha\gamma}\delta_{\beta\delta}(\delta Z_{q})_{ik}\delta_{jl}+\frac{1}{2}\delta_{\alpha\gamma}\delta_{\beta\delta}\delta_{ik}(\delta Z_{u})_{lj}\right][C_{\ell equ}^{(3)}]_{\gamma\delta kl}^{(0)}+\nonumber\\
 & +\frac{\lambda_{\alpha i}^{1L\dagger}\lambda_{j\beta}^{1R}}{M_{1}^{2}}\left\{ -\frac{1}{4}(1+L_{1})c_{1}-\frac{9}{8}(1+L_{3})c_{13}^{(1)}\frac{M_{3}^{2}}{M_{1}^{2}}+\right.\nonumber \\
 & \left.-\frac{1}{8}\left(\frac{3}{2}+L_{1}\right)\left[(Y_{q}-Y_{\ell})(Y_{u}-Y_{e})g^{\prime2}+\frac{N_c^2-1}{2 N_c}g_{s}^{2}\right]\right\} .
\end{align}

\section{Conclusions}
\label{sec:conclusions}

In this work we have presented the complete one-loop matching conditions, up to dimension-six SMEFT operators, for the $S_1 + S_3$ leptoquark model. This is one of the few available examples of a complete one-loop matching onto the SMEFT, and is substantially richer than previous ones due to the presence of two heavy fields charged under the SM gauge groups, coupled to SM fermions with a non-trivial flavour structure, and with potential couplings with the Higgs boson as well as themselves.
The matching was performed diagrammatically, by direct comparison of full theory and EFT 1LPI off-shell Green's functions, and can serve as a cross-check for functional or computer methods devoted to the same task.

As a by-product of this work, we have extended the Warsaw basis of dimension-six SMEFT operators, to a full Green's basis, where only integration by parts (without SM EOMs) are used to reduce the number of independent operators. This set provides an operator basis for off-shell 1PI Green's functions. We have provided the complete reduction equations from Green's to Warsaw basis, which we believe to be of general interest for any kind of SMEFT matching or computation beyond the leading order. All relevant information related to the Green's basis is contained in the appendices.

The model studied in the present paper has been known for a while to provide a good candidate combined explanation of neutral- and charged-current B-physics anomalies. The tools developed in the present paper allows a thorough and complete study of the model's phenomenology, which we will explore in a separate contribution.

\subsection*{Acknowledgements}

We thank D. Sutherland and A. Crivellin for useful discussions, and K. Mantzaropoulos for pointing out some errors present in the original version of this paper, after a comparison with results obtained via functional methods (see \cite{Dedes:2021abc}).\footnote{The detailed list of corrections, all regarding semileptonic operators, can be found in the erratum of the published version.}
We also thank Matthias K\"{o}nig for spotting the error in Eqs.(C.17-C.19), Mikael Chala for finding the error in Eqs.(B.2,B.3) (see \cite{Chala:2021pll}), and Aneesh Manohar for a discussion on the matching of the dipole operators and for pointing out the error in the signs of Eqs.(C.50,C.55) (see \cite{Aebischer:2021uvt}).
DM acknowledges support by the INFN grant SESAMO and MIUR grant PRIN\_2017L5W2PT. D.M is also partially supported by the European Research Council (ERC) under the European Union’s Horizon 2020 research and innovation programme, grant agreement 833280 (FLAY). EV has been partially supported by the DFG Cluster of Excellence 2094 ORIGINS, the Collaborative Research Center SFB1258 and the BMBF grant 05H18WOCA1 and thanks the Munich Institute for Astro- and Particle Physics (MIAPP) for hospitality.

\subsubsection*{Correction notes for the v.5}

Compared to the v.4 we specified the sign convention for the Levi-Civita tensor to $\epsilon_{0123} = +1$.
Following this convention, in the reduction from the Green's to the Warsaw basis we changed the sign of the following terms: $G^\prime_{\widetilde G q}$ in Eqs.(B.12,B.15), $G^\prime_{\widetilde W q}$ in Eqs.(B.13,B.16), $G^\prime_{\widetilde B q}$ in Eqs.(B.14,B.17), $G^\prime_{\widetilde W \ell}$ in Eq.(B.18), $G^\prime_{\widetilde B \ell}$ in Eq.(B.19). This convention also implied a change in the overall sign in the Green's basis matching for operators in the same $G^\prime_{\widetilde V \psi}$ class: Eqs.(C.36,C.38,C.41,C.43,C.45,C.47).

Furthermore, we removed the $G^\prime_{HD}$ contribution from Eq.(B.3) and added it to Eq.(B.2), multiplied by a factor of $1/2$ the Green's basis matching in Eqs.(C.17-C.19), and changed the overall sign of Eqs.(C.50,C.55).

Finally, the matching to the Warsaw basis has been corrected reflecting all the changes listed above. The same corrections have been applied to the ancillary files.

{\small
\bibliography{Biblio}{}
\bibliographystyle{JHEP}}


\appendix

\section{A Green's basis for the SMEFT}
\label{app:GreenBasis}

In this Appendix we present a basis of dimension-six IBP independent SMEFT operators, i.e. the Green's basis, extending the Warsaw basis of IBP and EOM independent operators.
The operator basis is given in Tables \ref{tab:GreenBosonic} (bosonic operators), \ref{tab:GreenSingleFermion} (single fermionic current operators), and \ref{tab:4FermionBconserving},~\ref{tab:4FermionBviolating} (four fermion operators), in which Warsaw basis operators are highlighted in blue; we count 132 independent operators for a single generation of SM fermions, including baryon number violating ones.
The following discussion is mainly a re-adaptation of the line of reasoning of Ref.~\cite{Grzadkowski:2010es} (to which we refer the reader for further clarification), with the important exception that we are not allowed to use of EOMs.
The strategy is to simply examine all possible Lorentz-invariant combinations of gauge field strengths, covariant derivatives, standard model fermions and the Higgs field, denoted $X$, $D$, $\psi$ and $H$ respectively.

In Tables \ref{tab:GreenBosonic}, \ref{tab:GreenSingleFermion}, \ref{tab:4FermionBconserving} and \ref{tab:4FermionBviolating}, we list all Green's basis operators (Warsaw basis \cite{Grzadkowski:2010es} operators in blue color).

\subsection*{Bosonic operators}

\subsubsection*{$X^3$}
All independent operators are contained in Warsaw basis.
\subsubsection*{$X^2 D^2$}
If we allow $X$ to be possibly dual, there is no need to consider contractions involving the $\epsilon$ tensor. Thus, the indices of the two derivatives must either be contracted (a) between themselves, (b) with the indices of a single tensor or (c) with the indices of the two different tensors. In the case (b), antisymmetry of $X$ and $[D_\mu,D_\nu]\sim X_{\mu\nu}$ brings us to $X^3$ class. In the case (a), we first note that taking both tensors to be dual is equivalent to considering no dual tensor. For the other two possibilities, we can take all derivatives to act on a single tensor and use Bianchi identities for the Yang-Mills tensors to obtain (here $Y$ is possibly dual, but $X$ is not): $$Y^{\mu \nu}D^\rho D_\rho X_{\mu \nu}=-Y^{\mu \nu}D^\rho (D_\mu X_{\nu \rho}+D_\nu X_{\rho \mu}),$$
which is equivalent to case (c).
We are thus left with case (c) where, due to Bianchi identities $D^\mu \widetilde F_{\mu\nu} = 0$, there exist only the following three possibilities:
\be
   \OO_{2B} = - \frac{1}{2}(\partial^\mu B_{\mu\nu})^2 \, ,
   \qquad
   \OO_{2W} = - \frac{1}{2}(D^\mu W_{\mu\nu})^2 \, ,   
   \qquad
   \OO_{2G} = - \frac{1}{2}(D^\mu G_{\mu\nu})^2 \, .
\ee

\subsubsection*{$X^2H^2$}
All independent operators are contained in Warsaw basis.
\subsubsection*{$XD^4$}
Because of its antisymmetry, the indices of $X$ must be contracted with two derivatives. This brings us to class $X^2 D^2$ (see above).


\subsubsection*{$X H^2 D^2$}
By hypercharge conservation, these operators must involve one field $H$ and one conjugate $H^\dagger$. Using the IBP freedom, we can assume the two derivatives to act either on $H$ or $X$.
Moreover, since the indices of $X$ must be contracted with the two derivatives, the latters get antisymmetrized. Thus, if the two derivatives both act on $X$ or $H$, we 
are moved to $X^2 H^2$ class (see above). The remaining possibilities are two operators which, modulo a total divergence and $X^2 H^2$ class operators, can be taken as:
\be\begin{split}
\OO_{BDH} &=  \partial^\nu B_{\mu\nu}(H^\dagger i\lra{D^\mu} H) ~, \\  
   	\OO_{WDH} &=  (D^\nu W_{\mu\nu})^I (H^\dagger i\lra{D^{\mu I}} H) ~ .
\end{split}\ee
Another possibility sometimes used in the literature is to use instead the operators
$\OO_{HHB} = i (D_\mu H)^\dagger (D_\nu H) B^{\mu\nu} $ and $\OO_{HHW} = i (D_\mu H)^\dagger \sigma^I (D_\nu H) W^{I \, \mu\nu}$, related via
\begin{equation}\begin{split}
	\OO_{BDH} &= 2 \OO_{HHB} + g^\prime Y_H \OO_{HB} + \frac{g}{2} \OO_{HWB} ~,\\
	\OO_{WDH} &= 2 \OO_{HHW} + g^\prime Y_H \OO_{HWB} + \frac{g}{2} \OO_{HW}~.
\end{split}\end{equation}
\subsubsection*{$H^2 D^4$}
The covariant derivatives must be contracted between themselves, either through the metric $\eta_{\mu \nu}$ or the volume form $\epsilon$. Contracting them through the $\epsilon$ tensor moves us to $H^2 X^2$ class. Hence, modulo total divergencies, the only operator in this class is:
\be
\OO_{DH}=(D_{\mu}D^{\mu}H)^{\dagger}(D_{\nu}D^{\nu}H) ~ .
\ee

\subsubsection*{$H^4 D^2$}
Because of hypercharge conservation, these operators involve exactly two fields $H$ and two conjugate fields $H^*$. Moreover, the two derivatives must be contracted together. We must thus form Lorentz and $\text{SU(2)}_L$ singlets by choosing four fields out of the four independent scalars: $$H,\,H^*,\,D_\mu D^\mu H,\,D_\mu D^\mu H^*,$$ and the two independent vectors $$D_\mu H, D^\mu H^*,$$ and, of course, by taking exactly two derivatives. We explore all possible field contents:

\begin{enumerate}[(a)]
\item $D_\mu H,D^\mu H,H^*,H^*$. The two $H^*$'s must necessarily form a triplet, so that we have a single $\text{SU(2)}_L$ contraction. Given the field content, such operator is clearly non-hermitian.
\item $D_\mu H^*,D^\mu H^*,H,H$. The two $H$'s must necessarily form a triplet, so that we have a single $\text{SU(2)}_L$ contraction. This is the hermitian conjugate of the previous case (a).
\item $D_\mu H,H,D^\mu H^*,H^*$. If $H$ and $H^*$ are contracted in a singlet (triplet), so must be $D_\mu H$ and $D^\mu H^*$, so that we have two independent $\text{SU(2)}_L$ contractions, which are readily verified to be hermitian.
\item $D_\mu D^\mu H,H,H^*,H^*$. By integration by parts, we can reduce this to cases (a, b, c) above.
\item $H,H,H^*,D_\mu D^\mu H^*$. By integration by parts, we can reduce this to cases (a, b, c) above.
\end{enumerate}

Thus we have, modulo divergencies, four independent hermitian singlets. We can complement the two Warsaw basis ones as follows:
\be\begin{split}
\OO_{H\square} & =(H^{\dagger}H)\square(H^{\dagger}H) ~,\\
\OO_{HD} & =(H^{\dagger}D_{\mu}H)^{\dagger}(H^{\dagger}D^{\mu}H)~,\\
\OO_{HD}^{\prime} & =(H^{\dagger}H)(D_{\mu}H)^{\dagger}(D^{\mu}H)~,\\
\OO_{HD}^{\prime\prime} & =(H^{\dagger}H)D_{\mu} \left( H{}^{\dagger}i\overleftrightarrow{D}^{\mu}H \right)~.
\end{split}\ee
Other operators sometimes encountered in the literature are:
\be\begin{split}
\OO_{T}^{(1)} & =\frac{1}{2}(H^{\dagger}\overleftrightarrow{D_{\mu}}H)(H^{\dagger}\overleftrightarrow{D}^{\mu}H)=-2\OO_{HD}-\frac{1}{2}\OO_{H\square}~,\\
\OO_{T}^{(3)} & =\frac{1}{2}(H^{\dagger}\overleftrightarrow{D_{\mu}}^{I}H)(H^{\dagger}\overleftrightarrow{D}^{\mu I}H)=-2\OO_{HD}^{\prime}-\frac{1}{2}\OO_{H\square}~.
\end{split}\ee

\subsubsection*{$H^6$}
This operator is contained in Warsaw basis.

\subsection*{Two-fermion operators}
Before considering the various cases, let us make a preliminary observation: these operators are obtained by contracting a scalar, vector or tensor current built out of two fermionic fields, with a corresponding current obtained from bosonic fields; it is easy to see that the only scalar (tensor) two-fermion currents allowed are $\overline q (\sigma _{\mu \nu})u,\,\overline q (\sigma _{\mu \nu})d,\,\overline \ell (\sigma _{\mu \nu})e$ and their conjugates, while the only vector currents allowed are $\overline \psi \gamma ^\mu \psi$, with $\psi = q,\,u,\,d,\,\ell,\,e$, and $\overline u \gamma ^\mu d$ with its conjugate.
This is so because these operators conserve both lepton and baryon numbers. In fact, since $B$ changes by integer units only, clearly $\Delta B=0$. Moreover, since $\Delta (B-L)=0$ at dimension-six level \cite{Kobach:2016ami}, this also implies $\Delta L =0$.

\subsubsection*{$\Psi^2 D^3$}

By hypercharge invariance, from the list of allowed two-fermion currents above, we can only pick $J^\mu = \overline \psi \gamma ^\mu \psi$, with $\psi = q,\,u,\,d,\,\ell,\,e$. If we contract $J^\mu$ with the three covariant derivatives through an $\epsilon$ tensor, we move to $\psi ^2 D X$ class (see below). Thus, at least two derivatives must be contracted with each other. Taking all derivatives to act on $\psi$, the only remaining possibility is: 
\[
\overline{\psi}i\slashed DD^{2}\psi=\frac{1}{2}\overline{\psi}\left\{ i\slashed D,D^{2}\right\} \psi+(\psi^{2}DX).
\]
We are thus left with the five hermitian operators: 
\be
\OO_{\psi D}=\frac{i}{2}\overline{\psi}\left\{ D_{\mu}D^{\mu},\slashed D\right\} \psi~, \qquad\psi=q,\,u,\,d,\,\ell,\,e ~ .
\ee

\subsubsection*{$\Psi^2 X D $}
By hypercharge invariance, the two fermions must pair in a vector current $J^\mu = \overline \psi \gamma ^\mu \psi$, with $\psi = q,\,u,\,d,\,\ell,\,e$. If we allow $X$ to be dual, we do not need to consider contractions through the $\epsilon$ tensor, and the Lorentz structure is completely specified by $J^\mu X_{\mu \nu} D^\nu $. Using the IBP freedom, we can arrange that $D^\mu$ never acts on $\overline \psi$. We are left with the possibilities summarized by:

\be\begin{split}
\OO_{X\psi} & =(\overline{\psi}\, t_{X\psi}^{a}\gamma^{\mu}\psi)D^{\nu}X_{\mu\nu}^{a},\quad\psi\in\left\{ q,\,u,\,d,\,\ell,\,e\right\} ,\quad X\in\left\{ G,\,W,\,B\right\},\\
\OO_{X\psi}^{\prime} & =(\overline{\psi} \, t_{X\psi}^{a}\gamma^{\mu}iD^{\nu}\psi)X_{\mu\nu}^{a},\quad\psi\in\left\{ q,\,u,\,d,\,\ell,\,e\right\} ,\quad X\in\left\{ G,\,\widetilde{G},\,W,\,\widetilde{W},\,B,\,\widetilde{B}\right\}.
\end{split}\ee
Here $t_{B\psi }\equiv1$, while $t_{ W \psi}$ and $t_{ G \psi}$ are the  $\text{SU(2)}$ and $\text{SU(3)}$ generators in the representation of $\psi$ (possibly zero). These three combinations are independent, since:
\begin{itemize}
\item The Feynman rules of the pure derivative parts of $\mathcal O _{X\psi} ^{(\prime)}$ vanish at some special kinematical configuration. These configurations are distinct for $\mathcal O _{X\psi}$ and $\mathcal O _{X\psi} ^{\prime}$.
\item $\mathcal O _{X\psi} ^\prime $ is CP-odd for dual $X$, and CP-even otherwise.
\end{itemize}
 Notice that $\OO_{X\psi}$  vanishes for dual $X$, by Bianchi identities.

\subsubsection*{$\Psi^2 D^2 H$}
By gauge invariance, the allowed field contents are those appearing in the Yukawa lagrangian (i.e. $\overline q u \widetilde H$, $\overline q d H$ and $\overline \ell e H$, plus hermitian conjugates). Let us take, for the sake of clarity, $\overline \ell e H$.
Integrating by parts, we can make the two derivatives act either on $e$ or $H$. If they both act on $e$ or $H$, combining $\overline \ell$  and $e$ into a tensor current moves us to class $\psi ^2 XH$ (see below). We must then combine $\overline \ell$ and $e$ into a scalar current, and we are left with: 
\be\begin{split}
\OO_{eHD1} & =(\overline{\ell}e)D_{\mu}D^{\mu}H,\\
\OO_{eHD3} & =(\overline{\ell}D_{\mu}D^{\mu}e)H  .
\end{split}\ee
If, instead, one derivative acts on $e$ and the other on $H$, we get other two possibilities:
\be\begin{split}
\OO_{eHD2} & =(\overline{\ell}i\sigma_{\mu\nu}D^{\mu}e)D^{\nu}H,\\
\OO_{eHD4} & =(\overline{\ell}D_{\mu}e)D^{\mu}H.
\end{split}\ee
\subsubsection*{$\Psi^2 X H$}
All independent operators are contained in Warsaw basis.
\subsubsection*{$\Psi^2 D H^2$}
By Lorentz invariance, the two fermions must combine into a vector current, that is $J^\mu = \overline \psi \gamma ^\mu \psi$, with $\psi = q,\,u,\,d,\,\ell,\,e$, or $J^\mu = \overline u \gamma ^\mu d$ together with its conjugate. Moreover, currents involving the two $\text {SU}(2)$ doublets $q$ or $\ell$, can either form an $\text {SU}(2)$ singlet or triplet, to be coupled to the corresponding Higgs singlet or triplet current (cf.~Eqs.~\eqref{eq:Higgs currents}). Finally, since the external fields have three independent momenta, for a given current $J^\mu = \overline \psi _1 \gamma ^\mu \psi _2$, we can form at most three independent Lorentz singlets, which we conveniently choose as: 
\be\begin{split}
\mathcal O & =[\overline \psi _1 \gamma ^\mu \psi _2]\cdot [H^\dagger i \overleftrightarrow D_\mu H]\\
\mathcal O^\prime & =[\overline \psi _1 i \overleftrightarrow D_\mu \gamma ^\mu \psi _2]\cdot[H^\dagger H]\\
\mathcal O^{\prime\prime} & =[\overline \psi _1 \gamma ^\mu \psi _2]\cdot[D_\mu (H^\dagger H)]
\end{split}\ee
(the objects in square brackets are either $\text{SU}(2)$ singlets or triplets, when possible). This results in the operators listed in the $\psi ^2 D H^2$ box of Table~\ref{tab:GreenSingleFermion} (notice that for the non-hermitian $\overline u d$ current, the operators $\mathcal O^{\prime}_{Hud}$ and $\mathcal O^{\prime\prime}_{Hud}$ vanish identically, as $(\widetilde H ^\dagger H)=0$).

\subsubsection*{$\Psi^2 H^3$}
All independent operators are contained in Warsaw basis.

\subsection*{Four-fermion operators}
All independent operators are contained in Warsaw basis.

\section{Reduction of Green's basis to Warsaw basis}
\label{app:GreenToWarsaw}

We give in this Appendix the reduction equations from the Green's basis to the Warsaw basis, which can be obtained by applying the SM EOMs to the operator basis derived in the previous Appendix. The SM EOM are:
\be\begin{split}
	(D_\mu D^\mu H)^a &= m^2 H^a - \lambda (H^\dagger H) H^a - \bar e y_E^\dagger \ell^a + i (\sigma_2)^{ab} \bar q^b y_U u - \bar d y_D^\dagger q^a ~, \\
	(D^\nu G_{\mu\nu})^A &= g_s (\bar q_i \gamma_\mu T^A q_i + \bar u_i \gamma_\mu T^A u_i + \bar d_i \gamma_\mu T^A d_i) ~, \\
	(D^\nu W_{\mu\nu})^I &= \frac{g}{2} (H^\dagger i \Dlr_\mu^I H + \bar \ell_\alpha \gamma_\mu \sigma^I \ell_\alpha  + \bar q_i \gamma_\mu \sigma^I q_i ) ~, \\
	(\partial^\nu B_{\mu\nu}) &= g^\prime (Y_H H^\dagger i \Dlr_\mu H + \sum_f Y_f \bar f \gamma_\mu f ) ~, \\
	i \Dslash \ell &= y_E e H ~, \\
	i \Dslash e &= y_E^\dagger H^\dagger \ell ~, \\
	i \Dslash q &= y_U u \widetilde H + y_D d H ~, \\
	i \Dslash u &= y_U^\dagger \widetilde H^\dagger q ~, \\
	i \Dslash d &= y_D^\dagger H^\dag	 q ~.
\label{eq:SMEOM}\end{split}\ee
Schematically, the change of basis formulae are given in the form $C_i = \sum _{j} a_{i j} \, G_{j}$, where $C_{i}$ and $G_{j}$ are the Warsaw and Green's basis WCs, respectively, and $a_{ij}$ is a function of SM couplings. All quantities are understood to be evaluated at the same scale.


\subsection{Renormalizable operators}

\[
Z_{\Phi}=Z_{\Phi}^{\text{G}}\quad\Phi=H,\,q,\,u,\,d,\,\ell,\,e,\,G,\,W,\,B
\]

\begin{align}
\delta\lambda= & \delta\lambda^{\text{G}}-g^{2}m^{2}G_{2W}+4\lambda m^{2}G_{DH}+4gm^{2}G_{WDH} {-2m^{2}G_{HD}^{\prime}} ,\\
\delta m^{2}= & (\delta m^{2})^{\text{G}}+m^{4}G_{DH},
\end{align}

\begin{align}
(\delta y_{E}){}_{\alpha\beta}= & (\delta y_{E}^{\text{G}})_{\alpha\beta}+m^{2}G_{DH}(y_{E}){}_{\alpha\beta}-m^{2}[G_{eHD1}]_{\alpha\beta}-\frac{1}{2}m^{2}[G_{eHD2}]_{\alpha\beta}+\frac{1}{2}m^{2}[G_{eHD4}]_{\alpha\beta},\\
(\delta y_{U}){}_{ij}= & (\delta y_{U}^{\text{G}}){}_{ij}+m^{2}G_{DH}(y_{U}){}_{ij}-m^{2}[G_{uHD1}]_{ij}-\frac{1}{2}m^{2}[G_{uHD2}]_{ij}+\frac{1}{2}m^{2}[G_{uHD4}]_{ij},\\
(\delta y_{D}){}_{ij}= & (\delta y_{D}^{\text{G}}){}_{ij}+m^{2}G_{DH}(y_{D}){}_{ij}-m^{2}[G_{dHD1}]_{ij}-\frac{1}{2}m^{2}[G_{dHD2}]_{ij}+\frac{1}{2}m^{2}[G_{dHD4}]_{ij}.
\end{align}

\subsection{Purely bosonic operators}

\paragraph{$X^{3}$}

\be\begin{split}
C_{3G} & =G_{3G},\\
C_{\widetilde{3G}} & =G_{\widetilde{3G}},\\
C_{3W} & =G_{3W},\\
C_{3\widetilde{W}} & =G_{3\widetilde{W}}.
\end{split}\ee

\paragraph{$X^{2}H^{2}$}

\be\begin{split}
C_{HG} & =G_{HG},\\
C_{H\widetilde{G}} & =G_{H\widetilde{G}},\\
C_{HW} & =G_{HW},\\
C_{H\widetilde{W}} & =G_{H\widetilde{W}},\\
C_{HB} & =G_{HB},\\
C_{H\widetilde{B}} & =G_{H\widetilde{B}},\\
C_{HWB} & =G_{HWB},\\
C_{H\widetilde{W}B} & =G_{H\widetilde{W}B}.
\end{split}\ee

\paragraph{$H^{4}D^{2}$}

\begin{align}
C_{H\square} & =-\frac{3}{8}g^{2}G_{2W}-\frac{1}{2}g^{\prime}{}^{2}Y_{H}^{2}G_{2B}+\frac{3}{2}gG_{WDH}+g^{\prime}Y_{H}G_{BDH}+G_{H\square}+\frac{1}{2}G_{HD}^{\prime},\\
C_{HD} & =-2g^{\prime}{}^{2}Y_{H}^{2}G_{2B}+4g^{\prime}Y_{H}G_{BDH}+G_{HD}.
\end{align}

\paragraph{$H^{6}$}

\be
C_{H}=-\frac{1}{2}g^{2}\lambda G_{2W}+2g\lambda G_{WDH}+\lambda^{2}G_{DH}+\lambda G_{HD}^{\prime}+G_{H}.
\ee

\subsection{Two-fermion operators}

\paragraph{$\psi^{2}XH$}

\begin{align}
[C_{uG}]_{ij}= & \frac{1}{4}g_{s}(y_{U})_{lj}[G_{qD}]_{il}+\frac{1}{4}g_{s}(y_{U})_{il}[G_{uD}]_{lj}+ \nonumber \\
 & -\frac{i}{4}(y_{U})_{lj}[G{}_{Gq}^{\prime}]_{il} {+} \frac{1}{4}(y_{U})_{lj}[G{}_{\widetilde{G}q}^{\prime}]_{il}+\frac{i}{4}(y_{U})_{il}[G_{Gu}^{\prime}]_{lj}-\frac{1}{4}(y_{U})_{il}[G_{\widetilde{G}u}^{\prime}]_{lj}+ \nonumber \\
 & +\frac{1}{2}g_{s}\left[G_{uHD3}\right]_{ij}+ \nonumber \\
 & +[G_{uG}]_{ij},\\{}
[C_{uW}]_{ij}= & \frac{1}{8}g(y_{U})_{lj}[G_{qD}]_{il}+ \nonumber \\
 & - \frac{i}{4}(y_{U})_{lj}[G_{Wq}^{\prime}]_{il} {+} \frac{1}{4}(y_{U})_{lj}[G_{\widetilde{W}q}^{\prime}]_{il}+ \nonumber \\
 & -\frac{1}{8}g\left[G_{uHD2}\right]_{ij}+\frac{1}{8}g\left[G_{uHD4}\right]_{ij}+ \nonumber \\
 & +[G_{uW}]_{ij},\\{}
[C_{uB}]_{ij}= & \frac{1}{4}g'Y_{q}(y_{U})_{lj}[G_{qD}]_{il}+\frac{1}{4}g'Y_{u}(y_{U})_{il}[G_{uD}]_{lj}+ \nonumber \\
 & -\frac{i}{4}(y_{U})_{lj}[G{}_{Bq}^{\prime}]_{il} {+} \frac{1}{4}(y_{U})_{lj}[G{}_{\widetilde{B}q}^{\prime}]_{il}+\frac{i}{4}(y_{U})_{il}[G{}_{Bu}^{\prime}]_{lj}-\frac{1}{4}(y_{U})_{il}[G{}_{\widetilde{B}u}^{\prime}]_{lj}+ \nonumber \\
 & +\frac{1}{4}g^{\prime}Y_{H}\left[G_{uHD2}\right]_{ij}+\frac{1}{2}g^{\prime}Y_{u}\left[G_{uHD3}\right]_{ij}-\frac{1}{4}g^{\prime}Y_{H}\left[G_{uHD4}\right]_{ij}+ \nonumber \\
 & +[G_{uB}]_{ij},
\end{align}
\begin{align}
[C_{dG}]_{ij}= & +\frac{1}{4}g_{s}(y_{D})_{lj}[G_{qD}]_{il}+\frac{1}{4}g_{s}(y_{D})_{il}[G_{dD}]_{lj}+ \nonumber \\
 & -\frac{i}{4}(y_{D})_{lj}[G_{Gq}^{\prime}]_{il} {+} \frac{1}{4}(y_{D})_{lj}[G_{\widetilde{G}q}^{\prime}]_{il}+\frac{i}{4}(y_{D})_{il}[G{}_{Gd}^{\prime}]_{lj}-\frac{1}{4}(y_{D})_{il}[G{}_{\widetilde{G}d}^{\prime}]_{lj}+ \nonumber \\
 & +\frac{1}{2}g_{s}\left[G_{dHD3}\right]_{ij}+ \nonumber \\
 & +[G_{dG}]_{ij},\\{}
[C_{dW}]{}_{ij}= & +\frac{1}{8}g(y_{D})_{lj}[G_{qD}]_{il}+ \nonumber \\
 & -\frac{i}{4}(y_{D})_{lj}[G_{Wq}^{\prime}]_{il} {+} \frac{1}{4}(y_{D})_{lj}[G_{\widetilde{W}q}^{\prime}]_{il}+ \nonumber \\
 & -\frac{1}{8}g\left[G_{dHD2}\right]_{ij}+\frac{1}{8}g\left[G_{dHD4}\right]_{ij}+ \nonumber \\
 & +[G_{dW}]_{ij},\\{}
[C_{dB}]{}_{ij}= & +\frac{1}{4}g'Y_{q}(y_{D})_{lj}[G_{qD}]_{il}+\frac{1}{4}g'Y_{d}(y_{D})_{il}[G_{dD}]_{lj}+ \nonumber \\
 & -\frac{i}{4}(y_{D})_{lj}[G{}_{Bq}^{\prime}]_{il} {+} \frac{1}{4}(y_{D})_{lj}[G{}_{\widetilde{B}q}^{\prime}]_{il}+\frac{i}{4}(y_{D})_{il}[G{}_{Bd}^{\prime}]_{lj}-\frac{1}{4}(y_{D})_{il}[G{}_{\widetilde{B}d}^{\prime}]_{lj}+ \nonumber \\
 & -\frac{1}{4}g^{\prime}Y_{H}\left[G_{dHD2}\right]_{ij}+\frac{1}{2}g^{\prime}Y_{d}\left[G_{dHD3}\right]_{ij}+\frac{1}{4}g^{\prime}Y_{H}\left[G_{dHD4}\right]_{ij}+ \nonumber \\
 & +[G_{dB}]_{ij},
\end{align}
\begin{align}
[C_{eW}]_{\alpha\beta}= & +\frac{1}{8}g(y_{E})_{\delta\beta}[G_{\ell D}]_{\alpha\delta}+ \nonumber \\
 & -\frac{i}{4}(y_{E})_{\delta\beta}[G_{W\ell}^{\prime}]_{\alpha\delta} {+} \frac{1}{4}(y_{E})_{\delta\beta}[G_{\widetilde{W}\ell}^{\prime}]_{\alpha\delta}+ \nonumber \\
 & -\frac{1}{8}g[G_{eHD2}]_{\alpha\beta}+\frac{1}{8}g[G_{eHD4}]_{\alpha\beta}+ \nonumber \\
 & +[G_{eW}]_{\alpha\beta},\\{}
[C_{eB}]{}_{\alpha\beta}= & \frac{1}{4}g'Y_{\ell}(y_{E})_{\delta\beta}[G_{\ell D}]_{\alpha\delta}+\frac{1}{4}g'Y_{e}(y_{E})_{\alpha\delta}[G_{eD}]_{\delta\beta}+ \nonumber \\
 & -\frac{i}{4}(y_{E})_{\delta\beta}[G{}_{B\ell}^{\prime}]_{\alpha\delta} {+} \frac{1}{4}(y_{E})_{\delta\beta}[G{}_{\widetilde{B}\ell}^{\prime}]_{\alpha\delta}+\frac{i}{4}(y_{E})_{\alpha\delta}[G{}_{Be}^{\prime}]_{\delta\beta}-\frac{1}{4}(y_{E})_{\alpha\delta}[G{}_{\widetilde{B}e}^{\prime}]_{\delta\beta}+ \nonumber \\
 & -\frac{1}{4}g^{\prime}Y_{H}[G_{eHD2}]_{\alpha\beta}+\frac{1}{2}g^{\prime}Y_{e}[G_{eHD3}]_{\alpha\beta}+\frac{1}{4}g^{\prime}Y_{H}[G_{eHD4}]_{\alpha\beta}+ \nonumber \\
 & +[G_{eB}]_{\alpha\beta}.
\end{align}

\paragraph{$\psi^{2}H^{2}D$}

\begin{align}
[C_{Hq}^{(1)}]_{ij}= & -g^{\prime2}Y_{H}Y_{q}\delta_{ij}G_{2B}+g^{\prime}Y_{q}\delta_{ij}G_{BDH}+ \nonumber \\
 & +\frac{1}{4}(y_{U})_{ik}(y_{U}^{\dagger})_{lj}[G_{uD}]_{kl}-\frac{1}{4}(y_{D})_{ik}(y_{D}^{\dagger})_{lj}[G_{dD}]_{kl}+ \nonumber \\
 & +g^{\prime}Y_{H}[G_{Bq}]_{ij}-\frac{1}{2}g^{\prime}Y_{H}[G_{\widetilde{B}q}^{\prime}]_{ij}+ \nonumber \\
 & +\frac{1}{8}(y_{U}^{\dagger})_{lj}[G_{uHD2}]_{il}+\frac{1}{8}(y_{U})_{il}[G_{uHD2}]_{jl}^{*}+ \nonumber \\
 & -\frac{1}{4}(y_{U}^{\dagger})_{lj}[G_{uHD3}]_{il}-\frac{1}{4}(y_{U})_{il}[G_{uHD3}]_{jl}^{*}+ \nonumber \\
 & +\frac{1}{8}(y_{U}^{\dagger})_{lj}[G_{uHD4}]_{il}+\frac{1}{8}(y_{U})_{il}[G_{uHD4}]_{jl}^{*}+ \nonumber \\
 & -\frac{1}{8}(y_{D}^{\dagger})_{lj}\left[G_{dHD2}\right]_{il}-\frac{1}{8}(y_{D}^{\dagger})_{il}\left[G_{dHD2}\right]_{jl}^{*}+ \nonumber \\
 & +\frac{1}{4}(y_{D}^{\dagger})_{lj}\left[G_{dHD3}\right]_{il}+\frac{1}{4}(y_{D})_{il}\left[G_{dHD3}\right]_{jl}^{*}+ \nonumber \\
 & -\frac{1}{8}(y_{D}^{\dagger})_{lj}\left[G_{dHD4}\right]_{il}-\frac{1}{8}(y_{D})_{il}\left[G_{dHD4}\right]_{jl}^{*}+ \nonumber \\
 & +[G_{Hq}^{(1)}]_{ij},\\{}
[C_{Hq}^{(3)}]{}_{ij}= & -\frac{1}{4}g^{2}\delta_{ij}G_{2W}+\frac{1}{2}g\delta_{ij}G_{WDH}+ \nonumber \\
 & -\frac{1}{4}(y_{U})_{ik}(y_{U}^{\dagger})_{lj}[G_{uD}]_{kl}-\frac{1}{4}(y_{D})_{ik}(y_{D}^{\dagger})_{lj}[G_{dD}]_{kl}+ \nonumber \\
 & +\frac{1}{2}g[G_{Wq}]_{ij}-\frac{1}{4}g[G_{\widetilde{W}q}^{\prime}]_{ij}+ \nonumber \\
 & -\frac{1}{8}(y_{U}^{\dagger})_{lj}[G_{uHD2}]_{il}-\frac{1}{8}(y_{U})_{il}[G_{uHD2}]_{jl}^{*}+ \nonumber \\
 & +\frac{1}{4}(y_{U}^{\dagger})_{lj}[G_{uHD3}]_{il}+\frac{1}{4}(y_{U})_{il}[G_{uHD3}]_{jl}^{*}+ \nonumber \\
 & -\frac{1}{8}(y_{U}^{\dagger})_{lj}[G_{uHD4}]_{il}-\frac{1}{8}(y_{U})_{il}[G_{uHD4}]_{jl}^{*}+ \nonumber \\
 & -\frac{1}{8}(y_{D}^{\dagger})_{lj}\left[G_{dHD2}\right]_{il}-\frac{1}{8}(y_{D}^{\dagger})_{il}\left[G_{dHD2}\right]_{jl}^{*}+ \nonumber \\
 & +\frac{1}{4}(y_{D}^{\dagger})_{lj}\left[G_{dHD3}\right]_{il}+\frac{1}{4}(y_{D})_{il}\left[G_{dHD3}\right]_{jl}^{*}+ \nonumber \\
 & -\frac{1}{8}(y_{D}^{\dagger})_{lj}\left[G_{dHD4}\right]_{il}-\frac{1}{8}(y_{D})_{il}\left[G_{dHD4}\right]_{jl}^{*}+ \nonumber \\
 & +[G_{Hq}^{(3)}]_{ij},
\end{align}

\begin{align}
[C_{Hu}]{}_{ij}= & -g^{\prime2}Y_{H}Y_{u}\delta_{ij}G_{2B}+g^{\prime}Y_{u}\delta_{ij}G_{BDH}+ \nonumber \\
 & -\frac{1}{2}(y_{U}^{\dagger})_{ik}(y_{U})_{lj}[G_{qD}]_{kl}+ \nonumber \\
 & +g^{\prime}Y_{H}[G_{Bu}]_{ij}-\frac{1}{2}g^{\prime}Y_{H}[G_{\widetilde{B}u}^{\prime}]_{ij}+ \nonumber \\
 & -\frac{1}{4}(y_{U}^{\dagger})_{il}\left[G_{uHD2}\right]_{lj}-\frac{1}{4}(y_{U})_{lj}\left[G_{uHD2}\right]_{li}^{*}+ \nonumber \\
 & +\frac{1}{4}(y_{U}^{\dagger})_{il}\left[G_{uHD4}\right]_{lj}+\frac{1}{4}(y_{U})_{lj}\left[G_{uHD4}\right]_{li}^{*}+ \nonumber \\
 & +[G_{Hu}]_{ij},\\{}
[C_{Hd}]{}_{ij}= & -g^{\prime2}Y_{H}Y_{d}\delta_{ij}G_{2B}+g^{\prime}Y_{d}\delta_{ij}G_{BDH}+ \nonumber \\
 & -\frac{1}{2}(y_{D}^{\dagger})_{ik}(y_{D})_{lj}[G_{qD}]_{kl}+ \nonumber \\
 & +g^{\prime}Y_{H}[G_{Bd}]_{ij}-\frac{1}{2}g^{\prime}Y_{H}[G_{\widetilde{B}d}^{\prime}]_{ij}+ \nonumber \\
 & +\frac{1}{4}(y_{D}^{\dagger})_{il}\left[G_{dHD2}\right]_{lj}+\frac{1}{4}(y_{D})_{lj}\left[G_{dHD2}\right]_{li}^{*}+ \nonumber \\
 & -\frac{1}{4}(y_{D}^{\dagger})_{il}\left[G_{dHD4}\right]_{lj}-\frac{1}{4}(y_{D})_{lj}\left[G_{dHD4}\right]_{li}^{*}+ \nonumber \\
 & +[G_{Hd}]_{ij},\\{}
[C_{Hud}]{}_{ij}= & -\frac{1}{2}(y_{U}^{\dagger})_{ik}(y_{D})_{lj}[G_{qD}]_{kl}+ \nonumber \\
 &-\frac{1}{2}(y_{D})_{lj}\left[G_{uHD2}\right]_{li}^{*}+\frac{1}{2}(y_{D})_{lj}\left[G_{uHD4}\right]_{li}^{*}+ \nonumber \\
 & +\frac{1}{2}(y_{U}^{\dagger})_{il}\left[G_{dHD2}\right]_{lj}-\frac{1}{2}(y_{U}^{\dagger})_{il}\left[G_{dHD4}\right]_{lj}+ \nonumber \\
 & +[G_{Hud}]_{ij},
\end{align}
\begin{align}
[C_{H\ell}^{(1)}]_{\alpha\beta}= & -g^{\prime2}Y_{H}Y_{\ell}\delta_{\alpha\beta}G_{2B}+g^{\prime}Y_{\ell}\delta_{\alpha\beta}G_{BDH}+ \nonumber \\
 & -\frac{1}{4}(y_{E})_{\alpha\gamma}(y_{E}^{\dagger})_{\delta\beta}[G_{eD}]_{\gamma\delta}+ \nonumber \\
 & +g^{\prime}Y_{H}[G_{B\ell}]_{\alpha\beta}-\frac{1}{2}g^{\prime}Y_{H}[G_{\widetilde{B}\ell}^{\prime}]_{\alpha\beta}+ \nonumber \\
 & -\frac{1}{8}(y_{E}^{\dagger})_{\delta\beta}[G_{eHD2}]_{\alpha\delta}-\frac{1}{8}(y_{E})_{\alpha\delta}[G_{eHD2}]_{\beta\delta}^{*}+ \nonumber \\
 & +\frac{1}{4}(y_{E}^{\dagger})_{\delta\beta}[G_{eHD3}]_{\alpha\delta}+\frac{1}{4}(y_{E})_{\alpha\delta}[G_{eHD3}]_{\beta\delta}^{*}+ \nonumber \\
 & -\frac{1}{8}(y_{E}^{\dagger})_{\delta\beta}[G_{eHD4}]_{\alpha\delta}-\frac{1}{8}(y_{E})_{\alpha\delta}[G_{eHD4}]_{\beta\delta}^{*}+ \nonumber \\
 & +[G_{H\ell}^{(1)}]_{\alpha\beta},\\{}
[C_{H\ell}^{(3)}]_{\alpha\beta}= & -\frac{1}{4}g^{2}\delta_{\alpha\beta}G_{2W}+\frac{1}{2}g\delta_{\alpha\beta}G_{WDH}+ \nonumber \\
 & -\frac{1}{4}(y_{E})_{\alpha\gamma}(y_{E}^{\dagger})_{\delta\beta}[G_{eD}]_{\gamma\delta}+ \nonumber \\
 & +\frac{g}{2}[G_{W\ell}]_{\alpha\beta}-\frac{g}{4}[G_{\widetilde{W}\ell}^{\prime}]_{\alpha\beta}+ \nonumber \\
 & -\frac{1}{8}(y_{E}^{\dagger})_{\delta\beta}[G_{eHD2}]_{\alpha\delta}-\frac{1}{8}(y_{E})_{\alpha\delta}[G_{eHD2}]_{\beta\delta}^{*}+ \nonumber \\
 & +\frac{1}{4}(y_{E}^{\dagger})_{\delta\beta}[G_{eHD3}]_{\alpha\delta}+\frac{1}{4}(y_{E})_{\alpha\delta}[G_{eHD3}]_{\beta\delta}^{*}+ \nonumber \\
 & -\frac{1}{8}(y_{E}^{\dagger})_{\delta\beta}[G_{eHD4}]_{\alpha\delta}-\frac{1}{8}(y_{E})_{\alpha\delta}[G_{eHD4}]_{\beta\delta}^{*}+ \nonumber \\
 & +[G_{H\ell}^{(3)}]_{\alpha\beta},\\{}
[C_{He}]{}_{\alpha\beta}= & -g^{\prime2}Y_{H}Y_{e}\delta_{\alpha\beta}G_{2B}+g^{\prime}Y_{e}\delta_{\alpha\beta}G_{BDH}+ \nonumber \\
 & -\frac{1}{2}(y_{E}^{\dagger})_{\alpha\gamma}(y_{E})_{\delta\beta}[G_{\ell D}]_{\gamma\delta}+ \nonumber \\
 & +g^{\prime}Y_{H}[G_{Be}]_{\alpha\beta}-\frac{1}{2}g^{\prime}Y_{H}[G_{\widetilde{B}e}^{\prime}]_{\alpha\beta}+ \nonumber \\
 & +\frac{1}{4}(y_{E}^{\dagger})_{\alpha\delta}[G_{eHD2}]_{\delta\beta}+\frac{1}{4}(y_{E})_{\delta\beta}[G_{eHD2}]_{\delta\alpha}^{*}+ \nonumber \\
 & -\frac{1}{4}(y_{E}^{\dagger})_{\alpha\delta}[G_{eHD4}]_{\delta\beta}-\frac{1}{4}(y_{E})_{\delta\beta}[G_{eHD4}]_{\delta\alpha}^{*}+ \nonumber \\
 & +[G_{He}]_{\alpha\beta}.
\end{align}

\paragraph{$\psi^{2}H^{3}$}

\begin{align}
[C_{uH}]_{ij}= & -\frac{1}{4}g^{2}(y_{U})_{ij}G_{2W}+g(y_{U})_{ij}G_{WDH}+\lambda(y_{U})_{ij}G_{DH}+\frac{1}{2}(y_{U})_{ij}G_{HD}^{\prime}-i(y_{U})_{ij}G_{HD}^{\prime\prime}+ \nonumber \\
 & -\frac{1}{2}(y_{U})_{lj}(y_{U}y_{U}^{\dagger})_{ik}[G_{qD}]_{kl}-\frac{1}{2}(y_{U})_{ik}(y_{U}^{\dagger}y_{U})_{lj}[G_{uD}]_{kl}+ \nonumber \\
 & -\lambda[G_{uHD1}]_{ij}+ \nonumber \\
 & +\frac{1}{2}\left(\frac{1}{2}(y_{U}^{\dagger}y_{U})_{lj}\delta_{ik}+\frac{1}{2}(y_{U}y_{U}^{\dagger})_{ik}\delta_{lj}-\lambda\delta_{ik}\delta_{lj}\right)[G_{uHD2}]_{kl}-\frac{1}{2}(y_{U})_{il}(y_{U})_{kj}\left[G_{uHD2}\right]_{kl}^{*} \nonumber \\
 & -\frac{1}{2}(y_{U}^{\dagger}y_{U})_{lj}\delta_{ik}\left[G_{uHD3}\right]_{kl}-\frac{1}{2}(y_{U})_{il}(y_{U})_{kj}\left[G_{uHD3}\right]_{kl}^{*}+ \nonumber \\
 & +\frac{1}{2}\left(\frac{1}{2}(y_{U}^{\dagger}y_{U})_{lj}\delta_{ik}-\frac{1}{2}(y_{U}y_{U}^{\dagger})_{ik}\delta_{lj}+\lambda\delta_{ik}\delta_{lj}\right)[G_{uHD4}]_{kl}+ \nonumber \\
 & +(y_{U})_{lj}[G_{Hq}^{\prime(1)}]_{il}+i(y_{U})_{lj}[G_{Hq}^{\prime\prime(1)}]_{il}-(y_{U})_{lj}[G_{Hq}^{\prime(3)}]_{il}-i(y_{U})_{lj}[G_{Hq}^{\prime\prime(3)}]_{il}+ \nonumber \\
 & +(y_{U})_{ik}[G_{Hu}^{\prime}]_{kj}-i(y_{U})_{ik}[G_{Hu}^{\prime\prime}]_{kj}+ \nonumber \\
 & +[G_{uH}]_{ij},\\{}
[C_{dH}]{}_{ij}= & -\frac{1}{4}g^{2}(y_{D})_{ij}G_{2W}+g(y_{D})_{ij}G_{WDH}+\lambda(y_{D}){}_{ij}G_{DH}+\frac{1}{2}(y_{D})_{ij}G_{HD}^{\prime}+i(y_{D})_{ij}G_{HD}^{\prime\prime}+ \nonumber \\
 & -\frac{1}{2}(y_{D})_{lj}(y_{D}y_{D}^{\dagger})_{ik}[G_{qD}]_{kl}-\frac{1}{2}(y_{D})_{ik}(y_{D}^{\dagger}y_{D})_{lj}[G_{dD}]_{kl}+ \nonumber \\
 & -\lambda\left[G_{dHD1}\right]_{ij}+ \nonumber \\
 & +\frac{1}{2}\left(\frac{1}{2}\delta_{ik}(y_{D}^{\dagger}y_{D})_{lj}+\frac{1}{2}(y_{D}y_{D}^{\dagger})_{ik}\delta_{lj}-\lambda\delta_{ik}\delta_{lj}\right)\left[G_{dHD2}\right]_{kl}-\frac{1}{2}(y_{D})_{il}(y_{D})_{kj}\left[G_{dHD2}\right]_{kl}^{*}+ \nonumber \\
 & -\frac{1}{2}(y_{D}^{\dagger}y_{D})_{lj}\left[G_{dHD3}\right]_{il}-\frac{1}{2}(y_{D})_{il}(y_{D})_{kj}\left[G_{dHD3}\right]_{kl}^{*}+ \nonumber \\
 & +\frac{1}{2}\left(\frac{1}{2}\delta_{ik}(y_{D}^{\dagger}y_{D})_{lj}-\frac{1}{2}(y_{D}y_{D}^{\dagger})_{ik}\delta_{lj}+\lambda\delta_{ik}\delta_{lj}\right)\left[G_{dHD4}\right]_{kl}+ \nonumber \\
 & +(y_{D})_{lj}[G_{Hq}^{\prime(1)}]_{il}+i(y_{D})_{lj}[G_{Hq}^{\prime\prime(1)}]_{il}+(y_{D})_{lj}[G_{Hq}^{\prime(3)}]_{il}+i(y_{D})_{lj}[G_{Hq}^{\prime\prime(3)}]_{il}+ \nonumber \\
 & +(y_{D})_{ik}[G_{Hd}^{\prime}]_{kj}-i(y_{D})_{ik}[G_{Hd}^{\prime\prime}]_{kj}+ \nonumber \\
 & +[G_{dH}]_{ij},\\{}
[C_{eH}]_{\alpha\beta}= & -\frac{1}{4}g^{2}(y_{E})_{\alpha\beta}G_{2W}+g(y_{E})_{\alpha\beta}G_{WDH}+\lambda(y_{E}){}_{\alpha\beta}G_{DH}+\frac{1}{2}(y_{E}){}_{\alpha\beta}G_{HD}^{\prime}+i(y_{E}){}_{\alpha\beta}G_{HD}^{\prime\prime}+ \nonumber \\
 & -\frac{1}{2}(y_{E})_{\delta\beta}(y_{E}y_{E}^{\dagger})_{\alpha\gamma}[G_{\ell D}]_{\gamma\delta}-\frac{1}{2}(y_{E})_{\alpha\gamma}(y_{E}^{\dagger}y_{E})_{\delta\beta}[G_{eD}]_{\gamma\delta}+ \nonumber \\
 & -\lambda[G_{eHD1}]_{\alpha\beta}+ \nonumber \\
 & +\frac{1}{2}\left(\frac{1}{2}\delta_{\alpha\gamma}(y_{E}^{\dagger}y_{E})_{\delta\beta}+\frac{1}{2}(y_{E}y_{E}^{\dagger})_{\alpha\gamma}\delta_{\delta\beta}-\lambda\delta_{\alpha\gamma}\delta_{\delta\beta}\right)[G_{eHD2}]_{\gamma\delta}-\frac{1}{2}(y_{E})_{\alpha\delta}(y_{E})_{\gamma\beta}[G_{eHD2}]_{\gamma\delta}^{*}+ \nonumber \\
 & -\frac{1}{2}(y_{E}^{\dagger}y_{E})_{\delta\beta}[G_{eHD3}]_{\alpha\delta}-\frac{1}{2}(y_{E})_{\alpha\delta}(y_{E})_{\gamma\beta}[G_{eHD3}]_{\gamma\delta}^{*}+ \nonumber \\
 & +\frac{1}{2}\left(\frac{1}{2}\delta_{\alpha\gamma}(y_{E}^{\dagger}y_{E})_{\delta\beta}-\frac{1}{2}(y_{E}y_{E}^{\dagger})_{\alpha\gamma}\delta_{\delta\beta}+\lambda\delta_{\alpha\gamma}\delta_{\delta\beta}\right)[G_{eHD4}]_{\gamma\delta}+ \nonumber \\
 & +(y_{E})_{\delta\beta}[G_{H\ell}^{\prime(1)}]_{\alpha\delta}+i(y_{E})_{\delta\beta}[G_{H\ell}^{\prime\prime(1)}]_{\alpha\delta}+(y_{E})_{\delta\beta}[G_{H\ell}^{\prime(3)}]_{\alpha\delta}+i(y_{E})_{\delta\beta}[G_{H\ell}^{\prime\prime(3)}]_{\alpha\delta}+ \nonumber \\
 & +(y_{E})_{\alpha\gamma}[G_{He}^{\prime}]_{\gamma\beta}-i(y_{E})_{\alpha\gamma}[G_{He}^{\prime\prime}]_{\gamma\beta}+ \nonumber \\
 & +[G_{eH}]_{\alpha\beta}.
\end{align}

\subsection{Four-fermion operators}

\paragraph{Four-quark}

\begin{align}
[C_{qq}^{(1)}]_{ijkl}= & -\frac{1}{4}g_{s}^{2}(\frac{1}{2}\delta_{il}\delta_{kj}-\frac{1}{3}\delta_{ij}\delta_{kl})G_{2G}-\frac{1}{2}g^{\prime2}Y_{q}^{2}\delta_{ij}\delta_{kl}G_{2B}+\nonumber \\
 & +\frac{1}{4}g_{s}\left(\frac{1}{2}\delta_{kj}\delta_{im}\delta_{ln}+\frac{1}{2}\delta_{il}\delta_{km}\delta_{jn}-\frac{1}{3}\delta_{kl}\delta_{im}\delta_{jn}-\frac{1}{3}\delta_{ij}\delta_{km}\delta_{ln}\right)[G_{Gq}]_{mn}+\nonumber \\
 & -\frac{1}{8}g_{s}\left(\frac{1}{2}\delta_{kj}\delta_{im}\delta_{ln}+\frac{1}{2}\delta_{il}\delta_{km}\delta_{jn}-\frac{1}{3}\delta_{kl}\delta_{im}\delta_{jn}-\frac{1}{3}\delta_{ij}\delta_{km}\delta_{ln}\right)[G_{\widetilde{G}q}^{\prime}]_{mn}+\nonumber \\
 & +\frac{1}{2}g^{\prime}Y_{q}\left(\delta_{kl}\delta_{im}\delta_{jn}+\delta_{ij}\delta_{km}\delta_{ln}\right)[G_{Bq}]_{mn}+\nonumber \\
 & -\frac{1}{4}g^{\prime}Y_{q}\left(\delta_{kl}\delta_{im}\delta_{jn}+\delta_{ij}\delta_{km}\delta_{ln}\right)[G_{\widetilde{B}q}^{\prime}]_{mn}+\nonumber \\
 & +[G_{qq}^{(1)}]_{ijkl},\\{}
[C_{qq}^{(3)}]{}_{ijkl}= & -\frac{1}{8}g_{s}^{2}\delta_{il}\delta_{kj}G_{2G}-\frac{1}{8}g^{2}\delta_{ij}\delta_{kl}G_{2W}+\nonumber \\
 & +\frac{1}{8}g_{s}\left(\delta_{kj}\delta_{im}\delta_{ln}+\delta_{il}\delta_{km}\delta_{jn}\right)[G_{Gq}]_{mn}+\nonumber \\
 & -\frac{1}{16}g_{s}\left(\delta_{kj}\delta_{im}\delta_{ln}+\delta_{il}\delta_{km}\delta_{jn}\right)[G_{\widetilde{G}q}^{\prime}]_{mn}+\nonumber \\
 & +\frac{1}{4}g\left(\delta_{kl}\delta_{im}\delta_{jn}+\delta_{ij}\delta_{km}\delta_{ln}\right)[G_{Wq}]_{mn}+\nonumber \\
 & -\frac{1}{8}g\left(\delta_{kl}\delta_{im}\delta_{jn}+\delta_{ij}\delta_{km}\delta_{ln}\right)[G_{\widetilde{W}q}^{\prime}]_{mn}+\nonumber \\
 & +[G_{qq}^{(3)}]_{ijkl},
\end{align}

\begin{align}
[C_{uu}]_{ijkl}= & -\frac{1}{4}g_{s}^{2}(\delta_{il}\delta_{kj}-\frac{1}{3}\delta_{ij}\delta_{kl})G_{2G}-\frac{1}{2}g^{\prime2}Y_{u}^{2}\delta_{ij}\delta_{kl}G_{2B}+\nonumber \\
 & +\frac{1}{4}g_{s}\left(-\frac{1}{3}\delta_{kl}\delta_{im}\delta_{jn}-\frac{1}{3}\delta_{ij}\delta_{km}\delta_{ln}+\delta_{kj}\delta_{im}\delta_{ln}+\delta_{il}\delta_{km}\delta_{jn}\right)[G_{Gu}]_{mn}+\nonumber \\
 & -\frac{1}{8}g_{s}\left(-\frac{1}{3}\delta_{kl}\delta_{im}\delta_{jn}-\frac{1}{3}\delta_{ij}\delta_{km}\delta_{ln}+\delta_{kj}\delta_{im}\delta_{ln}+\delta_{il}\delta_{km}\delta_{jn}\right)[G_{\widetilde{G}u}^{\prime}]_{mn}+\nonumber \\
 & +\frac{1}{2}g^{\prime}Y_{u}\left(\delta_{kl}\delta_{im}\delta_{jn}+\delta_{ij}\delta_{km}\delta_{ln}\right)[G_{Bu}]_{mn}+\nonumber \\
 & -\frac{1}{4}g^{\prime}Y_{u}\left(\delta_{kl}\delta_{im}\delta_{jn}+\delta_{ij}\delta_{km}\delta_{ln}\right)[G_{\widetilde{B}u}^{\prime}]_{mn}+\nonumber \\
 & +[G_{uu}]_{ijkl},\\{}
[C_{dd}]_{ijkl}= & -\frac{1}{4}g_{s}^{2}(\delta_{il}\delta_{kj}-\frac{1}{3}\delta_{ij}\delta_{kl})G_{2G}-\frac{1}{2}g^{\prime2}Y_{d}^{2}\delta_{ij}\delta_{kl}G_{2B}+\nonumber \\
 & +\frac{1}{4}g_{s}\left(-\frac{1}{3}\delta_{kl}\delta_{im}\delta_{jn}-\frac{1}{3}\delta_{ij}\delta_{km}\delta_{ln}+\delta_{kj}\delta_{im}\delta_{ln}+\delta_{il}\delta_{km}\delta_{jn}\right)[G_{Gd}]_{mn}+\nonumber \\
 & -\frac{1}{8}g_{s}\left(-\frac{1}{3}\delta_{kl}\delta_{im}\delta_{jn}-\frac{1}{3}\delta_{ij}\delta_{km}\delta_{ln}+\delta_{kj}\delta_{im}\delta_{ln}+\delta_{il}\delta_{km}\delta_{jn}\right)[G_{\widetilde{G}d}^{\prime}]_{mn}+\nonumber \\
 & +\frac{1}{2}g^{\prime}Y_{d}\left(\delta_{kl}\delta_{im}\delta_{jn}+\delta_{ij}\delta_{km}\delta_{ln}\right)[G_{Bd}]_{mn}+\nonumber \\
 & -\frac{1}{4}g^{\prime}Y_{d}\left(\delta_{kl}\delta_{im}\delta_{jn}+\delta_{ij}\delta_{km}\delta_{ln}\right)[G_{\widetilde{B}d}^{\prime}]_{mn}+\nonumber \\
 & +[G_{dd}]_{ijkl},\\{}
[C_{ud}^{(1)}]_{ijkl}= & -g^{\prime2}Y_{u}Y_{d}\delta_{ij}\delta_{kl}G_{2B}+\nonumber \\
 & +g^{\prime}Y_{d}\delta_{kl}[G_{Bu}]_{ij}-\frac{1}{2}g^{\prime}Y_{d}\delta_{kl}[G_{\widetilde{B}u}^{\prime}]_{ij}+\nonumber \\
 & +g^{\prime}Y_{u}\delta_{ij}[G_{Bd}]_{kl}-\frac{1}{2}g^{\prime}Y_{u}\delta_{ij}[G_{\widetilde{B}d}^{\prime}]_{kl}+\nonumber \\
 & +[G_{ud}^{(1)}]_{ijkl},\\{}
[C_{ud}^{(8)}]_{ijkl}= & -g_{s}^{2}\delta_{ij}\delta_{kl}G_{2G}+\nonumber \\
 & +g_{s}\delta_{kl}[G_{Gu}]_{ij}-\frac{1}{2}g_{s}\delta_{kl}[G_{\widetilde{G}u}^{\prime}]_{ij}+g_{s}\delta_{ij}[G_{Gd}]_{kl}-\frac{1}{2}g_{s}\delta_{ij}[G_{\widetilde{G}d}^{\prime}]_{kl}+\nonumber \\
 & +[G_{ud}^{(8)}]_{ijkl},
\end{align}

\begin{align}
[C_{qu}^{(1)}]_{ijkl}= & -g^{\prime2}Y_{q}Y_{u}\delta_{ij}\delta_{kl}G_{2B}-\frac{1}{6}(y_{U})_{il}(y_{U}^{\dagger})_{kj}G_{DH}+\nonumber \\
 & +g^{\prime}Y_{u}\delta_{kl}[G_{Bq}]_{ij}-\frac{1}{2}g^{\prime}Y_{u}\delta_{kl}[G_{\widetilde{B}q}^{\prime}]_{ij}+\nonumber \\
 & +g^{\prime}Y_{q}\delta_{ij}[G_{Bu}]_{kl}-\frac{1}{2}g^{\prime}Y_{q}\delta_{ij}[G_{\widetilde{B}u}^{\prime}]_{kl}+\nonumber \\
 & +\frac{1}{6}(y_{U}^{\dagger})_{kj}[G_{uHD1}]_{il}+\frac{1}{6}(y_{U})_{il}[G_{uHD1}]_{jk}^{*}+\nonumber \\
 & +\frac{1}{12}(y_{U}^{\dagger})_{kj}[G_{uHD2}]_{il}+\frac{1}{12}(y_{U})_{il}[G_{uHD2}]_{jk}^{*}+\nonumber \\
 & -\frac{1}{12}(y_{U}^{\dagger})_{kj}[G_{uHD4}]_{il}-\frac{1}{12}(y_{U})_{il}[G_{uHD4}]_{jk}^{*} \nonumber \\
 & +[G_{qu}^{(1)}]_{ijkl},\\{}
[C_{qu}^{(8)}]_{ijkl}= & -g_{s}^{2}\delta_{ij}\delta_{kl}G_{2G}-(y_{U})_{il}(y_{U}^{\dagger})_{kj}G_{DH}+\nonumber \\
 & +g_{s}\delta_{kl}[G_{Gq}]_{ij}-\frac{1}{2}g_{s}\delta_{kl}[G_{\widetilde{G}q}^{\prime}]_{ij}+g_{s}\delta_{ij}[G_{Gu}]_{kl}-\frac{1}{2}g_{s}\delta_{ij}[G_{\widetilde{G}u}^{\prime}]_{kl}+\nonumber \\
 & +(y_{U}^{\dagger})_{kj}[G_{uHD1}]_{il}+(y_{U})_{il}[G_{uHD1}]_{jk}^{*}+\nonumber \\
 & +\frac{1}{2}(y_{U}^{\dagger})_{kj}[G_{uHD2}]_{il}+\frac{1}{2}(y_{U})_{il}[G_{uHD2}]_{jk}^{*}+\nonumber \\
 & -\frac{1}{2}(y_{U}^{\dagger})_{kj}[G_{uHD4}]_{il}-\frac{1}{2}(y_{U})_{il}[G_{uHD4}]_{jk}^{*}+ \nonumber \\
 & +[G_{qu}^{(8)}]_{ijkl} ,\\
[C_{qd}^{(1)}]_{ijkl}= & -g^{\prime2}Y_{q}Y_{d}\delta_{ij}\delta_{kl}G_{2B}-\frac{1}{6}(y_{D})_{il}(y_{D}^{\dagger})_{kj}G_{DH}+\nonumber \\
 & +g'Y_{d}\delta_{kl}[G_{Bq}]_{ij}-\frac{1}{2}g'Y_{d}\delta_{kl}[G_{\widetilde{B}q}^{\prime}]_{ij}+\nonumber \\
 & +g'Y_{q}\delta_{ij}[G_{Bd}]_{kl}-\frac{1}{2}g'Y_{q}\delta_{ij}[G_{\widetilde{B}d}^{\prime}]_{kl}+\nonumber \\
 & +\frac{1}{6}(y_{D}^{\dagger})_{kj}\left[G_{dHD1}\right]_{il}+\frac{1}{6}(y_{D})_{il}\left[G_{dHD1}\right]_{jk}^{*}+\nonumber \\
 & +\frac{1}{12}(y_{D}^{\dagger})_{kj}\left[G_{dHD2}\right]_{il}+\frac{1}{12}(y_{D})_{il}\left[G_{dHD2}\right]_{jk}^{*}+\nonumber \\
 & -\frac{1}{12}(y_{D}^{\dagger})_{kj}\left[G_{dHD4}\right]_{il}-\frac{1}{12}(y_{D})_{il}\left[G_{dHD4}\right]_{jk}^{*}+\nonumber \\
 & +[G_{qd}^{(1)}]_{ijkl},\\{}
[C_{qd}^{(8)}]_{ijkl}= & -g_{s}^{2}\delta_{ij}\delta_{kl}G_{2G}-(y_{D})_{il}(y_{D}^{\dagger})_{kj}G_{DH}+\nonumber \\
 & +g_{s}\delta_{kl}[G_{Gq}]_{ij}-\frac{1}{2}g_{s}\delta_{kl}[G_{\widetilde{G}q}^{\prime}]_{ij}+g_{s}\delta_{ij}[G_{Gd}]_{kl}-\frac{1}{2}g_{s}\delta_{ij}[G_{\widetilde{G}d}^{\prime}]_{kl}+\nonumber \\
 & +(y_{D}^{\dagger})_{kj}\left[G_{dHD1}\right]_{il}+(y_{D})_{il}\left[G_{dHD1}\right]_{jk}^{*}+\nonumber \\
 & +\frac{1}{2}(y_{D}^{\dagger})_{kj}\left[G_{dHD2}\right]_{il}+\frac{1}{2}(y_{D})_{il}\left[G_{dHD2}\right]_{jk}^{*}+\nonumber \\
 & -\frac{1}{2}(y_{D}^{\dagger})_{kj}\left[G_{dHD4}\right]_{il}-\frac{1}{2}(y_{D})_{il}\left[G_{dHD4}\right]_{jk}^{*}+\nonumber \\
 & +[G_{qd}^{(8)}]_{ijkl},
\end{align}

\begin{align}
[C_{quqd}^{(1)}]_{ijkl}= & (y_{U})_{ij}(y_{D})_{kl}G_{DH}+\nonumber \\
 & -(y_{D})_{kl}[G_{uHD1}]_{ij}-\frac{1}{2}(y_{D})_{kl}[G_{uHD2}]_{ij}+\frac{1}{2}(y_{D})_{kl}[G_{uHD4}]_{ij}+\nonumber \\
 & -(y_{U})_{ij}[G_{dHD1}]_{kl}-\frac{1}{2}(y_{U})_{ij}[G_{dHD2}]_{kl}+\frac{1}{2}(y_{U})_{ij}[G_{dHD4}]_{kl}+\nonumber \\
 & +[G_{quqd}^{(1)}]_{ijkl},\\{}
[C_{quqd}^{(8)}]{}_{ijkl}= & [G_{quqd}^{(8)}]_{ijkl}.
\end{align}

\paragraph{Four-lepton }

\begin{align}
[C_{\ell\ell}]_{\alpha\beta\gamma\delta}= & -\frac{1}{8}g^{2}(2\delta_{\alpha\delta}\delta_{\gamma\beta}-\delta_{\alpha\beta}\delta_{\gamma\delta})G_{2W}-\frac{1}{2}g^{\prime2}Y_{\ell}^{2}\delta_{\alpha\beta}\delta_{\gamma\delta}G_{2B}+\nonumber \\
 & +\frac{1}{4}g\left(-\delta_{\gamma\delta}\delta_{\alpha\rho}\delta_{\beta\sigma}+2\delta_{\gamma\beta}\delta_{\alpha\rho}\delta_{\delta\sigma}-\delta_{\alpha\beta}\delta_{\gamma\rho}\delta_{\delta\sigma}+2\delta_{\alpha\delta}\delta_{\gamma\rho}\delta_{\beta\sigma}\right)[G_{W\ell}]_{\rho\sigma}+\nonumber \\
 & -\frac{1}{8}g\left(-\delta_{\gamma\delta}\delta_{\alpha\rho}\delta_{\beta\sigma}+2\delta_{\gamma\beta}\delta_{\alpha\rho}\delta_{\delta\sigma}-\delta_{\alpha\beta}\delta_{\gamma\rho}\delta_{\delta\sigma}+2\delta_{\alpha\delta}\delta_{\gamma\rho}\delta_{\beta\sigma}\right)[G_{\widetilde{W}\ell}^{\prime}]_{\rho\sigma}+\nonumber \\
 & +\frac{1}{2}g^{\prime}Y_{\ell}\left(\delta_{\gamma\delta}\delta_{\alpha\rho}\delta_{\beta\sigma}+\delta_{\alpha\beta}\delta_{\gamma\rho}\delta_{\delta\sigma}\right)[G_{B\ell}]_{\rho\sigma}+\nonumber \\
 & -\frac{1}{4}g^{\prime}Y_{\ell}\left(\delta_{\gamma\delta}\delta_{\alpha\rho}\delta_{\beta\sigma}+\delta_{\alpha\beta}\delta_{\gamma\rho}\delta_{\delta\sigma}\right)[G_{\widetilde{B}\ell}^{\prime}]_{\rho\sigma}+\nonumber \\
 & +[G_{\ell\ell}]_{\alpha\beta\gamma\delta},
\end{align}

\begin{align}
[C_{ee}]_{\alpha\beta\gamma\delta}= & -\frac{g^{\prime2}}{2}Y_{e}^{2}\delta_{\alpha\beta}\delta_{\gamma\delta}G_{2B}+\nonumber \\
 & +\frac{1}{2}g^{\prime}Y_{e}\left(\delta_{\gamma\delta}\delta_{\alpha\rho}\delta_{\beta\sigma}+\delta_{\alpha\beta}\delta_{\gamma\rho}\delta_{\delta\sigma}\right)[G_{Be}]_{\rho\sigma}+\nonumber \\
 & -\frac{1}{4}g^{\prime}Y_{e}\left(\delta_{\gamma\delta}\delta_{\alpha\rho}\delta_{\beta\sigma}+\delta_{\alpha\beta}\delta_{\gamma\rho}\delta_{\delta\sigma}\right)[G_{\widetilde{B}e}^{\prime}]_{\rho\sigma}+\nonumber \\
 & +[G_{ee}]_{\alpha\beta\gamma\delta},
\end{align}

\begin{align}
[C_{\ell e}]_{\alpha\beta\gamma\delta}= & -g^{\prime2}Y_{\ell}Y_{e}\delta_{\alpha\beta}\delta_{\gamma\delta}G_{2B}-\frac{1}{2}(y_{E})_{\alpha\delta}(y_{E}^{\dagger})_{\gamma\beta}G_{DH}+\nonumber \\
 & +g^{\prime}Y_{e}\delta_{\gamma\delta}([G_{B\ell}]_{\alpha\beta}-\frac{1}{2}[G_{\widetilde{B}\ell}^{\prime}]_{\alpha\beta})+\nonumber \\
 & +g^{\prime}Y_{\ell}\delta_{\alpha\beta}([G_{Be}]_{\gamma\delta}-\frac{1}{2}[G_{\widetilde{B}e}^{\prime}]_{\gamma\delta})+\nonumber \\
 & +\frac{1}{2}(y_{E}^{\dagger})_{\gamma\beta}[G_{eHD1}]_{\alpha\delta}+\frac{1}{2}(y_{E})_{\alpha\delta}[G_{eHD1}]_{\beta\gamma}^{*}+\nonumber \\
 & +\frac{1}{4}(y_{E}^{\dagger})_{\gamma\beta}[G_{eHD2}]_{\alpha\delta}+\frac{1}{4}(y_{E})_{\alpha\delta}[G_{eHD2}]_{\beta\gamma}^{*}+\nonumber \\
 & -\frac{1}{4}(y_{E}^{\dagger})_{\gamma\beta}[G_{eHD4}]_{\alpha\delta}-\frac{1}{4}(y_{E})_{\alpha\delta}[G_{eHD4}]_{\beta\gamma}^{*}+\nonumber \\
 & +[G_{\ell e}]_{\alpha\beta\gamma\delta}.
\end{align}

\paragraph{Semileptonic}

\begin{align}
[C_{\ell q}^{(1)}]{}_{\alpha\beta ij}= & -g^{\prime2}Y_{\ell}Y_{q}\delta_{\alpha\beta}\delta_{ij}G_{2B}+\nonumber \\
 & +g^{\prime}Y_{\ell}\delta_{\alpha\beta}[G_{Bq}]_{ij}-\frac{1}{2}g^{\prime}Y_{\ell}\delta_{\alpha\beta}[G_{\widetilde{B}q}^{\prime}]_{ij}+\nonumber \\
 & +g^{\prime}Y_{q}\delta_{ij}[G_{B\ell}]_{\alpha\beta}-\frac{1}{2}g^{\prime}Y_{q}\delta_{ij}[G_{\widetilde{B}\ell}^{\prime}]_{\alpha\beta}+\nonumber \\
 & +[G_{\ell q}^{(1)}]_{\alpha\beta ij},\\{}
[C_{\ell q}^{(3)}]_{\alpha\beta ij}= & -\frac{1}{4}g^{2}\delta_{\alpha\beta}\delta_{ij}G_{2W}+\nonumber \\
 & +\frac{1}{2}g\delta_{\alpha\beta}[G_{Wq}]_{ij}-\frac{1}{4}g\delta_{\alpha\beta}[G_{\widetilde{W}q}^{\prime}]_{ij}+\nonumber \\
 & +\frac{1}{2}g\delta_{ij}[G_{W\ell}]_{\alpha\beta}-\frac{1}{4}g\delta_{ij}[G_{\widetilde{W}\ell}^{\prime}]_{\alpha\beta}+\nonumber \\
 & +[G_{\ell q}^{(3)}]_{\alpha\beta ij},
\end{align}

\begin{align}
[C_{eu}]_{\alpha\beta ij}= & -g^{\prime2}Y_{e}Y_{u}\delta_{\alpha\beta}\delta_{ij}G_{2B}+\nonumber \\
 & +g^{\prime}Y_{e}\delta_{\alpha\beta}[G_{Bu}]_{ij}-\frac{1}{2}g^{\prime}Y_{e}\delta_{\alpha\beta}[G_{\widetilde{B}u}^{\prime}]_{ij}+\nonumber \\
 & +g^{\prime}Y_{u}\delta_{ij}[G_{Be}]_{\alpha\beta}-\frac{1}{2}g^{\prime}Y_{u}\delta_{ij}[G_{\widetilde{B}e}^{\prime}]_{\alpha\beta}+\nonumber \\
 & +[G_{eu}]_{\alpha\beta ij},\\{}
[C_{ed}]{}_{\alpha\beta ij}= & -g^{\prime2}Y_{e}Y_{d}\delta_{\alpha\beta}\delta_{ij}G_{2B}+\nonumber \\
 & +g^{\prime}Y_{e}\delta_{\alpha\beta}[G_{Bd}]_{ij}-\frac{1}{2}g^{\prime}Y_{e}\delta_{\alpha\beta}[G_{\widetilde{B}d}^{\prime}]_{ij}+\nonumber \\
 & +g^{\prime}Y_{d}\delta_{ij}[G_{Be}]_{\alpha\beta}-\frac{1}{2}g^{\prime}Y_{d}\delta_{ij}[G_{\widetilde{B}e}^{\prime}]_{\alpha\beta}+\nonumber \\
 & +[G_{ed}]_{\alpha\beta ij},
\end{align}

\begin{align}
[C_{qe}]_{ij\alpha\beta}= & -g^{\prime2}Y_{q}Y_{e}\delta_{ij}\delta_{\alpha\beta}G_{2B}+\nonumber \\
 & +g^{\prime}Y_{e}\delta_{\alpha\beta}[G_{Bq}]_{ij}-\frac{1}{2}g^{\prime}Y_{e}\delta_{\alpha\beta}[G_{\widetilde{B}q}^{\prime}]_{ij}+\nonumber \\
 & +g^{\prime}Y_{q}\delta_{ij}[G_{Be}]_{\alpha\beta}-\frac{1}{2}g^{\prime}Y_{q}\delta_{ij}[G_{\widetilde{B}e}^{\prime}]_{\alpha\beta}+\nonumber \\
 & +[G_{qe}]_{ij\alpha\beta},\\{}
[C_{\ell u}]_{\alpha\beta ij}= & -g^{\prime2}Y_{\ell}Y_{u}\delta_{\alpha\beta}\delta_{ij}G_{2B}+\nonumber \\
 & +g^{\prime}Y_{\ell}\delta_{\alpha\beta}[G_{Bu}]_{ij}-\frac{1}{2}g^{\prime}Y_{\ell}\delta_{\alpha\beta}[G_{\widetilde{B}u}^{\prime}]_{ij}+\nonumber \\
 & +g^{\prime}Y_{u}\delta_{ij}[G_{B\ell}]_{\alpha\beta}-\frac{1}{2}g^{\prime}Y_{u}\delta_{ij}[G_{\widetilde{B}\ell}^{\prime}]_{\alpha\beta}+\nonumber \\
 & +[G_{\ell u}]_{\alpha\beta ij},\\{}
[C_{\ell d}]_{\alpha\beta ij}= & -g^{\prime2}Y_{\ell}Y_{d}\delta_{\alpha\beta}\delta_{ij}G_{2B}+\nonumber \\
 & +g^{\prime}Y_{\ell}\delta_{\alpha\beta}[G_{Bd}]_{ij}-\frac{1}{2}g^{\prime}Y_{\ell}\delta_{\alpha\beta}[G_{\widetilde{B}d}^{\prime}]_{ij}+\nonumber \\
 & +g^{\prime}Y_{d}\delta_{ij}[G_{B\ell}]_{\alpha\beta}-\frac{1}{2}g^{\prime}Y_{d}\delta_{ij}[G_{\widetilde{B}\ell}^{\prime}]_{\alpha\beta}+\nonumber \\
 & +[G_{\ell d}]_{\alpha\beta ij},
\end{align}

\begin{align}
[C_{\ell edq}^{(1)}]_{\alpha\beta ij}= & (y_{E})_{\alpha\beta}(y_{D}^{\dagger})_{ij}G_{DH}+\nonumber \\
 & -(y_{E})_{\alpha\beta}[G_{dHD1}]_{ji}^{*}-\frac{1}{2}(y_{E})_{\alpha\beta}[G_{dHD2}]_{ji}^{*}+\frac{1}{2}(y_{E})_{\alpha\beta}[G_{dHD4}]_{ji}^{*}+\nonumber \\
 & -(y_{D}^{\dagger})_{ij}[G_{eHD1}]_{\alpha\beta}-\frac{1}{2}(y_{D}^{\dagger})_{ij}[G_{eHD2}]_{\alpha\beta}+\frac{1}{2}(y_{D}^{\dagger})_{ij}[G_{eHD4}]_{\alpha\beta}+\nonumber \\
 & +[G_{\ell edq}^{(1)}]_{\alpha\beta ij},
\end{align}

\begin{align}
[C_{\ell equ}^{(1)}]_{\alpha\beta ij}= & -(y_{E})_{\alpha\beta}(y_{U})_{ij}G_{DH}+\nonumber \\
 & +(y_{E})_{\alpha\beta}[G_{uHD1}]_{ij}+\frac{1}{2}(y_{E})_{\alpha\beta}[G_{uHD2}]_{ij}-\frac{1}{2}(y_{E})_{\alpha\beta}[G_{uHD4}]_{ij}+\nonumber \\
 & +(y_{U})_{ij}[G_{eHD1}]_{\alpha\beta}+\frac{1}{2}(y_{U})_{ij}[G_{eHD2}]_{\alpha\beta}-\frac{1}{2}(y_{U})_{ij}[G_{eHD4}]_{\alpha\beta}+\nonumber \\
 & +[G_{\ell equ}^{(1)}]_{\alpha\beta ij},\\{}
[C_{\ell equ}^{(3)}]_{\alpha\beta ij}= & [G_{\ell equ}^{(3)}]_{\alpha\beta ij}.
\end{align}

\paragraph{$B$-violating}

\begin{align}
 [C_{duq}]_{ijk\alpha} & =[G_{duq}]_{ijk\alpha},\\
 [C_{qqu}]_{ijk\alpha} & =[G_{qqu}]_{ijk\alpha},\\
 [C_{qqq}]_{ijk\alpha} & =[G_{qqq}]_{ijk\alpha},\\
 [C_{duu}]_{ijk\alpha} & =[G_{duu}]_{ijk\alpha}.\\ 
\end{align}

\section{One-loop matching conditions in the Green's basis}
\label{app:MatchToGreen}

We report in this Section the complete one-loop SMEFT matching contributions for the $S_1 + S_3$ model in the Green's basis. 
Green's basis WCs are denoted by $G$. We absorb the loop factor defining
$G_i = \frac{1}{(4\pi)^2} G_i^{(1)}$.



\subsection{Renormalizable operators}

\begin{align}
(\delta Z_{\ell}^{\text{G}})_{\alpha\beta} & =\frac{N_{c}}{2}\left[(\frac{1}{2}+L_{1})(\Lambda_{\ell}^{(1)})_{\alpha\beta}+3(\frac{1}{2}+L_{3})(\Lambda_{\ell}^{(3)})_{\alpha\beta}\right],\\
(\delta Z_{e}^{\text{G}})_{\alpha\beta} & =\frac{N_{c}}{2}(\frac{1}{2}+L_{1})(\Lambda_{e})_{\alpha\beta},\\
(\delta Z_{q}^{\text{G}})_{ij} & =\frac{1}{2}\left[(\frac{1}{2}+L_{1})(\Lambda_{q}^{(1)})_{ij}+3(\frac{1}{2}+L_{3})(\Lambda_{q}^{(3)})_{ij}\right],\\
(\delta Z_{u}^{\text{G}})_{ij} & =\frac{1}{2}(\frac{1}{2}+L_{1})(\Lambda_{u})_{ij}, \\
(\delta Z_{d}^{\text{G}})_{ij} & = 0,
\end{align}
\begin{align}
(\delta y_{E}^{\text{G}})_{\alpha\beta} & =-N_{c}(1+L_{1})(X_{1U}^{1L})_{\alpha\beta},\\
(\delta y_{U}^{\text{G}})_{ij} & =-(1+L_{1})(X_{1E}^{1L})_{ij}, \\
(\delta y_{D}^{\text{G}})_{ij} & =0,
\end{align}
\begin{align}
\delta \lambda ^\text{G} &= -N_c \left[ \lambda _{H1}^2 L_1 + (3\lambda _{H3}^2+2\lambda _{\epsilon H3}^2) L_3+2\vert \lambda _{H13} \vert ^2\left(1+\dfrac{M_3 ^2 L_3 -M_1 ^2 L_1}{M_3^2-M_1^2}\right) \right],\\
(\delta m^2)^{\text G}&=N_c\left[\lambda _{H1}(1+L_1)M_1^2+3\lambda _{H3}(1+L_3)M_3^2 \right].
\end{align}

\subsection{Purely bosonic operators}

\paragraph{$X^{3}$}
\begin{align}
G^{(1)}_{3G} & =\frac{g_{s}^{3}}{360}\left(\frac{3}{M_{3}^{2}}+\frac{1}{M_{1}^{2}}\right),\\
G^{(1)}_{3W} & =\frac{g^{3}N_{c}}{90M_{3}^{2}},\\
G^{(1)}_{3 \widetilde W} & = G^{(1)}_{3\widetilde G} = 0.
\end{align}

\paragraph{$X^{2}D^{2}$}
\begin{align}
G^{(1)}_{2G} & =\frac{g_{s}^{2}}{60}\left(\frac{3}{M_{3}^{2}}+\frac{1}{M_{1}^{2}}\right),\\
G^{(1)}_{2W} & =\frac{g^{2}N_{c}}{15M_{3}^{2}},\\
G^{(1)}_{2B} & =\frac{g^{\prime2}N_{c}}{30}\left(\frac{3 Y_{S_3}^{2}}{M_{3}^{2}}+\frac{Y_{S_1}^{2}}{M_{1}^{2}}\right).
\end{align}

\paragraph{$X^{2}H^{2}$}
\begin{align}
G^{(1)}_{HG} & =\frac{g_{s}^{2}}{{24}}\left(\frac{3\lambda_{H3}}{M_{3}^{2}}+\frac{\lambda_{H1}}{M_{1}^{2}}\right),\\
G^{(1)}_{HW} & =\frac{g^{2}N_{c}\lambda_{H3}}{{6}M_{3}^{2}},\\
G^{(1)}_{HB} & =\frac{g^{\prime2}N_{c}}{ {12} }\left(3\frac{\lambda_{H3}Y_{S_{3}}^{2}}{M_{3}^{2}}+\frac{\lambda_{H1}Y_{S_{1}}^{2}}{M_{1}^{2}}\right),\\
G^{(1)}_{HWB} & = -N_{c}\frac{gg^\prime Y_{S_3}\lambda_{\epsilon H3}}{3 M_{3}^{2}}, \\
G^{(1)}_{H\widetilde G} & = G^{(1)}_{H\widetilde W} = G^{(1)}_{H\widetilde B}  = G^{(1)}_{H\widetilde W B} = 0.
\end{align}

\paragraph{$H^{2} X D^2$}
\be
G^{(1)}_{WDH} = G^{(1)}_{BDH} = 0.
\ee

\paragraph{$H^{2} D^4$}
\be
G^{(1)}_{DH}  = 0.
\ee

\paragraph{$H^{4}D^{2}$}
\begin{align}
G^{(1)}_{H\Box} & =-\frac{N_{c}}{12}\left(\frac{(3\lambda_{H3}^{2}+2\lambda _{\epsilon H3}^2)}{M_{3}^{2}}+\frac{\lambda_{H1}^{2}}{M_{1}^{2}}\right)-\frac{N_{c}}{2}\left|\lambda_{H13}\right|{}^{2}h(M_{1},M_{3}),\\
G^{(1)}_{HD} & =-N_{c}\frac{2\lambda _{\epsilon H3}^2}{3M_{3}^{2}}-2N_{c}\left|\lambda_{H13}\right|{}^{2}h(M_{1},M_{3}),\\
G^{\prime (1)}_{HD} & =+N_{c}\frac{2\lambda _{\epsilon H3}^2}{3M_{3}^{2}}+2N_{c}\left|\lambda_{H13}\right|{}^{2}h(M_{1},M_{3}), \\
G^{\prime\prime (1)}_{HD} & = 0.
\end{align}

\paragraph{$H^{6}$}
\begin{align}
G^{(1)}_{H} &= \frac{N_{c}}{6}\left\{ -\frac{3\lambda_{H3}^{3}+6\lambda_{\epsilon H3}^{2}\lambda_{H3}}{M_{3}^{2}}-\frac{\lambda_{H1}^{3}}{M_{1}^{2}} + \right. \nonumber\\
	& \left. + \frac{6\left|\lambda_{H13}\right|{}^{2}}{M_{1}^{2}-M_{3}^{2}}\left[\lambda_{H3}-\lambda_{H1}+\frac{\log\left(\frac{M_{1}^{2}}{M_{3}^{2}}\right)}{M_{1}^{2}-M_{3}^{2}}(\lambda_{H1}M_{3}^{2}-\lambda_{H3}M_{1}^{2})\right]\right\} .
\end{align}

\subsection{Two-fermion operators}

\paragraph{$\psi^{2}D^{3}$}
\begin{align}
[G_{qD}]^{(1)}_{ij}= & -\frac{1}{6}\left(3\frac{(\Lambda_{q}^{(3)})_{ij}}{M_{3}^{2}}+\frac{(\Lambda_{q}^{(1)})_{ij}}{M_{1}^{2}}\right),\\
[G_{uD}]^{(1)}_{ij}= & -\frac{1}{6}\frac{(\Lambda_{u})_{ij}}{M_{1}^{2}},\\
[G_{dD}]^{(1)}_{ij}= & \, 0 ,\\
[G_{\ell D}]^{(1)}_{\alpha\beta}= & -\frac{N_{c}}{6}\left(3\frac{(\Lambda_{\ell}^{(3)})_{\alpha\beta}}{M_{3}^{2}}+\frac{(\Lambda_{\ell}^{(1)})_{\alpha\beta}}{M_{1}^{2}}\right),\\
[G_{eD}]^{(1)}_{\alpha\beta}= & -\frac{N_{c}}{6}\frac{(\Lambda_{e})_{\alpha\beta}}{M_{1}^{2}}.
\end{align}

\paragraph{$\psi^{2}XD$}
\begin{align}
[G_{Gq}]^{(1)}_{ij} & =\frac{1}{18}g_{s}\left(3\frac{(\Lambda_{q}^{(3)})_{ij}}{M_{3}^{2}}+\frac{(\Lambda_{q}^{(1)})_{ij}}{M_{1}^{2}}\right)~,\\
[G_{Wq}]^{(1)}_{ij} & =\frac{1}{6}g\left((L_{3}+\frac{5}{4})\frac{(\Lambda_{q}^{(3)})_{ij}}{M_{3}^{2}}-(L_{1}+\frac{7}{12})\frac{(\Lambda_{q}^{(1)})_{ij}}{M_{1}^{2}}\right),\\
[G_{\widetilde{W}q}^{\prime}]^{(1)}_{ij} & = {-} \frac{1}{4}g\left(\frac{(\Lambda_{q}^{(3)})_{ij}}{M_{3}^{2}}-\frac{(\Lambda_{q}^{(1)})_{ij}}{M_{1}^{2}}\right),\\
[G_{Bq}]^{(1)}_{ij} & =\frac{1}{3}g^{\prime}\left\{ 3\left(\frac{7Y_{\ell}-2Y_{S_{3}}}{12}+Y_{\ell}L_{3}\right)\frac{(\Lambda_{q}^{(3)})_{ij}}{M_{3}^{2}}+\left(\frac{7Y_{\ell}-2Y_{S_{1}}}{12}+Y_{\ell}L_{1}\right)\frac{(\Lambda_{q}^{(1)})_{ij}}{M_{1}^{2}}\right\} ,\\
[G_{\widetilde{B}q}^{\prime}]^{(1)}_{ij} & = {-} \frac{1}{2}g^{\prime}Y_{\ell}\left(3\frac{(\Lambda_{q}^{(3)})_{ij}}{M_{3}^{2}}+\frac{(\Lambda_{q}^{(1)})_{ij}}{M_{1}^{2}}\right),\\
[G_{Gu}]^{(1)}_{ij} & =\frac{1}{18}g_{s}\frac{(\Lambda_{u})_{ij}}{M_{1}^{2}},\\
[G_{Bu}]^{(1)}_{ij} & =\frac{1}{3}g^{\prime}\left(\frac{7Y_{e}-2Y_{S_{1}}}{12}+Y_{e}L_{1}\right)\frac{(\Lambda_{u})_{ij}}{M_{1}^{2}},\\
[G_{\widetilde{B}u}^{\prime}]^{(1)}_{ij} & = {+} \frac{1}{2}g^{\prime}Y_{e}\frac{(\Lambda_{u})_{ij}}{M_{1}^{2}},
\end{align}
\begin{align}
[G_{W\ell}]^{(1)}_{\alpha\beta} & =\frac{N_{c}}{6}g\left((L_{3}+\frac{5}{4})\frac{(\Lambda_{\ell}^{(3)})_{\alpha\beta}}{M_{3}^{2}}-(L_{1}+\frac{7}{12})\frac{(\Lambda_{\ell}^{(1)})_{\alpha\beta}}{M_{1}^{2}}\right),\\
[G_{\widetilde{W}\ell}^{\prime}]^{(1)}_{\alpha\beta} & ={-}\frac{N_{c}}{4}g\left(\frac{(\Lambda_{\ell}^{(3)})_{\alpha\beta}}{M_{3}^{2}}-\frac{(\Lambda_{\ell}^{(1)})_{\alpha\beta}}{M_{1}^{2}}\right),\\
[G_{B\ell}]^{(1)}_{\alpha\beta} & =\frac{N_{c}}{3}g^{\prime}\left(3\left(\frac{7Y_{q}-2Y_{S_{3}}}{12}+Y_{q}L_{3}\right)\frac{(\Lambda_{\ell}^{(3)})_{\alpha\beta}}{M_{3}^{2}}+\left(\frac{7Y_{q}-2Y_{S_{1}}}{12}+Y_{q}L_{1}\right)\frac{(\Lambda_{\ell}^{(1)})_{\alpha\beta}}{M_{1}^{2}}\right),\\
[G_{\widetilde{B}\ell}^{\prime}]^{(1)}_{\alpha\beta} & = {-} \frac{N_{c}}{2}g^{\prime}Y_{q}\left(3\frac{(\Lambda_{\ell}^{(3)})_{\alpha\beta}}{M_{3}^{2}}+\frac{(\Lambda_{\ell}^{(1)})_{\alpha\beta}}{M_{1}^{2}}\right),\\
[G_{Be}]^{(1)}_{\alpha\beta} & =\frac{N_{c}}{3}g^{\prime}\left(\frac{7Y_{u}-2Y_{S_{1}}}{12}+Y_{u}L_{1}\right)\frac{(\Lambda_{e})_{\alpha\beta}}{M_{1}^{2}},\\
[G_{\widetilde{B}e}^{\prime}]^{(1)}_{ij} & ={+} \frac{N_{c}}{2}g^{\prime}Y_{u}\frac{(\Lambda_{e})_{\alpha\beta}}{M_{1}^{2}},
\end{align}
\be\begin{split}
	[G^{\prime}_{G q}]^{(1)}_{ij} &= [G^{\prime}_{\widetilde G q}]^{(1)}_{ij} = [G_{Wq}^{\prime}]^{(1)}_{ij} = [G_{B q}^{\prime}]^{(1)}_{ij} = 0 ,\\
	[G^{\prime}_{G u}]^{(1)}_{ij} &= [G^{\prime}_{\widetilde G u}]^{(1)}_{ij} = [G^{\prime}_{B u}]^{(1)}_{ij} = 0 ,\\
	[G_{G(B) d}]^{(1)}_{ij} &= [G^{\prime}_{G(B) d}]^{(1)}_{ij} = [G^{\prime}_{\widetilde G( \widetilde B) d}]^{(1)}_{ij} = 0 ,\\
	[G^{\prime}_{W \ell}]^{(1)}_{\alpha\beta} &= [G^{\prime}_{B \ell}]^{(1)}_{\alpha\beta} = [G^{\prime}_{B e}]^{(1)}_{\alpha\beta} = 0.
\end{split}\ee

\paragraph{$\psi^{2}HD^{2}$}
\begin{align}
[G_{uHD1}]^{(1)}_{ij} & =+\frac{1}{2}(L_{1}+\frac{1}{2})\frac{(X_{1E}^{1L})_{ij}}{M_{1}^{2}},\\
[G_{uHD2}]^{(1)}_{ij} & = {+} \frac{1}{2}\frac{(X_{1E}^{1L})_{ij}}{M_{1}^{2}},\\
[G_{uHD3}]^{(1)}_{ij} & =- \frac{1}{2}\frac{(X_{1E}^{1L})_{ij}}{M_{1}^{2}},\\
[G_{uHD4}]^{(1)}_{ij} & =- \frac{1}{2}\frac{(X_{1E}^{1L})_{ij}}{M_{1}^{2}},\\
[G_{dHDn}]^{(1)}_{ij} & = 0 \; (n = 1,2,3,4),\\
[G_{eHD1}]^{(1)}_{\alpha\beta} & =+\frac{N_{c}}{2}(L_{1}+\frac{1}{2})\frac{(X_{1U}^{1L})_{\alpha\beta}}{M_{1}^{2}},\\
[G_{eHD2}]^{(1)}_{\alpha\beta} & = {+} \frac{N_{c}}{2}\frac{(X_{1U}^{1L})_{\alpha\beta}}{M_{1}^{2}},\\
[G_{eHD3}]^{(1)}_{\alpha\beta} & =- \frac{N_{c}}{2}\frac{(X_{1U}^{1L})_{\alpha\beta}}{M_{1}^{2}},\\
[G_{eHD4}]^{(1)}_{\alpha\beta} & =- \frac{N_{c}}{2}\frac{(X_{1U}^{1L})_{\alpha\beta}}{M_{1}^{2}}.
\end{align}

\paragraph{$\psi^{2}XH$}
\begin{align}
[G_{uG}]^{(1)}_{ij} & = 0 ,\\
[G_{uW}]^{(1)}_{ij} & =-\frac{1}{8}g(L_{1}+\frac{1}{2})\frac{(X_{1E}^{1L})_{ij}}{M_{1}^{2}},\\
[G_{uB}]^{(1)}_{ij} & =\frac{1}{4}g^{\prime}\left[(Y_{l}+Y_{e})L_{1}+\frac{1}{2}Y_{l}+\frac{3}{2}Y_{e}\right]\frac{(X_{1E}^{1L})_{ij}}{M_{1}^{2}},\\
[G_{dG}]^{(1)}_{ij} & = [G_{dW}]^{(1)}_{ij} = [G_{dB}]^{(1)}_{ij} =0 ,\\
[G_{eW}]^{(1)}_{\alpha\beta} & =-\frac{N_{c}}{8}g(L_{1}+\frac{1}{2})\frac{(X_{1U}^{1L})_{\alpha\beta}}{M_{1}^{2}},\\
[G_{eB}]^{(1)}_{\alpha\beta} & = \frac{N_{c}}{4}g^{\prime}\left[(Y_{q}+Y_{u})L_{1}+\frac{1}{2}Y_{q}+\frac{3}{2}Y_{u}\right]\frac{(X_{1U}^{1L})_{\alpha\beta}}{M_{1}^{2}}.
\end{align}

\paragraph{$\psi^{2}DH^{2}$}
\begin{align}
[G_{Hq}^{(1)}]^{(1)}_{ij} & =-\frac{1}{4}\left( 3(1+L_{3})\frac{(X_{2E}^{3L})_{ij}}{M_{3}^{2}}+(1+L_{1})\frac{(X_{2E}^{1L})_{ij}}{M_{1}^{2}}\right),\\
[G_{Hq}^{\prime(1)}]^{(1)}_{ij} &=  -\frac{1}{8}\left(3\frac{(X_{2E}^{3L})_{ij}+2\lambda_{H3}(\Lambda_{q}^{(3)})_{ij}}{M_{3}^{2}}+\frac{(X_{2E}^{1L})_{ij}+2\lambda_{H1}(\Lambda_{q}^{(1)})_{ij}}{M_{1}^{2}}\right),\\
 [G_{Hq}^{\prime\prime(1)}]^{(1)}_{ij} &= 0 ,\\
[G_{Hq}^{(3)}]^{(1)}_{ij} & =-\frac{1}{4}\left(  (1+L_{3})\frac{(X_{2E}^{3L})_{ij}}{M_{3}^{2}}- (1+L_{1})\frac{(X_{2E}^{1L})_{ij}}{M_{1}^{2}}\right),\\
[G_{Hq}^{\prime(3)}]^{(1)}_{ij}&=  -\frac{1}{8}\left(\frac{(X_{2E}^{3L})_{ij}+4\lambda_{\epsilon H3}(\Lambda_{q}^{(3)})_{ij}}{M_{3}^{2}}-\frac{(X_{2E}^{1L})_{ij}}{M_{1}^{2}}\right)+ \nonumber\\
	& + \frac{1}{4}\frac{\left[\lambda_{H13}^{*} \Lambda_{q}^{(31)}+\lambda_{H13} \Lambda_{q}^{(31) \dagger}\right]_{ij}\log\frac{M_{3}^{2}}{M_{1}^{2}}}{M_{3}^{2}-M_{1}^{2}},\\
 [G_{Hq}^{\prime\prime(3)}]^{(1)}_{ij} &= -\frac{1}{2} \left[ i \lambda_{H13}^{*}\Lambda_{q}^{(31)}- i \lambda_{H13}\Lambda_{q}^{(31)\dagger}\right]_{ij} \frac{M_1^2 - M_3^2 + \frac{1}{2} (M_1^2 + M_3^2) \log \frac{M_3^2}{M_1^2}}{(M_1^2 - M_3^2)^2}~,  \\
[G_{Hu}]^{(1)}_{ij} & = \frac{1}{2}(1+L_{1})\frac{(X_{2E}^{1R})_{ij}}{M_{1}^{2}},\\
[G_{Hu}^{\prime}]^{(1)}_{ij} &=  -\frac{1}{4}\frac{(X_{2E}^{1R})_{ij}+\lambda_{H1}(\Lambda_{u})_{ij}}{M_{1}^{2}},\\
[G_{Hu}^{\prime\prime}]^{(1)}_{ij} &= 0~,\\
[G_{Hd}]^{(1)}_{ij} &= [G_{Hd}^{\prime}]^{(1)}_{ij} = [G_{Hd}^{\prime\prime}]^{(1)}_{ij} = 0,
\end{align}
\begin{align}
[G_{H\ell}^{(1)}]^{(1)}_{\alpha\beta} & =-\frac{N_{c}}{4}\left\{ 3(1+L_{3})\frac{-(X_{2U}^{3L})_{\alpha\beta}+(X_{2D}^{3L})_{\alpha\beta}}{M_{3}^{2}}+(1+L_{1})\frac{-(X_{2U}^{1L})_{\alpha\beta}+(X_{2D}^{1L})_{\alpha\beta}}{M_{1}^{2}}\right\}, \\
[G_{H\ell}^{\prime(1)}]^{(1)}_{\alpha\beta} &= -\frac{N_{c}}{8}\left\{ 3\frac{(X_{2U}^{3L})_{\alpha\beta}+(X_{2D}^{3L})_{\alpha\beta}+2\lambda_{H3}(\Lambda_{\ell}^{(3)})_{\alpha\beta}}{M_{3}^{2}}+\right. \nonumber\\
 & \left.+\frac{(X_{2U}^{1L})_{\alpha\beta}+(X_{2D}^{1L})_{\alpha\beta}+2\lambda_{H1}(\Lambda_{\ell}^{(1)})_{\alpha\beta}}{M_{1}^{2}}\right\} ,\\
  [G_{H\ell}^{\prime\prime(1)}]^{(1)}_{\alpha\beta} &= 0, \\
[G_{H\ell}^{(3)}]^{(1)}_{\alpha\beta} & =-\frac{N_{c}}{4}\left\{ (1+L_{3})\frac{(X_{2U}^{3L})_{\alpha\beta}+(X_{2D}^{3L})_{\alpha\beta}}{M_{3}^{2}}-(1+L_{1})\frac{(X_{2U}^{1L})_{\alpha\beta}+(X_{2D}^{1L})_{\alpha\beta}}{M_{1}^{2}}\right\} ,\\
[G_{H\ell}^{\prime(3)}]^{(1)}_{\alpha\beta} &= +N_{c}\left\{ -\frac{1}{8}\left(\frac{-(X_{2U}^{3L})_{\alpha\beta}+(X_{2D}^{3L})_{\alpha\beta}+4\lambda_{\epsilon H3}(\Lambda_{\ell}^{(3)})_{\alpha\beta}}{M_{3}^{2}}-\frac{-(X_{2U}^{1L})_{\alpha\beta}+(X_{2D}^{1L})_{\alpha\beta}}{M_{1}^{2}}\right)+\right. \nonumber\\
 & \left.-\frac{1}{4}\frac{\left[\lambda_{H13}^{*}\Lambda_{\ell}^{(31)}+\lambda_{H13}\Lambda_{\ell}^{(31)\dagger}\right]_{\alpha\beta}\log\frac{M_{3}^{2}}{M_{1}^{2}}}{M_{3}^{2}-M_{1}^{2}}\right\} ,\\
 [G_{H\ell}^{\prime\prime(3)}]^{(1)}_{\alpha\beta} &= \frac{N_c}{2} \left[ i \lambda_{H13}^{*}\Lambda_{\ell}^{(31)}- i \lambda_{H13}\Lambda_{\ell}^{(31)\dagger}\right]_{\alpha\beta} \frac{M_1^2 - M_3^2 + \frac{1}{2} (M_1^2 + M_3^2) \log \frac{M_3^2}{M_1^2}}{(M_1^2 - M_3^2)^2}~,  \\
[G_{He}]^{(1)}_{\alpha\beta} & =-\frac{N_{c}}{2}(1+L_{1})\frac{(X_{2U}^{1R})_{\alpha\beta}}{M_{1}^{2}},\\
[G_{He}^{\prime}]^{(1)}_{\alpha\beta} &= -\frac{N_{c}}{4}\frac{(X_{2U}^{1R})_{\alpha\beta}+\lambda_{H1}(\Lambda_{e})_{\alpha\beta}}{M_{1}^{2}}, \\
 [G_{He}^{\prime\prime(3)}]^{(1)}_{\alpha\beta} &= 0~.
\end{align}

\paragraph{$\psi^{2}H^{3}$}
\begin{align}
[G_{uH}]^{(1)}_{ij} & =\frac{(1+L_1)(X_{3E}^{1L})_{ij}-\lambda_{H1}(X_{1E}^{1L})_{ij}}{M_{1}^{2}} -\frac{\lambda_{H13}^* (X_{1E}^{3L})_{ij}\log \frac{M_1^2}{M_3^2}}{M_1^2-M_3^2},\\
[G_{dH}]^{(1)}_{ij} & = 0 ,\\
[G_{eH}]^{(1)}_{\alpha\beta} & = N_{c}\frac{(1+L_1)(X_{3U}^{1L})_{\alpha\beta}-\lambda_{H1}(X_{1U}^{1L})_{\alpha\beta}}{M_{1}^{2}}- N_c \frac{\lambda_{H13}^* (X_{1U}^{3L})_{\alpha\beta}\log \frac{M_1^2}{M_3^2}}{M_1^2-M_3^2}. \\
\end{align}

\subsection{Four-fermion operators}

\paragraph{Four-quark}
\begin{align}
[G_{qq}^{(1)}]^{(1)}_{ijkl}= & -\frac{1}{16}\left(9\frac{(\Lambda_{q}^{(3)})_{il}(\Lambda_{q}^{(3)})_{kj}}{M_{3}^{2}}+\frac{(\Lambda_{q}^{(1)})_{il}(\Lambda_{q}^{(1)})_{kj}}{M_{1}^{2}}+3\frac{\log\frac{M_{3}^{2}}{M_{1}^{2}}\left[(\Lambda_{q}^{31})_{il}(\Lambda_{q}^{31\dagger})_{kj}+(\Lambda_{q}^{31\dagger})_{il}(\Lambda_{q}^{31})_{kj}\right]}{M_{3}^{2}-M_{1}^{2}}\right),\\
[G_{qq}^{(3)}]^{(1)}_{ijkl}= & -\frac{1}{16}\left(\frac{(\Lambda_{q}^{(3)})_{il}(\Lambda_{q}^{(3)})_{kj}}{M_{3}^{2}}+\frac{(\Lambda_{q}^{(1)})_{il}(\Lambda_{q}^{(1)})_{kj}}{M_{1}^{2}}-\frac{\log\frac{M_{3}^{2}}{M_{1}^{2}}\left[(\Lambda_{q}^{31})_{il}(\Lambda_{q}^{31\dagger})_{kj}+(\Lambda_{q}^{31\dagger})_{il}(\Lambda_{q}^{31})_{kj}\right]}{M_{3}^{2}-M_{1}^{2}}\right),\\
[G_{uu}]^{(1)}_{ijkl} = & -\frac{1}{8}\frac{(\Lambda_{u})_{il}(\Lambda_{u})_{kj}}{M_{1}^{2}},\\
[G_{dd}]^{(1)}_{ijkl} = & [G_{ud}^{(1)}]^{(1)}_{ijkl} = [G_{ud}^{(8)}]^{(1)}_{ijkl} = 0 , \\
[G_{qu}^{(1)}]^{(1)}_{ijkl}= & -\frac{1}{12M_{1}^{2}}(\Lambda_{q}^{(1)})_{ij}(\Lambda_{u})_{kl},\\
[G_{qu}^{(8)}]^{(1)}_{ijkl}= & -\frac{1}{2M_{1}^{2}}(\Lambda_{q}^{(1)})_{ij}(\Lambda_{u})_{kl}, \\
[G_{qd}^{(1)}]^{(1)}_{ijkl} = & [G_{qd}^{(8)}]^{(1)}_{ijkl} = [G_{quqd}^{(1)}]^{(1)}_{ijkl} = [G_{quqd}^{(8)}]^{(1)}_{ijkl} = 0.
\end{align}

\paragraph{Four-lepton}
\begin{align}
[G_{\ell\ell}]^{(1)}_{\alpha\beta\gamma\delta}= & -\frac{N_{c}}{8}\left\{ \frac{(\Lambda_{\ell}^{(3)})_{\alpha\beta}(\Lambda_{\ell}^{(3)})_{\gamma\delta}}{M_{3}^{2}}+4\frac{(\Lambda_{\ell}^{(3)})_{\alpha\delta}(\Lambda_{\ell}^{(3)})_{\gamma\beta}}{M_{3}^{2}}+\frac{(\Lambda_{\ell}^{(1)})_{\alpha\beta}(\Lambda_{\ell}^{(1)})_{\gamma\delta}}{M_{1}^{2}}+\right.\nonumber\\
 & -\frac{\log\frac{M_{3}^{2}}{M_{1}^{2}}}{M_{3}^{2}-M_{1}^{2}}\left[(\Lambda_{\ell}^{(31)})_{\alpha\beta}(\Lambda_{\ell}^{(31)\dagger})_{\gamma\delta}+(\Lambda_{\ell}^{(31)\dagger})_{\alpha\beta}(\Lambda_{\ell}^{(31)})_{\gamma\delta}+\right.\nonumber\\
 & \left.\left.-2(\Lambda_{\ell}^{(31)})_{\alpha\delta}(\Lambda_{\ell}^{(31)\dagger})_{\gamma\beta}-2(\Lambda_{\ell}^{(31)\dagger})_{\alpha\delta}(\Lambda_{\ell}^{(31)})_{\gamma\beta}\right]\right\},\\
[G_{ee}]^{(1)}_{\alpha\beta\gamma\delta}= & -\frac{N_{c}}{8}\frac{(\Lambda_{e})_{\alpha\beta}(\Lambda_{e})_{\gamma\delta}}{M_{1}^{2}},\\
[G_{\ell e}]^{(1)}_{\alpha\beta\gamma\delta}= & -\frac{N_{c}}{4}\dfrac{(\Lambda_{\ell}^{(1)})_{\alpha\beta}(\Lambda_{e})_{\gamma\delta}}{M_{1}^{2}}.
\end{align}

\paragraph{Semileptonic}
\begin{align}
[G_{\ell q}^{(1)}]^{(1)}_{\alpha\beta ij} & =\frac{1}{4}(\frac{1}{2}+a_{\text{ev}})\left[g_{s}^{2}\dfrac{N_c^2 -1}{2N_c}+g^{\prime2}(Y_{q}-Y_{\ell})^{2}\right]\left(\frac{3\lambda_{i\alpha}^{3L*}\lambda_{j\beta}^{3L}}{M_{3}^{2}}+\frac{\lambda_{i\alpha}^{1L*}\lambda_{j\beta}^{1L}}{M_{1}^{2}}\right)+\nonumber\\
&+\frac{1}{4}(\frac{1}{2}+a_{\text{ev}}) g^2 \, 3\, \left(\frac{\lambda_{i\alpha}^{3L*}\lambda_{j\beta}^{3L}}{M_{3}^{2}}+\frac{\lambda_{i\alpha}^{1L*}\lambda_{j\beta}^{1L}}{M_{1}^{2}}\right)+\nonumber\\
&-\frac{1}{4}\left(3\frac{(\Lambda_{\ell}^{(3)})_{\alpha\beta}(\Lambda_{q}^{(3)})_{ij}}{M_{3}^{2}}+\frac{(\Lambda_{\ell}^{(1)})_{\alpha\beta}(\Lambda_{q}^{(1)})_{ij}}{M_{1}^{2}}\right)+\nonumber\\
& +\left(c_{1}(1+L_{1})+\frac{9}{4}(1+L_{3})c_{13}^{(1)}\frac{M_{3}^{2}}{M_{1}^{2}}\right)\frac{\lambda_{\alpha i}^{1L\dagger}\lambda_{j\beta}^{1L}}{M_{1}^{2}}+\nonumber\\
 & +\left(\frac{9}{4}(1+L_{1})c_{13}^{(1)}\frac{M_{1}^{2}}{M_{3}^{2}}+\frac{3}{2}(1+L_{3})\left[5c_{3}^{(1)}-c_{3}^{(3)}+\frac{5}{6}c_{3}^{(5)}\right]\right)\frac{\lambda_{\alpha i}^{3L\dagger}\lambda_{j\beta}^{3L}}{M_{3}^{2}},\\
[G_{\ell q}^{(3)}]^{(1)}_{\alpha\beta ij} & = \frac{1}{4}(\frac{1}{2}+a_{\text{ev}})\left[g_{s}^{2}\dfrac{N_c^2 -1}{2N_c}+g^{\prime2}(Y_{q}-Y_{\ell})^{2}\right]\left(\frac{\lambda_{i\alpha}^{3L*}\lambda_{j\beta}^{3L}}{M_{3}^{2}}-\frac{\lambda_{i\alpha}^{1L*}\lambda_{j\beta}^{1L}}{M_{1}^{2}}\right)+\nonumber\\
&+\frac{1}{4}(\frac{1}{2}+a_{\text{ev}}) g^2 \left(-\frac{3\lambda_{i\alpha}^{3L*}\lambda_{j\beta}^{3L}}{M_{3}^{2}}+\frac{\lambda_{i\alpha}^{1L*}\lambda_{j\beta}^{1L}}{M_{1}^{2}}\right)+\nonumber\\
&-\frac{1}{4}\left(2\frac{(\Lambda_{\ell}^{(3)})_{\alpha\beta}(\Lambda_{q}^{(3)})_{ij}}{M_{3}^{2}}+\frac{\log\frac{M_{3}^{2}}{M_{1}^{2}}\left[(\Lambda_{\ell}^{(31)})_{\alpha\beta}(\Lambda_{q}^{(31)\dagger})_{ij}+(\Lambda_{\ell}^{(31)\dagger})_{\alpha\beta}(\Lambda_{q}^{(31)})_{ij}\right]}{M_{3}^{2}-M_{1}^{2}}\right)+\nonumber\\
& -\left(c_{1}(1+L_{1})+\frac{9}{4}(1+L_{3})c_{13}^{(1)}\frac{M_{3}^{2}}{M_{1}^{2}}\right)\frac{\lambda_{\alpha i}^{1L\dagger}\lambda_{j\beta}^{1L}}{M_{1}^{2}}+\nonumber\\
 &+\left(\frac{3}{4}(1+L_{1})c_{13}^{(1)}\frac{M_{1}^{2}}{M_{3}^{2}}+\frac{1}{2}(1+L_{3})\left[5c_{3}^{(1)}-c_{3}^{(3)}+\frac{5}{6}c_{3}^{(5)}\right]\right)\frac{\lambda_{\alpha i}^{3L\dagger}\lambda_{j\beta}^{3L}}{M_{3}^{2}},\\
 [G_{eu}]^{(1)}_{\alpha\beta ij} & =\frac{1}{2}(\frac{1}{2}+a_{\text{ev}})\left[g_{s}^{2}\dfrac{N_c^2 -1}{2N_c}+g^{\prime2}(Y_{u}-Y_{e})^{2}\right]\frac{\lambda_{i\alpha}^{1R*}\lambda_{j\beta}^{1R}}{M_{1}^{2}}+\nonumber\\
 &-\frac{1}{4}\frac{(\Lambda_{e})_{\alpha\beta}(\Lambda_{u})_{ij}}{M_{1}^{2}}+\left(2c_{1}(1+L_{1})+\frac{9}{2}(1+L_{3})c_{13}^{(1)}\frac{M_{3}^{2}}{M_{1}^{2}}\right)\frac{\lambda_{\alpha i}^{1R\dagger}\lambda_{j\beta}^{1R}}{M_{1}^{2}},\\
[G_{qe}]^{(1)}_{ij\alpha\beta} & =-\frac{1}{4}\frac{(\Lambda_{q}^{(1)})_{ij}(\Lambda_{e})_{\alpha\beta}}{M_{1}^{2}}-\frac{3}{4}(\frac{3}{2}+L_{3})\frac{(\lambda^{3L*}y_{E}^{*})_{i\alpha}(\lambda^{3L}y_{E})_{j\beta}}{M_{3}^{2}}+\nonumber\\
 & -\frac{1}{4}(\frac{3}{2}+L_{1})\frac{(\lambda^{1L*}y_{E}^{*}-y_{U}\lambda^{1R*})_{i\alpha}(\lambda^{1L}y_{E}-y_{U}^{*}\lambda^{1R})_{j\beta}}{M_{1}^{2}}, \\
 [G_{\ell u}]^{(1)}_{\alpha\beta ij} & =-\frac{1}{4}\frac{(\Lambda_{\ell}^{(1)})_{\alpha\beta}(\Lambda_{u})_{ij}}{M_{1}^{2}}-\frac{3}{4}(\frac{3}{2}+L_{3})\frac{(\lambda^{3L\dagger}y_{U}^{*})_{\alpha i}(\lambda^{3L\,T}y_{U})_{\beta j}}{M_{3}^{2}}+\nonumber\\
 & -\frac{1}{4}(\frac{3}{2}+L_{1})\dfrac{(\lambda^{1L\,\dagger}y_{U}^{*}-y_{E}\lambda^{1R\dagger})_{\alpha i}(\lambda^{1L\,T}y_{U}-y_{E}^{*}\lambda^{1RT})_{\beta j}}{M_{1}^{2}},\\
 [G_{\ell d}]^{(1)}_{\alpha\beta ij} & =-\frac{1}{4}\left(3(\frac{3}{2}+L_{3})\frac{(\lambda^{3L\dagger}y_{D}^{*})_{\alpha i}(\lambda^{3L\,T}y_{D})_{\beta j}}{M_{3}^{2}}+(\frac{3}{2}+L_{1})\frac{(\lambda^{1L\,\dagger}y_{D}^{*})_{\alpha i}(\lambda^{1L\,T}y_{D})_{\beta j}}{M_{1}^{2}}\right),\nonumber\\
\end{align}
\begin{align}
 [G_{\ell edq}]^{(1)}_{\alpha\beta ij}=&-\frac{1}{2}\left(-3(\frac{3}{2}+L_{3})\frac{(\lambda^{3L\,\dagger}y_{D}^{*})_{\alpha i}(\lambda^{3L}y_{E})_{j\beta}}{M_{3}^{2}}+(\frac{3}{2}+L_{1})\frac{(\lambda^{1L\,\dagger}y_{D}^{*})_{\alpha i}(\lambda^{1L}y_{E})_{j\beta}}{M_{1}^{2}}\right) +\nonumber\\
&+\frac{1}{2}(\frac{3}{2}+L_{1})\frac{(\lambda^{1L\,\dagger}y_{D}^{*})_{\alpha i}(y_{U}^*\lambda^{1R})_{j\beta}}{M_{1}^{2}},\\
[G_{\ell equ}^{(1)}]^{(1)}_{\alpha\beta ij} & = \frac{\lambda_{\alpha i}^{1L\dagger}\lambda_{j\beta}^{1R}}{M_{1}^{2}}  \left\{ 2c_{1}(1+L_{1})+\frac{9}{2}(1+L_{3})c_{13}^{(1)}\frac{M_{3}^{2}}{M_{1}^{2}} + \right. \nonumber\\
	& \left. - \frac{3}{2}(\frac{3}{2}+L_{1})\left[(Y_{q}-Y_{\ell})(Y_{u}-Y_{e})g^{\prime2}+\frac{N_c^2 -1}{2 N_c}g_{s}^{2}\right]   \right\},\\
[G_{\ell equ}^{(3)}]^{(1)}_{\alpha\beta ij} & = \frac{\lambda_{\alpha i}^{1L\dagger}\lambda_{j\beta}^{1R}}{M_{1}^{2}} \left\{ -\frac{1}{2}c_{1}(1+L_{1})-\frac{9}{8}(1+L_{3})c_{13}^{(1)}\frac{M_{3}^{2}}{M_{1}^{2}} + \right. \nonumber\\
	& \left.  -\frac{1}{8}(\frac{3}{2}+L_{1})\left[(Y_{q}-Y_{\ell})(Y_{u}-Y_{e})g^{\prime2}+\frac{N_c^2 -1}{2 N_c}g_{s}^{2}\right] \right\rbrace.
\end{align}


\vspace{2cm}
\begin{table}[hb]
\begin{centering}
\begin{tabular}{|c|c|c|c|c|c|}
\hline 
\multicolumn{2}{|c|}{\textbf{\large{}$\boldsymbol{X^{3}}$}} & \multicolumn{2}{c|}{{\large{}$\text{\ensuremath{\boldsymbol{X^{2}H^{2}}}}$}} & \multicolumn{2}{c|}{{\large{}$\boldsymbol{H^{2}D^{4}}$}}\tabularnewline
\hline 
\textcolor{blue}{\cellcolor[gray]{0.92}}$\mathcal{O}_{3G}$  & \textcolor{blue}{\cellcolor[gray]{0.92}}$f^{ABC}G_{\mu}^{A\nu}G_{\nu}^{B\rho}G_{\rho}^{C\mu}$  & \textcolor{blue}{\cellcolor[gray]{0.92}}$\mathcal{O}_{HG}$  & \textcolor{blue}{\cellcolor[gray]{0.92}}$G_{\mu\nu}^{A}G^{A\mu\nu}(H^{\dagger}H)$  & $\mathcal{O}_{DH}$  & $(D_{\mu}D^{\mu}H)^{\dagger}(D_{\nu}D^{\nu}H)$\tabularnewline
\cline{5-6} \cline{6-6} 
\textcolor{blue}{\cellcolor[gray]{0.92}}$\mathcal{O}_{\widetilde{3G}}$  & \textcolor{blue}{\cellcolor[gray]{0.92}}$f^{ABC}\widetilde{G}_{\mu}^{A\nu}G_{\nu}^{B\rho}G_{\rho}^{C\mu}$  & \textcolor{blue}{\cellcolor[gray]{0.92}}$\mathcal{O}_{H\widetilde{G}}$  & \textcolor{blue}{\cellcolor[gray]{0.92}}$\widetilde{G}_{\mu\nu}^{A}G^{A\mu\nu}(H^{\dagger}H)$  & \multicolumn{2}{c|}{{\large{}$\text{\ensuremath{\boldsymbol{H^{4}D^{2}}}}$}}\tabularnewline
\cline{5-6} \cline{6-6} 
\textcolor{blue}{\cellcolor[gray]{0.92}}$\mathcal{O}_{3W}$  & \textcolor{blue}{\cellcolor[gray]{0.92}}$\epsilon^{IJK}W_{\mu}^{I\nu}W_{\nu}^{J\rho}W_{\rho}^{K\mu}$  & \textcolor{blue}{\cellcolor[gray]{0.92}}$\mathcal{O}_{HW}$  & \textcolor{blue}{\cellcolor[gray]{0.92}}$W_{\mu\nu}^{I}W^{I\mu\nu}(H^{\dagger}H)$  & \textcolor{blue}{\cellcolor[gray]{0.92}}$\mathcal{O}_{H\square}$  & \textcolor{blue}{\cellcolor[gray]{0.92}}$(H^{\dagger}H)\square(H^{\dagger}H)$\tabularnewline
\textcolor{blue}{\cellcolor[gray]{0.92}}$\mathcal{O}_{\widetilde{3W}}$  & \textcolor{blue}{\cellcolor[gray]{0.92}}$\epsilon^{IJK}\widetilde{W}_{\mu}^{I\nu}W_{\nu}^{J\rho}W_{\rho}^{K\mu}$  & \textcolor{blue}{\cellcolor[gray]{0.92}}$\mathcal{O}_{H\widetilde{W}}$  & \textcolor{blue}{\cellcolor[gray]{0.92}}$\widetilde{W}_{\mu\nu}^{I}W^{I\mu\nu}(H^{\dagger}H)$  & \textcolor{blue}{\cellcolor[gray]{0.92}}$\mathcal{O}_{HD}$  & \textcolor{blue}{\cellcolor[gray]{0.92}}$(H^{\dagger}D^{\mu}H)^{\dagger}(H^{\dagger}D_{\mu}H)$\tabularnewline
\cline{1-2} \cline{2-2} 
\multicolumn{2}{|c|}{{\large{}$\boldsymbol{X^{2}D^{2}}$}} & \textcolor{blue}{\cellcolor[gray]{0.92}}$\mathcal{O}_{HB}$  & \textcolor{blue}{\cellcolor[gray]{0.92}}$B_{\mu\nu}B^{\mu\nu}(H^{\dagger}H)$  & $\mathcal{O}_{HD}^{\prime}$  & $(H^{\dagger}H)(D_{\mu}H)^{\dagger}(D^{\mu}H)$\tabularnewline
\cline{1-2} \cline{2-2} 
$\mathcal{O}_{2G}$  & $-\frac{1}{2}(D_{\mu}G^{A\mu\nu})(D^{\rho}G_{\rho\nu}^{A})$  & \textcolor{blue}{\cellcolor[gray]{0.92}}$\mathcal{O}_{H\widetilde{B}}$  & \textcolor{blue}{\cellcolor[gray]{0.92}}$\widetilde{B}_{\mu\nu}B^{\mu\nu}(H^{\dagger}H)$  & $\mathcal{O}_{HD}^{\prime\prime}$  & $(H^{\dagger}H)D_{\mu}(H^{\dagger}i\overleftrightarrow{D}^{\mu}H)$\tabularnewline
\cline{5-6} \cline{6-6} 
$\mathcal{O}_{2W}$  & $-\frac{1}{2}(D_{\mu}W^{I\mu\nu})(D^{\rho}W_{\rho\nu}^{I})$  & \textcolor{blue}{\cellcolor[gray]{0.92}}$\mathcal{O}_{HWB}$  & \textcolor{blue}{\cellcolor[gray]{0.92}}$W_{\mu\nu}^{I}B^{\mu\nu}(H^{\dagger}\sigma^{I}H)$  & \multicolumn{2}{c|}{{\large{}$\boldsymbol{H^{6}}$}}\tabularnewline
\cline{5-6} \cline{6-6} 
$\mathcal{O}_{2B}$  & $-\frac{1}{2}(\partial_{\mu}B^{\mu\nu})(\partial^{\rho}B_{\rho\nu})$  & \textcolor{blue}{\cellcolor[gray]{0.92}}$\mathcal{O}_{H\widetilde{W}B}$  & \textcolor{blue}{\cellcolor[gray]{0.92}}$\widetilde{W}_{\mu\nu}^{I}B^{\mu\nu}(H^{\dagger}\sigma^{I}H)$  & \textcolor{blue}{\cellcolor[gray]{0.92}}$\mathcal{O}_{H}$  & \textcolor{blue}{\cellcolor[gray]{0.92}}$(H^{\dagger}H)^{3}$\tabularnewline
\cline{3-4} \cline{4-4} 
 &  & \multicolumn{2}{c|}{{\large{}$\boldsymbol{H^{2}XD^{2}}$}} &  & \tabularnewline
\cline{3-4} \cline{4-4} 
 &  & $\mathcal{O}_{WDH}$  & $D_{\nu}W^{I\mu\nu}(H^{\dagger}i\overleftrightarrow{D}_{\mu}^{I}H)$  &  & \tabularnewline
 &  & $\mathcal{O}_{BDH}$  & $\partial_{\nu}B^{\mu\nu}(H^{\dagger}i\overleftrightarrow{D}_{\mu}H)$  &  & \tabularnewline
\hline 
\end{tabular}
\par\end{centering}
\caption{\label{tab:GreenBosonic} Bosonic operators in the Green's basis.
Shaded ones are also included in Warsaw basis.}
\end{table}

\begin{table}[p]
\begin{centering}
\begin{tabular}{|c|c|c|c|c|c|}
\hline 
\multicolumn{2}{|c|}{{\large{}$\boldsymbol{\psi^{2}D^{3}}$}} & \multicolumn{2}{c|}{{\large{}$\boldsymbol{\psi^{2}XD}$}} & \multicolumn{2}{c|}{{\large{}$\boldsymbol{\psi^{2}DH^{2}}$}}\tabularnewline
\hline 
$\mathcal{O}_{qD}$  & $\frac{i}{2}\overline{q}\left\{ D_{\mu}D^{\mu},\slashed D\right\} q$  & $\mathcal{O}_{Gq}$  & $(\overline{q}T^{A}\gamma^{\mu}q)D^{\nu}G_{\mu\nu}^{A}$  & \textcolor{blue}{\cellcolor[gray]{0.92}}$\mathcal{O}_{Hq}^{(1)}$  & \textcolor{blue}{\cellcolor[gray]{0.92}}$(\overline{q}\gamma^{\mu}q)(H^{\dagger}i\overleftrightarrow{D}_{\mu}H)$\tabularnewline
$\mathcal{O}_{uD}$  & $\frac{i}{2}\overline{u}\left\{ D_{\mu}D^{\mu},\slashed D\right\} u$  & $\mathcal{O}_{Gq}^{\prime}$  & $\frac{1}{2}(\overline{q}T^{A}\gamma^{\mu}i\overleftrightarrow{D}^{\nu}q)G_{\mu\nu}^{A}$  & $\mathcal{O}_{Hq}^{\prime(1)}$  & $(\overline{q}i\overleftrightarrow{\slashed D}q)(H^{\dagger}H)$\tabularnewline
$\mathcal{O}_{dD}$  & $\frac{i}{2}\overline{d}\left\{ D_{\mu}D^{\mu},\slashed D\right\} d$  & $\mathcal{O}_{\widetilde{G}q}^{\prime}$  & $\frac{1}{2}(\overline{q}T^{A}\gamma^{\mu}i\overleftrightarrow{D}^{\nu}q)\widetilde{G}_{\mu\nu}^{A}$  & $\mathcal{O}_{Hq}^{\prime\prime(1)}$  & $(\overline{q}\gamma^{\mu}q)\partial_{\mu}(H^{\dagger}H)$\tabularnewline
$\mathcal{O}_{\ell D}$  & $\frac{i}{2}\overline{\ell}\left\{ D_{\mu}D^{\mu},\slashed D\right\} \ell$  & $\mathcal{O}_{Wq}$  & $(\overline{q}\sigma^{I}\gamma^{\mu}q)D^{\nu}W_{\mu\nu}^{I}$  & \textcolor{blue}{\cellcolor[gray]{0.92}}$\mathcal{O}_{Hq}^{(3)}$  & \textcolor{blue}{\cellcolor[gray]{0.92}}$(\overline{q}\sigma^{I}\gamma^{\mu}q)(H^{\dagger}i\overleftrightarrow{D}_{\mu}^{I}H)$\tabularnewline
$\mathcal{O}_{eD}$  & $\frac{i}{2}\overline{e}\left\{ D_{\mu}D^{\mu},\slashed D\right\} e$  & $\mathcal{O}_{Wq}^{\prime}$  & $\frac{1}{2}(\overline{q}\sigma^{I}\gamma^{\mu}i\overleftrightarrow{D}^{\nu}q)W_{\mu\nu}^{I}$  & $\mathcal{O}_{Hq}^{\prime(3)}$  & $(\overline{q}i\overleftrightarrow{\slashed D}^{I}q)(H^{\dagger}\sigma^{I}H)$\tabularnewline
\cline{1-2} \cline{2-2} 
\multicolumn{2}{|c|}{{\large{}$\boldsymbol{\psi^{2}HD^{2}}+\text{\textbf{h.c.}}$}} & $\mathcal{O}_{\widetilde{W}q}^{\prime}$  & $\frac{1}{2}(\overline{q}\sigma^{I}\gamma^{\mu}i\overleftrightarrow{D}^{\nu}q)\widetilde{W}_{\mu\nu}^{I}$  & $\mathcal{O}_{Hq}^{\prime\prime(3)}$  & $(\overline{q}\sigma^{I}\gamma^{\mu}q)D_{\mu}(H^{\dagger}\sigma^{I}H)$\tabularnewline
\cline{1-2} \cline{2-2} 
$\mathcal{O}_{uHD1}$  & $(\overline{q}u)D_{\mu}D^{\mu}\widetilde{H}$  & $\mathcal{O}_{Bq}$  & $(\overline{q}\gamma^{\mu}q)\partial^{\nu}B_{\mu\nu}$  & \textcolor{blue}{\cellcolor[gray]{0.92}}$\mathcal{O}_{Hu}$  & \textcolor{blue}{\cellcolor[gray]{0.92}}$(\overline{u}\gamma^{\mu}u)(H^{\dagger}i\overleftrightarrow{D}_{\mu}H)$\tabularnewline
$\mathcal{O}_{uHD2}$  & $(\overline{q}i\sigma_{\mu\nu}D^{\mu}u)D^{\nu}\widetilde{H}$  & $\mathcal{O}_{Bq}^{\prime}$  & $\frac{1}{2}(\overline{q}\gamma^{\mu}i\overleftrightarrow{D}^{\nu}q)B_{\mu\nu}$  & $\mathcal{O}_{Hu}^{\prime}$  & $(\overline{u}i\overleftrightarrow{\slashed D}u)(H^{\dagger}H)$\tabularnewline
$\mathcal{O}_{uHD3}$  & $(\overline{q}D_{\mu}D^{\mu}u)\widetilde{H}$  & $\mathcal{O}_{\widetilde{B}q}^{\prime}$  & $\frac{1}{2}(\overline{q}\gamma^{\mu}i\overleftrightarrow{D}^{\nu}q)\widetilde{B}_{\mu\nu}$  & $\mathcal{O}_{Hu}^{\prime\prime}$  & $(\overline{u}\gamma^{\mu}u)\partial_{\mu}(H^{\dagger}H)$\tabularnewline
$\mathcal{O}_{uHD4}$  & $(\overline{q}D_{\mu}u)D^{\mu}\widetilde{H}$  & $\mathcal{O}_{Gu}$  & $(\overline{u}T^{A}\gamma^{\mu}u)D^{\nu}G_{\mu\nu}^{A}$  & \textcolor{blue}{\cellcolor[gray]{0.92}}$\mathcal{O}_{Hd}$  & \textcolor{blue}{\cellcolor[gray]{0.92}}$(\overline{d}\gamma^{\mu}d)(H^{\dagger}i\overleftrightarrow{D}_{\mu}H)$\tabularnewline
$\mathcal{O}_{dHD1}$  & $(\overline{q}d)D_{\mu}D^{\mu}H$  & $\mathcal{O}_{Gu}^{\prime}$  & $\frac{1}{2}(\overline{u}T^{A}\gamma^{\mu}i\overleftrightarrow{D}^{\nu}u)G_{\mu\nu}^{A}$  & $\mathcal{O}_{Hd}^{\prime}$  & $(\overline{d}i\overleftrightarrow{\slashed D}d)(H^{\dagger}H)$\tabularnewline
$\mathcal{O}_{dHD2}$  & $(\overline{q}i\sigma_{\mu\nu}D^{\mu}d)D^{\nu}H$  & $\mathcal{O}_{\widetilde{G}u}^{\prime}$  & $\frac{1}{2}(\overline{u}T^{A}\gamma^{\mu}i\overleftrightarrow{D}^{\nu}u)\widetilde{G}_{\mu\nu}^{A}$  & $\mathcal{O}_{Hd}^{\prime\prime}$  & $(\overline{d}\gamma^{\mu}d)\partial_{\mu}(H^{\dagger}H)$\tabularnewline
$\mathcal{O}_{dHD3}$  & $(\overline{q}D_{\mu}D^{\mu}d)H$  & $\mathcal{O}_{Bu}$  & $(\overline{u}\gamma^{\mu}u)\partial^{\nu}B_{\mu\nu}$  & \textcolor{blue}{\cellcolor[gray]{0.92}}$\mathcal{O}_{Hud}$  & \textcolor{blue}{\cellcolor[gray]{0.92}}$(\overline{u}\gamma^{\mu}d)(\widetilde{H}^{\dagger}iD_{\mu}H)$\tabularnewline
$\mathcal{O}_{dHD4}$  & $(\overline{q}D_{\mu}d)D^{\mu}H$  & $\mathcal{O}_{Bu}^{\prime}$  & $\frac{1}{2}(\overline{u}\gamma^{\mu}i\overleftrightarrow{D}^{\nu}u)B_{\mu\nu}$  & \textcolor{blue}{\cellcolor[gray]{0.92}}$\mathcal{O}_{H\ell}^{(1)}$  & \textcolor{blue}{\cellcolor[gray]{0.92}}$(\overline{\ell}\gamma^{\mu}\ell)(H^{\dagger}i\overleftrightarrow{D}_{\mu}H)$\tabularnewline
$\mathcal{O}_{eHD1}$  & $(\overline{\ell}e)D_{\mu}D^{\mu}H$  & $\mathcal{O}_{\widetilde{B}u}^{\prime}$  & $\frac{1}{2}(\overline{u}\gamma^{\mu}i\overleftrightarrow{D}^{\nu}u)\widetilde{B}_{\mu\nu}$  & $\mathcal{O}_{H\ell}^{\prime(1)}$  & $(\overline{\ell}i\overleftrightarrow{\slashed D}\ell)(H^{\dagger}H)$\tabularnewline
$\mathcal{O}_{eHD2}$  & $(\overline{\ell}i\sigma_{\mu\nu}D^{\mu}e)D^{\nu}H$  & $\mathcal{O}_{Gd}$  & $(\overline{d}T^{A}\gamma^{\mu}d)D^{\nu}G_{\mu\nu}^{A}$  & $\mathcal{O}_{H\ell}^{\prime\prime(1)}$  & $(\overline{\ell}\gamma^{\mu}\ell)\partial_{\mu}(H^{\dagger}H)$\tabularnewline
$\mathcal{O}_{eHD3}$  & $(\overline{\ell}D_{\mu}D^{\mu}e)H$  & $\mathcal{O}_{Gd}^{\prime}$  & $\frac{1}{2}(\overline{d}T^{A}\gamma^{\mu}i\overleftrightarrow{D}^{\nu}d)G_{\mu\nu}^{A}$  & \textcolor{blue}{\cellcolor[gray]{0.92}}$\mathcal{O}_{H\ell}^{(3)}$  & \textcolor{blue}{\cellcolor[gray]{0.92}}$(\overline{\ell}\sigma^{I}\gamma^{\mu}\ell)(H^{\dagger}i\overleftrightarrow{D}_{\mu}^{I}H)$\tabularnewline
$\mathcal{O}_{eHD4}$  & $(\overline{\ell}D_{\mu}e)D^{\mu}H$  & $\mathcal{O}_{\widetilde{G}d}^{\prime}$  & $\frac{1}{2}(\overline{d}T^{A}\gamma^{\mu}i\overleftrightarrow{D}^{\nu}d)\widetilde{G}_{\mu\nu}^{A}$  & $\mathcal{O}_{H\ell}^{\prime(3)}$  & $(\overline{\ell}i\overleftrightarrow{\slashed D}^{I}\ell)(H^{\dagger}\sigma^{I}H)$\tabularnewline
\cline{1-2} \cline{2-2} 
\multicolumn{2}{|c|}{{\large{}$\boldsymbol{\psi^{2}XH}+\text{\textbf{h.c.}}$}} & $\mathcal{O}_{Bd}$  & $(\overline{d}\gamma^{\mu}d)\partial^{\nu}B_{\mu\nu}$  & $\mathcal{O}_{H\ell}^{\prime\prime(3)}$  & $(\overline{\ell}\sigma^{I}\gamma^{\mu}\ell)D_{\mu}(H^{\dagger}\sigma^{I}H)$\tabularnewline
\cline{1-2} \cline{2-2} 
\textcolor{blue}{\cellcolor[gray]{0.92}}$\mathcal{O}_{uG}$  & \textcolor{blue}{\cellcolor[gray]{0.92}}$(\overline{q}T^{A}\sigma^{\mu\nu}u)\widetilde{H}G_{\mu\nu}^{A}$  & $\mathcal{O}_{Bd}^{\prime}$  & $\frac{1}{2}(\overline{d}\gamma^{\mu}i\overleftrightarrow{D}^{\nu}d)B_{\mu\nu}$  & \textcolor{blue}{\cellcolor[gray]{0.92}}$\mathcal{O}_{He}$  & \textcolor{blue}{\cellcolor[gray]{0.92}}$(\overline{e}\gamma^{\mu}e)(H^{\dagger}i\overleftrightarrow{D}_{\mu}H)$\tabularnewline
\textcolor{blue}{\cellcolor[gray]{0.92}}$\mathcal{O}_{uW}$  & \textcolor{blue}{\cellcolor[gray]{0.92}}$(\overline{q}\sigma^{\mu\nu}u)\sigma^{I}\widetilde{H}W_{\mu\nu}^{I}$  & $\mathcal{O}_{\widetilde{B}d}^{\prime}$  & $\frac{1}{2}(\overline{d}\gamma^{\mu}i\overleftrightarrow{D}^{\nu}d)\widetilde{B}_{\mu\nu}$  & $\mathcal{O}_{He}^{\prime}$  & $(\overline{e}i\overleftrightarrow{\slashed D}e)(H^{\dagger}H)$\tabularnewline
\textcolor{blue}{\cellcolor[gray]{0.92}}$\mathcal{O}_{uB}$  & \textcolor{blue}{\cellcolor[gray]{0.92}}$(\overline{q}\sigma^{\mu\nu}u)\widetilde{H}B_{\mu\nu}$  & $\mathcal{O}_{W\ell}$  & $(\overline{\ell}\sigma^{I}\gamma^{\mu}\ell)D^{\nu}W_{\mu\nu}^{I}$  & $\mathcal{O}_{He}^{\prime\prime}$  & $(\overline{e}\gamma^{\mu}e)\partial_{\mu}(H^{\dagger}H)$\tabularnewline
\cline{5-6} \cline{6-6} 
\textcolor{blue}{\cellcolor[gray]{0.92}}$\mathcal{O}_{dG}$  & \textcolor{blue}{\cellcolor[gray]{0.92}}$(\overline{q}T^{A}\sigma^{\mu\nu}d)HG_{\mu\nu}^{A}$  & $\mathcal{O}_{W\ell}^{\prime}$  & $\frac{1}{2}(\overline{\ell}\sigma^{I}\gamma^{\mu}i\overleftrightarrow{D}^{\nu}\ell)W_{\mu\nu}^{I}$  & \multicolumn{2}{c|}{{\large{}$\boldsymbol{\psi^{2}H^{3}}+\text{\textbf{h.c.}}$}}\tabularnewline
\cline{5-6} \cline{6-6} 
\textcolor{blue}{\cellcolor[gray]{0.92}}$\mathcal{O}_{dW}$  & \textcolor{blue}{\cellcolor[gray]{0.92}}$(\overline{q}\sigma^{\mu\nu}d)\sigma^{I}HW_{\mu\nu}^{I}$  & $\mathcal{O}_{\widetilde{W}\ell}^{\prime}$  & $\frac{1}{2}(\overline{\ell}\sigma^{I}\gamma^{\mu}i\overleftrightarrow{D}^{\nu}\ell)\widetilde{W}_{\mu\nu}^{I}$  & \textcolor{blue}{\cellcolor[gray]{0.92}}$\mathcal{O}_{uH}$  & \textcolor{blue}{\cellcolor[gray]{0.92}}$(H^{\dagger}H)\overline{q}\widetilde{H}u$\tabularnewline
\textcolor{blue}{\cellcolor[gray]{0.92}}$\mathcal{O}_{dB}$  & \textcolor{blue}{\cellcolor[gray]{0.92}}$(\overline{q}\sigma^{\mu\nu}d)HB_{\mu\nu}$  & $\mathcal{O}_{B\ell}$  & $(\overline{\ell}\gamma^{\mu}\ell)\partial^{\nu}B_{\mu\nu}$  & \textcolor{blue}{\cellcolor[gray]{0.92}}$\mathcal{O}_{dH}$  & \textcolor{blue}{\cellcolor[gray]{0.92}}$(H^{\dagger}H)\overline{q}Hd$\tabularnewline
\textcolor{blue}{\cellcolor[gray]{0.92}}$\mathcal{O}_{eW}$  & \textcolor{blue}{\cellcolor[gray]{0.92}}$(\overline{\ell}\sigma^{\mu\nu}e)\sigma^{I}HW_{\mu\nu}^{I}$  & $\mathcal{O}_{B\ell}^{\prime}$  & $\frac{1}{2}(\overline{\ell}\gamma^{\mu}i\overleftrightarrow{D}^{\nu}\ell)B_{\mu\nu}$  & \textcolor{blue}{\cellcolor[gray]{0.92}}$\mathcal{O}_{eH}$  & \textcolor{blue}{\cellcolor[gray]{0.92}}$(H^{\dagger}H)\overline{\ell}He$\tabularnewline
\textcolor{blue}{\cellcolor[gray]{0.92}}$\mathcal{O}_{eB}$  & \textcolor{blue}{\cellcolor[gray]{0.92}}$(\overline{\ell}\sigma^{\mu\nu}e)HB_{\mu\nu}$  & $\mathcal{O}_{\widetilde{B}\ell}^{\prime}$  & $\frac{1}{2}(\overline{\ell}\gamma^{\mu}i\overleftrightarrow{D}^{\nu}\ell)\widetilde{B}_{\mu\nu}$  &  & \tabularnewline
 &  & $\mathcal{O}_{Be}$  & $(\overline{e}\gamma^{\mu}e)\partial^{\nu}B_{\mu\nu}$  &  & \tabularnewline
 &  & $\mathcal{O}_{Be}^{\prime}$  & $\frac{1}{2}(\overline{e}\gamma^{\mu}i\overleftrightarrow{D}^{\nu}e)B_{\mu\nu}$  &  & \tabularnewline
 &  & $\mathcal{O}_{\widetilde{B}e}^{\prime}$  & $\frac{1}{2}(\overline{e}\gamma^{\mu}i\overleftrightarrow{D}^{\nu}e)\widetilde{B}_{\mu\nu}$  &  & \tabularnewline
\hline 
\end{tabular}
\par\end{centering}
\caption{\label{tab:GreenSingleFermion} Two-fermion operators in the Green's
basis. Shaded ones are also included in Warsaw basis. Fermion family
indices are omitted.}
\end{table}

\pagebreak{} 
\begin{table}[t]
\begin{centering}
\begin{tabular}{|c|c|c|c|c|c|}
\hline 
\multicolumn{2}{|c|}{{\large{}Four-quark}} & \multicolumn{2}{c|}{{\large{}Four-lepton}} & \multicolumn{2}{c|}{{\large{}Semileptonic}}\tabularnewline
\hline 
\textcolor{blue}{\cellcolor[gray]{0.92}}$\mathcal{O}_{qq}^{(1)}$  & \textcolor{blue}{\cellcolor[gray]{0.92}}$(\overline{q}\gamma^{\mu}q)(\overline{q}\gamma_{\mu}q)$  & \textcolor{blue}{\cellcolor[gray]{0.92}}$\mathcal{O}_{\ell\ell}$  & \textcolor{blue}{\cellcolor[gray]{0.92}}$(\overline{\ell}\gamma^{\mu}\ell)(\overline{\ell}\gamma_{\mu}\ell)$  & \textcolor{blue}{\cellcolor[gray]{0.92}}$\mathcal{O}_{\ell q}^{(1)}$  & \textcolor{blue}{\cellcolor[gray]{0.92}}$(\overline{\ell}\gamma^{\mu}\ell)(\overline{q}\gamma_{\mu}q)$\tabularnewline
\textcolor{blue}{\cellcolor[gray]{0.92}}$\mathcal{O}_{qq}^{(3)}$  & \textcolor{blue}{\cellcolor[gray]{0.92}}$(\overline{q}\gamma^{\mu}\sigma^{I}q)(\overline{q}\gamma_{\mu}\sigma^{I}q)$  & \textcolor{blue}{\cellcolor[gray]{0.92}}$\mathcal{O}_{ee}$  & \textcolor{blue}{\cellcolor[gray]{0.92}}$(\overline{e}\gamma^{\mu}e)(\overline{e}\gamma_{\mu}e)$  & \textcolor{blue}{\cellcolor[gray]{0.92}}$\mathcal{O}_{\ell q}^{(3)}$  & \textcolor{blue}{\cellcolor[gray]{0.92}}$(\overline{\ell}\gamma^{\mu}\sigma^{I}\ell)(\overline{q}\gamma_{\mu}\sigma^{I}q)$\tabularnewline
\textcolor{blue}{\cellcolor[gray]{0.92}}$\mathcal{O}_{uu}$  & \textcolor{blue}{\cellcolor[gray]{0.92}}$(\overline{u}\gamma^{\mu}u)(\overline{u}\gamma_{\mu}u)$  & \textcolor{blue}{\cellcolor[gray]{0.92}}$\mathcal{O}_{\ell e}$  & \textcolor{blue}{\cellcolor[gray]{0.92}}$(\overline{\ell}\gamma^{\mu}\ell)(\overline{e}\gamma_{\mu}e)$  & \textcolor{blue}{\cellcolor[gray]{0.92}}$\mathcal{O}_{eu}$  & \textcolor{blue}{\cellcolor[gray]{0.92}}$(\overline{e}\gamma^{\mu}e)(\overline{u}\gamma_{\mu}u)$\tabularnewline
\textcolor{blue}{\cellcolor[gray]{0.92}}$\mathcal{O}_{dd}$  & \textcolor{blue}{\cellcolor[gray]{0.92}}$(\overline{d}\gamma^{\mu}d)(\overline{d}\gamma_{\mu}d)$  &  &  & \textcolor{blue}{\cellcolor[gray]{0.92}}$\mathcal{O}_{ed}$  & \textcolor{blue}{\cellcolor[gray]{0.92}}$(\overline{e}\gamma^{\mu}e)(\overline{d}\gamma_{\mu}d)$\tabularnewline
\textcolor{blue}{\cellcolor[gray]{0.92}}$\mathcal{O}_{ud}^{(1)}$  & \textcolor{blue}{\cellcolor[gray]{0.92}}$(\overline{u}\gamma^{\mu}u)(\overline{d}\gamma_{\mu}d)$  &  &  & \textcolor{blue}{\cellcolor[gray]{0.92}}$\mathcal{O}_{qe}$  & \textcolor{blue}{\cellcolor[gray]{0.92}}$(\overline{q}\gamma^{\mu}q)(\overline{e}\gamma_{\mu}e)$\tabularnewline
\textcolor{blue}{\cellcolor[gray]{0.92}}$\mathcal{O}_{ud}^{(8)}$  & \textcolor{blue}{\cellcolor[gray]{0.92}}$(\overline{u}\gamma^{\mu}T^{A}u)(\overline{d}\gamma_{\mu}T^{A}d)$  &  &  & \textcolor{blue}{\cellcolor[gray]{0.92}}$\mathcal{O}_{\ell u}$  & \textcolor{blue}{\cellcolor[gray]{0.92}}$(\overline{\ell}\gamma^{\mu}\ell)(\overline{u}\gamma_{\mu}u)$\tabularnewline
\textcolor{blue}{\cellcolor[gray]{0.92}}$\mathcal{O}_{qu}^{(1)}$  & \textcolor{blue}{\cellcolor[gray]{0.92}}$(\overline{q}\gamma^{\mu}q)(\overline{u}\gamma_{\mu}u)$  &  &  & \textcolor{blue}{\cellcolor[gray]{0.92}}$\mathcal{O}_{\ell d}$  & \textcolor{blue}{\cellcolor[gray]{0.92}}$(\overline{\ell}\gamma^{\mu}\ell)(\overline{d}\gamma_{\mu}d)$\tabularnewline
\textcolor{blue}{\cellcolor[gray]{0.92}}$\mathcal{O}_{qu}^{(8)}$  & \textcolor{blue}{\cellcolor[gray]{0.92}}$(\overline{q}\gamma^{\mu}T^{A}q)(\overline{u}\gamma_{\mu}T^{A}u)$  &  &  & \textcolor{blue}{\cellcolor[gray]{0.92}}$\mathcal{O}_{\ell edq}$  & \textcolor{blue}{\cellcolor[gray]{0.92}}$(\overline{\ell}e)(\overline{d}q)$\tabularnewline
\textcolor{blue}{\cellcolor[gray]{0.92}}$\mathcal{O}_{qd}^{(1)}$  & \textcolor{blue}{\cellcolor[gray]{0.92}}$(\overline{q}\gamma^{\mu}q)(\overline{d}\gamma_{\mu}d)$  &  &  & \textcolor{blue}{\cellcolor[gray]{0.92}}$\mathcal{O}_{\ell equ}^{(1)}$  & \textcolor{blue}{\cellcolor[gray]{0.92}}$(\overline{\ell}^{r}e)\epsilon_{rs}(\overline{q}^{s}u)$\tabularnewline
\textcolor{blue}{\cellcolor[gray]{0.92}}$\mathcal{O}_{qd}^{(8)}$  & \textcolor{blue}{\cellcolor[gray]{0.92}}$(\overline{q}\gamma^{\mu}T^{A}q)(\overline{d}\gamma_{\mu}T^{A}d)$  &  &  & \textcolor{blue}{\cellcolor[gray]{0.92}}$\mathcal{O}_{\ell equ}^{(3)}$  & \textcolor{blue}{\cellcolor[gray]{0.92}}$(\overline{\ell}^{r}\sigma^{\mu\nu}e)\epsilon_{rs}(\overline{q}^{s}\sigma_{\mu\nu}u)$\tabularnewline
\textcolor{blue}{\cellcolor[gray]{0.92}}$\mathcal{O}_{quqd}^{(1)}$  & \textcolor{blue}{\cellcolor[gray]{0.92}}$(\overline{q}^{r}u)\epsilon_{rs}(\overline{q}^{s}d)$  &  &  &  & \tabularnewline
\textcolor{blue}{\cellcolor[gray]{0.92}}$\mathcal{O}_{quqd}^{(8)}$  & \textcolor{blue}{\cellcolor[gray]{0.92}}$(\overline{q}^{r}T^{A}u)\epsilon_{rs}(\overline{q}^{s}T^{A}d)$  &  &  &  & \tabularnewline
\hline 
\end{tabular}
\par\end{centering}
\caption{\label{tab:4FermionBconserving} Baryon and lepton number conserving
four-fermion operators. All operators are included in Warsaw basis.
Fermion family indices are omitted. Indices $r,\,s,\,p,\,t,\,\dots$
denote the $\text{SU}(2)_{L}$ fundamental representations.}
\end{table}

\begin{table}[b]
\begin{centering}
\begin{tabular}{|c|c|}
\hline 
\multicolumn{2}{|c|}{{\large{}$B$ and $L$ violating}}\tabularnewline
\hline 
\textcolor{blue}{\cellcolor[gray]{0.92}}$\mathcal{O}_{duq}$  & \textcolor{blue}{\cellcolor[gray]{0.92}}$\varepsilon_{abc}\epsilon_{rs}\left[(d^{a})^{T}Cu^{b}\right]\left[(q^{cr})^{T}C\ell^{s}\right]$\tabularnewline
\textcolor{blue}{\cellcolor[gray]{0.92}}$\mathcal{O}_{qqu}$  & \textcolor{blue}{\cellcolor[gray]{0.92}}$\varepsilon_{abc}\epsilon_{rs}\left[(q^{ar})^{T}Cq^{bs}\right]\left[(u^{c})^{T}Ce\right]$\tabularnewline
\textcolor{blue}{\cellcolor[gray]{0.92}}$\mathcal{O}_{qqq}$  & \textcolor{blue}{\cellcolor[gray]{0.92}}$\varepsilon_{abc}\epsilon_{rs}\epsilon_{pt}\left[(q^{ar})^{T}Cq^{bs}\right]\left[(q^{cp})^{T}C\ell^{t}\right]$\tabularnewline
\textcolor{blue}{\cellcolor[gray]{0.92}}$\mathcal{O}_{duu}$  & \textcolor{blue}{\cellcolor[gray]{0.92}}$\varepsilon_{abc}\left[(d^{a})^{T}Cu^{b}\right]\left[(u^{c})^{T}Ce\right]$\tabularnewline
\hline 
\end{tabular}
\par\end{centering}
\caption{\label{tab:4FermionBviolating}Baryon and lepton number violating
four-fermion operators. All operators are included in Warsaw basis.
Fermion family indices are omitted. Indices $r,\,s,\,p,\,t,\,\dots$
and $a,\,b,\,c,\,\dots$ denote the $\text{SU}(2)_{L}$ and $\text{SU}(3)_{c}$
fundamental representations, respectively. $C$ is the Dirac charge
conjugation matrix.}
\end{table}

\end{document}